\documentclass[namedreferences]{SolarPhysics}
\usepackage[optionalrh,solaenum,natbib]{spr-sola-addons} 
\usepackage{graphicx}                    
\usepackage{color}                       
\usepackage{url}                         

\graphicspath{{./fig/}{./png/}}

\newcommand{\EQ}{\begin{equation}}
\newcommand{\EN}{\end{equation}}
\newcommand{\EQA}{\begin{eqnarray}}
\newcommand{\ENA}{\end{eqnarray}}
\newcommand{\eq}[1]{(\ref{#1})}

\newcommand{\Eq}[1]{Equation~(\ref{#1})}

\newcommand{\Sec}[1]{Section~\ref{#1}}

\newcommand{\Fig}[1]{Figure~\ref{#1}}

\newcommand{\Figss}[2]{Figures~\ref{#1}--\ref{#2}}
\newcommand{\Tab}[1]{Table~\ref{#1}}

\newcommand{\bra}[1]{\langle #1\rangle}

\newcommand{\meanrho}{\overline{\rho}}

{}
{}
{}

{}
{}
{}
{}
{}
{}
{}
{}
{}
{}
{}
{}
{}
{}
{}
{}
{}
{}

{}
{}
{}
\newcommand{\meanA}{\overline{A}}
\newcommand{\meanB}{\overline{B}}

\newcommand{\meanT}{\overline{T}}

{}

{}
{}

\newcommand{\Rsun}{R}
\newcommand{\Rc}{R_c}

\newcommand{\nablad}{\nabla_{\rm ad}}
\newcommand{\gggg}{\mbox{\boldmath $g$} {}}

\newcommand{\rr}{\mbox{\boldmath $r$} {}}

\newcommand{\uu}{\mbox{\boldmath $u$} {}}
\newcommand{\UU}{\mbox{\boldmath $U$} {}}

\newcommand{\BB}{\mbox{\boldmath $B$} {}}
\newcommand{\DA}{\mbox{\boldmath $D$} {}}

\newcommand{\JJ}{\mbox{\boldmath $J$} {}}

\newcommand{\AAA}{\mbox{\boldmath $A$} {}}

\newcommand{\nab}{\mbox{\boldmath $\nabla$} {}}
\newcommand{\OO}{\mbox{\boldmath $\Omega$} {}}
\newcommand{\oo}{\mbox{\boldmath $\omega$} {}}

\newcommand{\SSSS}{\mbox{\boldmath ${\sf S}$} {}}

\newcommand{\DD}{{\rm D} {}}

\newcommand{\dd}{{\rm d} {}}

\newcommand{\const}{{\rm const}  {}}

\def\degr{\hbox{$^\circ$}}

\def\Co{\mbox{\rm Co}}

\def\Pm{\mbox{\rm Pr}_M}
\def\Rm{\mbox{\rm Re}_M}

\def\Rey{\mbox{\rm Re}}

\def\Co{\mbox{\rm Co}}

\def\cs{c_{\rm s}}

\def\kf{k_{\rm f}}

\def\Brms{B_{\rm rms}}

\def\urms{u_{\rm rms}}

\def\rhob{\rho_{\rm b}}
\def\rhot{\rho_{\rm t}}

\def\Beq{B_{\rm eq}}

\def\half{{\textstyle{1\over2}}}

\def\onethird{{\textstyle{1\over3}}}

\newcommand{\etal}{{\em et al.}}
\begin{document}

\begin{article}

\begin{opening}

  \title{Ejections of magnetic structures above a spherical wedge
    driven by a convective dynamo with differential rotation}
%
\author{J\"orn~Warnecke{}$^{1,2}$\sep Petri~J.~K\"apyl\"a{}$^{1,3}$\sep
  Maarit~J.~Mantere{}$^{3}$\sep Axel~Brandenburg$^{1,2}$}
%
\runningauthor{J. Warnecke {\it et al.}}
\runningtitle{Ejections of magnetic structures}
%
  \institute{$^{1}$ Nordita, KTH Royal Institute of Technology and Stockholm University,
    Roslagstullsbacken 23, SE-10691 Stockholm, Sweden, email: joern@nordita.org \\
   $^{2}$ Department of Astronomy, Stockholm University, SE-10691
   Stockholm, Sweden\\
   $^{3}$ Department of Physics, Gustaf H\"allstr\"omin katu 2a (P.O.
   BOX 64), FI-00014 Helsinki University, Finland\\
 }
\begin{abstract}
We combine a convectively driven dynamo in a spherical shell with a nearly isothermal
density-stratified cooling layer that mimics some aspects of a stellar corona
to study the emergence and ejections of magnetic field structures.
This approach is an extension of earlier models, where forced
turbulence simulations were employed to generate magnetic fields.
A spherical wedge is used which consists of a convection zone and
an extended coronal region to $\approx1.5$ times the radius
of the sphere.  The wedge contains a quarter of the azimuthal extent
of the sphere and $150\degr$ in latitude.  The magnetic field is
self-consistently generated by the turbulent motions due to convection
beneath the surface.  Magnetic fields are found to emerge at the
surface and are ejected to the coronal part of the domain. These
ejections occur at irregular intervals and are weaker than in earlier
work.
We tentatively associate these events with coronal mass ejections on the Sun, even
though our model of the solar atmosphere is rather simplistic.
\end{abstract}
%
\end{opening}
%
\section{Introduction}

Recent observations of the {\it Solar Dynamics Observatory} \citep[SDO:][]{PTC12} have provided
us with a record of impressive solar eruptions.
These eruptions are mostly associated with coronal mass ejections
(CMEs).
These are events through which the Sun sheds hot plasma and magnetic
fields from the corona into the interplanetary space.
The energy causing such huge eruptions is stored in the magnetic
field and can be released {\it via} reconnection of field lines
\citep{St80,ADK99}.
Some of the CMEs are directed towards the Earth,
hitting its magnetosphere and causing phenomena like aurorae.
Furthermore, encounters with CMEs can cause
sudden outages of GPS signals due to ionospheric scintillation.
The resulting radiation dose from such events poses risks to astronauts.
This is now also of concern to airlines, because the radiation load during
polar flights can reach annual limits, especially for pregnant women.
This leads to great interest of scientists in many fields of
physics.
However, there is an additional motivation which comes along with
space weather effects.
The solar dynamo, which is broadly believed to be responsible for the 
generation of the solar magnetic
field, needs to be sustained by shedding magnetic helicity from the
Sun's interior \citep{BB03}.
Mean-field and direct numerical simulations have shown that the
magnetic field generation is catastrophically quenched at high
magnetic Reynolds numbers in closed systems \citep{VC92}
that do not allow magnetic
helicity fluxes out of the domain \citep{BF00a,BF00b,BS04},
or between different parts of it \citep{BCC09,MCCTB10,HB10}.
The magnetic Reynolds number, which quantifies the relative importance
of advective to
diffusive terms in the induction equation, is known to be very
large in the Sun, therefore implying the possibility of catastrophic quenching
in models of the solar dynamo, unless efficient magnetic helicity fluxes occur,
for example through CMEs \citep{BB03}.
Indeed, CMEs are well known to be closely
associated with magnetic helicity \citep{LO01}.
In particular observations \citep{PVSK00,RAK02} and a recent study by
\cite{TKT11}, where the observations are compared with numerical
models, suggest that CMEs have a twisted magnetic structure,
implying that CMEs transport helicity outwards.

There has been significant progress in the study of CMEs in recent
years.
In addition to improved observations from spacecrafts, \textit{e.g.} SDO or
the {\it Solar TErestical RElation Observatory} \citep[STEREO:][]{KKTD08}), there have also
been major advances in the field of numerical
modeling of CME events \citep{RFG03,AHSGD09}.
However, the formation and the origin of eruptive events like CMEs is
not yet completely understood.
Simulating CMEs and their formation is challenging.
Leaving the difficulties of modeling the interplanetary space aside, a
CME, after being ejected into the chromosphere or the lower corona, travels
over an extended radial distance to the upper corona.
In this environment, density and temperature vary by several
orders of magnitude, which is not easy to handle in numerical
models.
Additionally, the origin of the CMEs is assumed to relate to the magnetic
fields and the velocity pattern at the surface.
However, the surface magnetic and velocity fields are rooted
in the solar convection zone,
where convective motions, in interplay with differential rotation,
generate the magnetic field and the velocity patterns that are
observed at the surface.
The majority of researchers modeling CMEs do not include the
convection zone in their setup, and thereby neglect the effect of the
magnetic and velocity fields being rooted to this layer. 
Often the
initial conditions for the magnetic and velocity fields are prescribed
or taken from 2D observations; see for example \cite{ADK99} and \cite{ALML99}
as well as \cite{TK03}.

Another approach is to study the emergence of flux ropes from the
lower convection zone into the corona. 
In the presence of strong shear,
convection simulations have been showing the formation of flux tubes
\citep{GK11,Nelson11}, but such structures are similar to vortex tubes whose
diameter is known to relate with the visco-resistive scale \citep{BPS95}.
In other approaches flux ropes are inserted in a self-consistent model,
but their origin is left unexplained.  
In several recent papers \citep{MS08,JB09,FMAH10},
the focus lies on the emergence of magnetic flux and the resulting
features in the solar atmosphere.
However, eruptive events have not been investigated with this setup.
In earlier work \citep[][hereafter WB]{WB10} a different approach was developed.
The solar convection zone was combined with a simple model of the solar corona.
The magnetic field, which was here generated by dynamo action beneath the solar
surface, emerged through the surface and was ejected out of the domain.
The focus was on the connection of the dynamo-generated field and
eruptive events like CMEs through the dynamo-generated twist.
WB used a simplified coronal model and drove the dynamo
with forced turbulence.
These simplifications allowed them to study the emergence and a new
mechanism to drive ejections in great detail.
In subsequent work \citep[][hereafter WBM]{WBM11},
the setup of WB was improved by using a
spherical coordinate system and helical forcing with opposite signs in
each hemisphere to mimic the effects of rotation on inhomogeneous 
turbulence.
In addition, WBM included the stratification
resulting from radial gravity for an isothermal fluid.
To improve this model, we now employ convection to generate the 
velocity field. 
In a related approach, \cite{PB11} considered convective overshoot
into the chromosphere and the excitation of gravity waves therein,
but dynamo-generated twist seemed to be unimportant in their work.
The turbulent motions driving the generation of magnetic field are now
self-consistently generated by convective cells operating beneath the
surface.
The setup of the convection zone follows ideas of
\cite{KKB08}, \cite{KKBMT10,KKGBC11} and \cite{KMB11,KMB12}.
There are other approaches simulating convection in hot massive
stars, which have thin subsurface convection zones
\citep{CBBDKL11a}.
But we now use an extended cooling layer to
describe some properties of a solar corona.
The results of this work complement those of earlier work
and can be compared with observations.
The model of the solar atmosphere is still a very simplified one, but
can be regarded as a preliminary step, which will provide
a reference point for improved work in that direction.

\section{The model}
\label{model}
As in WB and WBM, a two-layer model is used, which represents the
convection zone and an extended corona-like layer in one and the
same model.  Our convection zone is similar to those of
\cite{KKBMT10,KKGBC11}.
The domain is a segment of the Sun and is
described in spherical polar coordinates $(r,\theta,\phi)$.  
We model the convection zone starting at radius $r=0.7\,\Rsun$
and the solar corona until $r=\Rc$, where $\Rc=1.5\,\Rsun$
in the present models, where $\Rsun$ corresponds to the solar radius.
In the latitudinal direction, our domain extends in colatitude from
$\theta=15^{\circ}$ to $165^{\circ}$ and in the azimuthal direction
from $\phi=0^{\circ}$ to $90^{\circ}$. We solve the following
equations of compressible magnetohydrodynamics:
\begin{eqnarray}
{\partial\AAA\over\partial t}&=&\UU\times\BB+\eta\nab^2\AAA,\\
{\DD\ln\rho\over \DD t} &=&-\nab\cdot\UU,\\
{\DD\UU\over\DD t}&=&  \gggg - 2\OO_0 \times \UU + {1\over\rho}
\left(\JJ\times\BB - \nab p+\nab\cdot 2\nu\rho\SSSS\right) -\DA(r,\theta,t),\\
T{\DD s\over\DD t}&=&{1\over\rho}\nab\cdot K\nab T +
2\nu\SSSS^2+{\mu_0\eta\over\rho}\JJ^2 - \Gamma_{\rm cool},
\label{entro}
\end{eqnarray}
where the magnetic field is given by $\BB=\nab\times\AAA$ and thus obeys
$\nab\cdot\BB=0$ at all times,
$\mu_0$ is the vacuum permeability,
$\eta$ and $\nu$ are the magnetic diffusivity and kinematic viscosity,
respectively,
$\DD/\DD t =\partial/\partial t+\UU\cdot\nab$ is the
advective time derivative, $\rho$ is the density, and $\UU$ is the
velocity.
The traceless rate-of-strain tensor is given by
\begin{equation}
{\mathsf
  S}_{ij}=\half(U_{i;j}+U_{j;i})-\onethird\delta_{ij}\nab\cdot\UU,
\end{equation}
where semicolons denote covariant differentiation; see \cite{MTBM09}
for details.
$\OO_0 =\Omega_0(\cos\theta,-\sin\theta,0)$ is the rotation vector,
$p$ is the pressure, $K$ is the radiative heat conductivity, and
$\DA(r,\theta,t)$ describes damping in the coronal region; see
\Sec{damp} for details.
The gravitational acceleration is given by
\begin{equation}
\gggg=-GM\rr/r^3,
\end{equation}
where $G$ is Newton's gravitational constant, and $M$ is the
mass of the star. 
The fluid obeys the ideal gas law, $p=(\gamma-1)\rho e$,
where $\gamma=c_p/c_v=5/3$ is the ratio of specific heats at constant
pressure and constant volume, respectively, and $e=c_v T$ is the internal
energy density, which defines the temperature $T$.
The cooling term $\Gamma_{\rm cool}$ will be explained in \Eq{cool}
below in more detail.
\subsection{Initial setup and boundary conditions}

For the thermal stratification in the convection zone, 
we consider a simple analytical setup instead of profiles
from solar structure models as in, e.g., \cite{Br04}.
The hydrodynamic temperature gradient is given by
\begin{equation}
{\partial T\over\partial r}={-|\gggg|\over c_v(\gamma-1)(m+1)},
\label{gradt}
\end{equation}
where $m=m(r)$ is the radially varying polytropic index,
for which we assume a stepwise constant profile.
We also use \Eq{gradt} as the lower boundary condition for the temperature.
This gives the logarithmic temperature gradient $\nabla$
(familiar to those working in stellar physics,
but not to be confused with the operator $\nab$) as
\begin{equation}
\nabla={\partial \ln T\over\partial \ln p}={1\over m+1}.
\end{equation}
The stratification is convectively unstable if $\nabla-\nablad >
0$, where $\nablad=1-1/\gamma$ is the adiabatic temperature gradient,
corresponding to $m<1.5$ for unstable stratification.
We choose $m=1$ in the convectively unstable layer beneath the
surface, $r<\Rsun$.
The region above $r=\Rsun$ is stably stratified and isothermal due 
to a cooling term $\Gamma_{\rm cool}$ with respect to a constant
reference temperature in the entropy equation.  
The density stratification is obtained by requiring the hydrostatic
equilibrium condition to be satisfied.

\begin{figure}[t!]
\begin{center}
\includegraphics[width=0.9\textwidth]{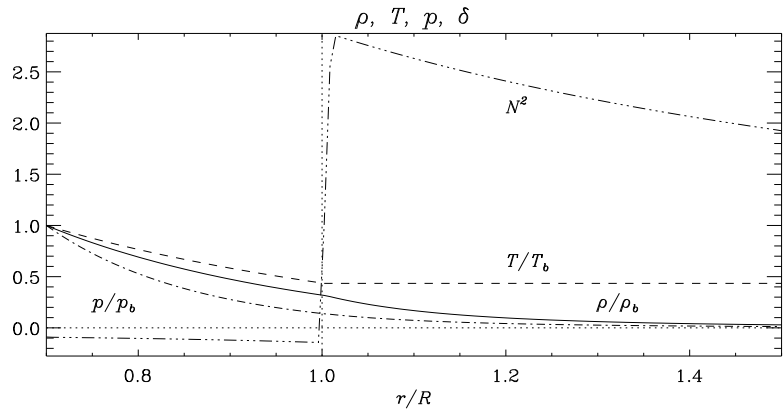}
\end{center}\caption[]{
Initial stratification of temperature (dashed line), density (solid),
pressure (dot-dashed) and the Brunt--V\"ais\"al\"a frequency $N^2=-(|\gggg|/H_p)
(\nabla-\nablad)$ (dash-triple-dotted) for Run~A5. 
The subscripts $b$ refers to the values at $r=0.7\,\Rsun$.
The dotted horizontal (vertical) line denotes the value of zero ($r=\Rsun$).
}
\label{strat}
\end{figure}

The thermal conductivity follows from the constancy of the radial
luminosity profile $L(r)=L_0=\const$ throughout the domain and is given by
\begin{equation}
K = {L_0\over 4\pi r^2\partial T/\partial r}.
\end{equation}
To speed up the thermal relaxation processes, we apply
shallower profiles, corresponding to $\rho\propto T^{1.4}$, for the thermal
variables within the convectively unstable layer.
The value $m=1$ is just used in the convection zone to determine the
thermal conductivity.
In \Fig{strat} we show the initial non-convecting stratification.
The radial temperature gradient at the bottom of the domain is set to a
constant value, which leads to a constant heat flux into the domain.
In the coronal part the gradient goes smoothly to $0$ by using the $r$
dependent cooling function $\Gamma_{\rm cool}$, which is included in the
entropy evolution \eq{entro}. 
The cooling term is given by
\begin{equation}
\Gamma_{\rm cool}=\Gamma_0 f(r)\left({\cs^2-c^2_{s0}\over
    c^2_{s0}}\right),
\label{cool}
\end{equation}
where $f(r)$ is a profile function equal to unity in $r>\Rsun$ and
smoothly connecting to zero in $r\leq\Rsun$, and $\Gamma_0$ is a
cooling luminosity chosen so that the sound speed in the coronal part
relaxes towards $c^2_{\rm s0}\equiv\cs^2(r=\Rc)$.
Whether the stratification is convectively stable or not depends on
the Brunt--V\"ais\"al\"a frequency $N$, defined through
\begin{equation}
N^2=|\gggg|\left({1\over\gamma}{\partial\ln p\over\partial r}
-{\partial\ln\rho\over\partial r}\right)
=-{|\gggg|\over H_p}\left(\nabla-\nablad\right),
\end{equation}
where $H_p=-\partial r/\partial\ln p$ is the pressure scale height. 
If $N^2$ is negative, the stratification is unstable.

We initialize the magnetic field as a weak, random, Gaussian-distributed seed
field in the whole domain.
In the coronal part the magnetic field diffuses after a short time.
We do not use a background coronal field, so the field is
self-consistently generated by the dynamo in the convective layer. 
We apply periodic boundary conditions in the azimuthal direction
over a $90^\circ$ fraction of the full circumference.
For the velocity we take stress-free boundary conditions on all
other boundaries.
As in WBM, the stress-free
boundary conditions prevent mass flux, so no stellar wind is possible.
Because no mass can escape, material will eventually
fall back from the boundary.
Thermodynamic variables have zero
gradients at the latitudinal boundaries.
We employ perfect conductor boundaries for the magnetic field at the
latitudinal and at the lower radial boundaries, and radial field
conditions at the outer radial boundary.
The latter is motivated by the fact that in the Sun,
the solar wind pushes the magnetic
field to open field lines and at a radius of $r=2.0\ldots2.5$ solar radii.
the field lines are mostly radial \citep{LSF82,HWS82}.
This choice has been substantiated by subsequent work of
\cite{WS92} as well as \cite{SDR03}.
While this choice might still be too restrictive for coronal holes and coronal
streamers, and given also that our radial extent in most of the simulations
is smaller than $r=2\Rsun$, 
we nevertheless choose the vertical field boundary condition because
it satisfies
our primary objective of letting magnetic helicity leave the domain,
which is believed to be crucial for the dynamo to operate
at large values of $\Rm$ \citep{BF00a,BF00b,BS04}.
However, we must be aware of the fact
that with this choice our description of the field in the exterior
layer is not a realistic one.

To describe the corona as an isothermal extended cooling
layer is a serious simplification,
in that the temperature inside the coronal layer is not
higher than in the convection zone as in a real stellar corona,
but it stays fixed at the surface value; see \Fig{strat}.
Besides the fact that a simple cooling layer is easy to handle numerically,
we emphasize the importance of facilitating comparison with previous
models of WBM.
It can also be seen as a step towards studying effects that are not
solely due to a low plasma $\beta$ corona,
for which the magnetic pressure, \textit{i.e.} the magnetic field, is
strong compared with the gas pressure
($\beta=2\mu_0 p/\BB^2$).
Indeed, given that our initial field is weak, the plasma $\beta$ is
necessarily large in the outer parts.
We note that it is not even clear whether a hot corona promotes
or hinders coronal ejections.
To understand the formation and evolution of magnetic ejections, 
studies that isolate these effects, such as the present one, may be important.

We use the 
{\sc Pencil Code}\footnote{\texttt{http://pencil-code.googlecode.com}}
with sixth-order centered finite differences in space and 
a third-order accurate Runge--Kutta scheme in time;
see \cite{MTBM09} for the extension of the {\sc Pencil Code} to
spherical coordinates.
We use a grid size of $128 \times 128 \times 64$ mesh points (Runs~A5
and Ar1), and $256 \times 256 \times 128$ (Run~A5a).

\subsection{Velocity damping in the corona}
\label{damp}

Whether the solar corona rotates like a solid body or differentially
coupled with the photosphere is unclear.
In recent work by \cite{WBHG10}, where SOHO-EIT data of the bright points
in the solar corona were used to estimate the rotation speeds, it was
found that the corona rotates similarly as the small magnetic
features in the photosphere.
Similar results have been obtained by \cite{B10}, where the coronal
rotation has been measured by analyzing the green Fe$^{\rm XIV}$
530.3 nm line.
This author finds also a variation pattern with the activity cycle.
However, the observations of the ``boot'' coronal hole by SKYLAB 
suggested rigid rotation \citep{TKV75}.
Recent work on coronal holes by \cite{LRLM05} claims that the rigid
rotation is only an apparent one.
The magnetic field is sheared by the differential rotation, but the
boundary of the hole remains relatively unchanged, due to
reconnection.
Owing to the low plasma $\beta$
in the solar corona, the fluid motions
are dominated by the magnetic fields whose footpoints
are anchored in the photosphere or even further down.
So the magnetic field might then be rigid enough to prevent
differential rotation of the solar corona.
However, the observed bright
points and other features in the corona are strongly correlated with the
magnetic field so they can give a misleading picture about the global
rotation of the corona.

In our simulations, the Coriolis force is included in the momentum
equation as a consequence of the rotation.
In the solar corona the density is more than 14 orders of magnitude
smaller than in the lower convection zone.
Because of the weak density stratification in our simulation, the
Coriolis force in our coronal part is too strong and can cause
possible artifacts such as the magnetorotational instability.
To avoid this---at least for runs with rapid rotation---we apply a
damping function $\DA(r,\theta)$ in the momentum equation, which is 
given by
\begin{equation}
\DA(r,\theta,t)={1\over\tau_{\rm D}}\Theta(r-R)\,\overline{\UU}(r,\theta,t),
\end{equation}
where
\begin{equation}
\Theta(r-R)= \half\left[1+\tanh{\left({r-\Rsun\over w}\right)}\right],
\end{equation}
with $\tau_{\rm D}$ being the damping time and $w$ the width of the
transition
layer from convection zone to the coronal part.
Here and elsewhere, the overbar denotes averaging over $\phi$, defined
as
$\overline{F}(r,\theta,t)$=$\int {F(r,\theta,\phi,t)\;\dd\phi/2\pi}$.
Occasionally we also use time averages denoted by $\bra{.}_t$.

\subsection{Units, nondimensional quantities, and parameters}
\label{nond}

Dimensionless quantities are obtained by setting
\begin{equation}
\Rsun=GM=\rhob=c_p=\mu_0=1,
\end {equation}
where $\rhob$ is the density at $r=0.7\,\Rsun$.
Below, we will describe the properties of the runs by the
following dimensionless parameters:
fluid Reynolds number $\Rey=\urms/\nu \kf$, magnetic Reynolds number
$\Rm=\urms/\eta\kf$, where $\kf=2\pi/0.3\Rsun$ is an estimate
for the typical wavenumber of the
energy-carrying eddies and $\urms=\sqrt{3/2\bra{U_r^2+U_{\theta}^2}}$ is
the volume-averaged rms
velocity in the convection zone ($r\leq \Rsun$).
In our definition of $\urms$ we omit the contribution from the
$\phi$-component of the velocity, because it is dominated by
contributions from the
large-scale differential rotation that develops when rotation is included
and would give an atypical estimate of the convective turnover time.
To compensate for this, and to have an estimate of $\urms$ comparable
with earlier work, we apply the 3/2 correction factor.
We also define the magnetic Prandtl number $\Pm=\nu/\eta=\Rm/\Rey$ and the Coriolis
number $\Co=2\Omega_0/\urms\kf$.
Time is expressed in units of $\tau = \left(\urms\kf\right)^{-1}$, which is
the eddy turnover time in the convection zone.
We measure the magnetic field strength as the rms value averaged over
the convection zone $\Brms$, where we often normalize this value with
the equipartition value of the magnetic field defined by
$\Beq^2=\mu_0\bra{\meanrho\urms^2)}_{r\leq \Rsun}$.
The relative kinetic helicity is
$h_{\rm rel}(r,t)=\overline{\oo\cdot\uu}/\omega_{\rm rms}\urms$, where
$\oo=\nab\times\uu$ is the vorticity and $\omega_{\rm rms}$ is its rms
value inside the convection zone.

\section{Results}
\label{results}

\begin{table*}[t!]\caption{
Summary of the runs.
$\Rey$ is the fluid Reynolds number,
$\urms=\sqrt{3/2(U_r^2+U_{\theta}^2)}$ is
the volume-averaged rms velocity in the convection zone normalized by
$\sqrt{GM/\Rsun}$, $\Pm$ is the magnetic Prandtl number, $\Co$
is the Coriolis number, and $h_{\rm rel}$ is the maximum value of the 
relative kinetic helicity using azimuthal averages
as defined in \Sec{nond}.
$\rhob \over \rho_{\rm s}$ and $\rhob \over \rhot$ give
the density ratios of the bottom of the convection zone to those at the surface
and the top of the domain, respectively.
In the right-most column we note if damping for velocity in the
coronal part is used (Y) or not (N); see \Sec{damp}.
}
\label{summary}
\begin{tabular}{lcccccccccc}
Run&Resolution&$\displaystyle{\urms\over\sqrt{GM/\Rsun}}$
&$\Rey$&$\Pm$&$\displaystyle{\Brms^2\over\Beq^2}$
&$\displaystyle{\rhob \over \rho_{\rm s}}$
&$\displaystyle{\rhob \over \rho_{\rm t}} $&$\Co$&$h_{\rm rel}$&$D$\\
\hline
A5&$128^2\times64$&0.0072&3.3&10&0.1--0.4&3.6&39& 7 &0.5&N\\
A5a&$256^2\times 128$&0.0105&100&1&0.2&3.6&39&4.5&0.3&N\\
Ar1&$128^2\times64$&0.0040&38&1&1.5--5.5&3.6&39&50&0.3&Y\\
\hline
\end{tabular}
\end{table*}
\subsection{Hydrodynamic phase of the simulations}

\begin{figure}[t!]
\begin{center}
\includegraphics[width=0.9\textwidth]{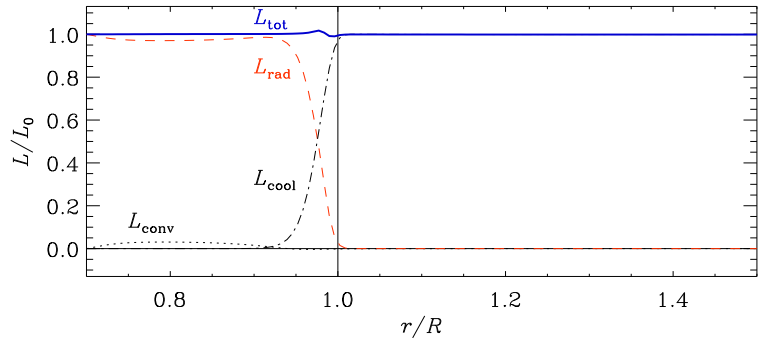}
\end{center}\caption[]{
Flux balance from Run~A5.
The different contributions to the total luminosity (solid blue line)
are due to radiative diffusion (dashed red line), resolved convection
(dotted black line) and cooling (dash-triple-dotted black line).
The black thin line denotes the zero level and the surface ($r=R$), respectively.
}
\label{pflux}
\end{figure}

After around 100 turnover times, convection has reached saturation and we
find convection cells as typical patterns in the radial velocity
just below the surface.
In our model, the dominant energy transport mechanisms are 
radiative and convective
fluxes in the bulk of the convection zone and an (optically thin)
cooling flux in the outer (coronal) parts.
The radiative and convective fluxes are defined as:
\begin{eqnarray}
F_{\rm rad} = -K{\partial \meanT \over \partial r},\quad
F_{\rm conv} = c_P\meanrho\overline{u_r^{\prime}T^{\prime}},
\end{eqnarray}
where the averages are taken over $\theta$ and $\phi$ and the prime
indicates fluctuations about the respective mean quantity.
In our present setup, however, the convective flux reaches barely
about 5\% in the convection zone; see \Fig{pflux} where we plot
the relevant contributions to the luminosity for Run~A5.
Above the surface the cooling takes over to maintain an approximately
isothermal atmosphere. 
The total flux is constant, except for small departures near the surface.
The kinetic energy and viscous fluxes are negligible in the present
runs.

\begin{figure}[t!]
\begin{center}
\includegraphics[width=0.327\textwidth]{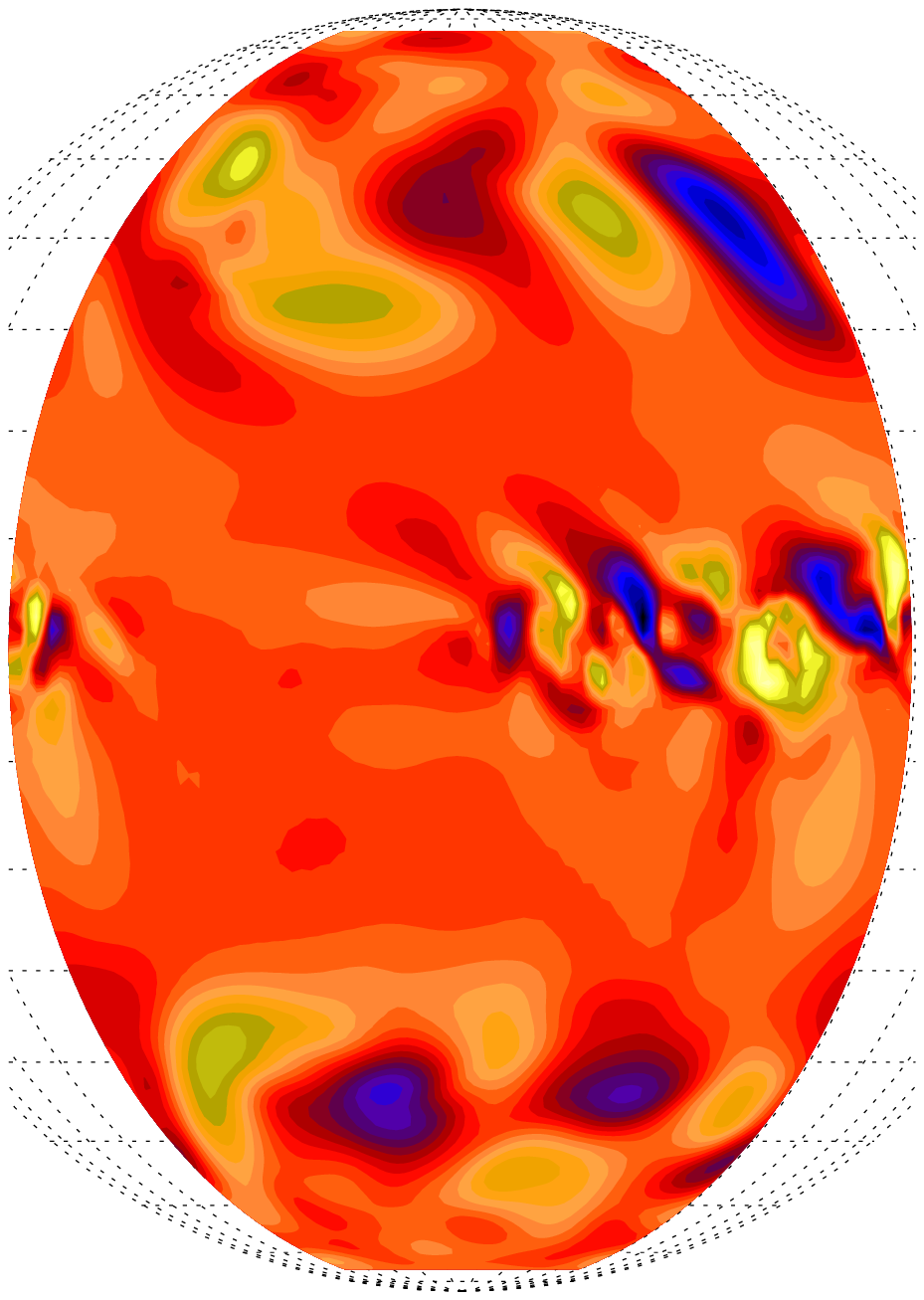}
\includegraphics[width=0.327\textwidth]{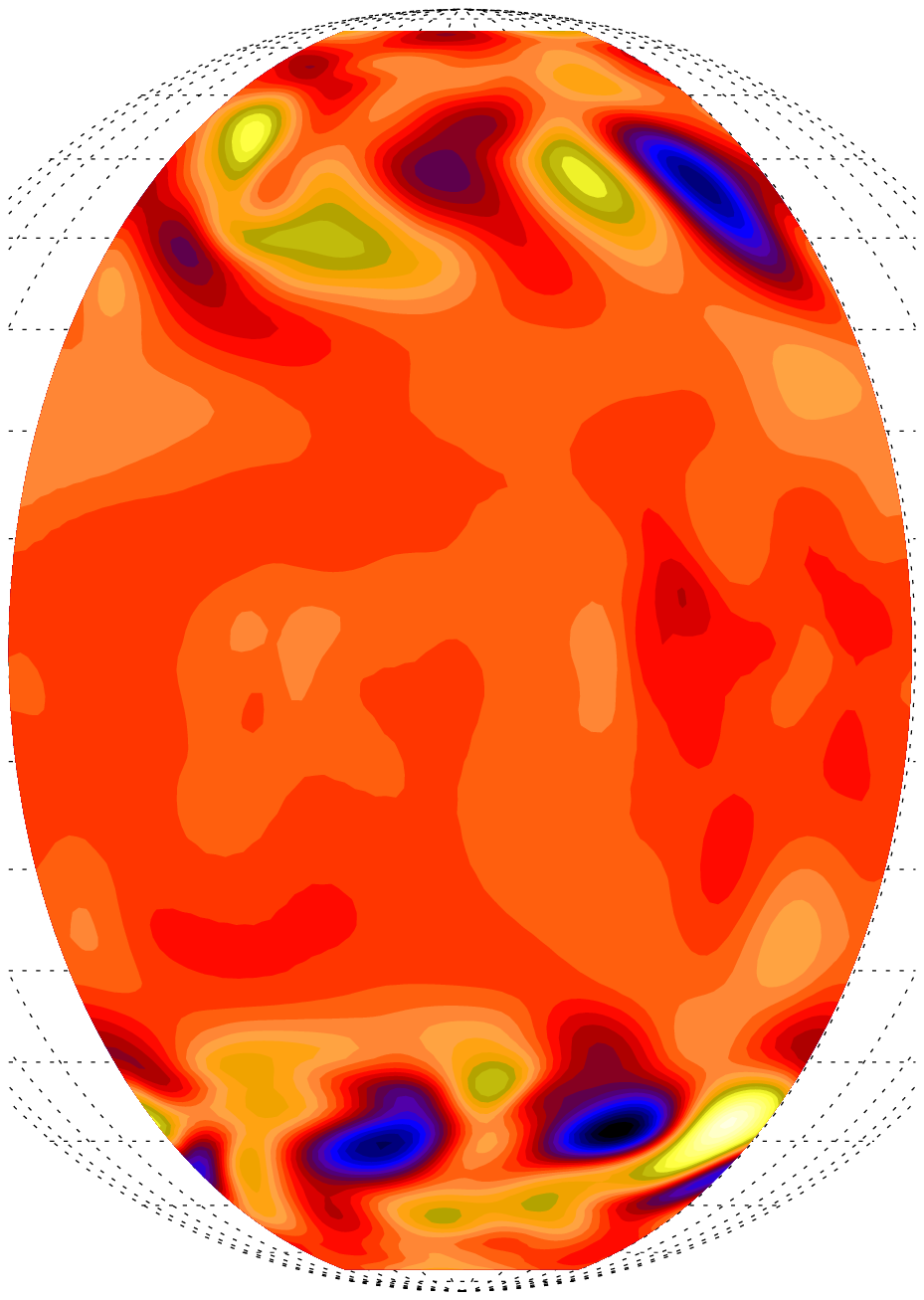}
\includegraphics[width=0.327\textwidth]{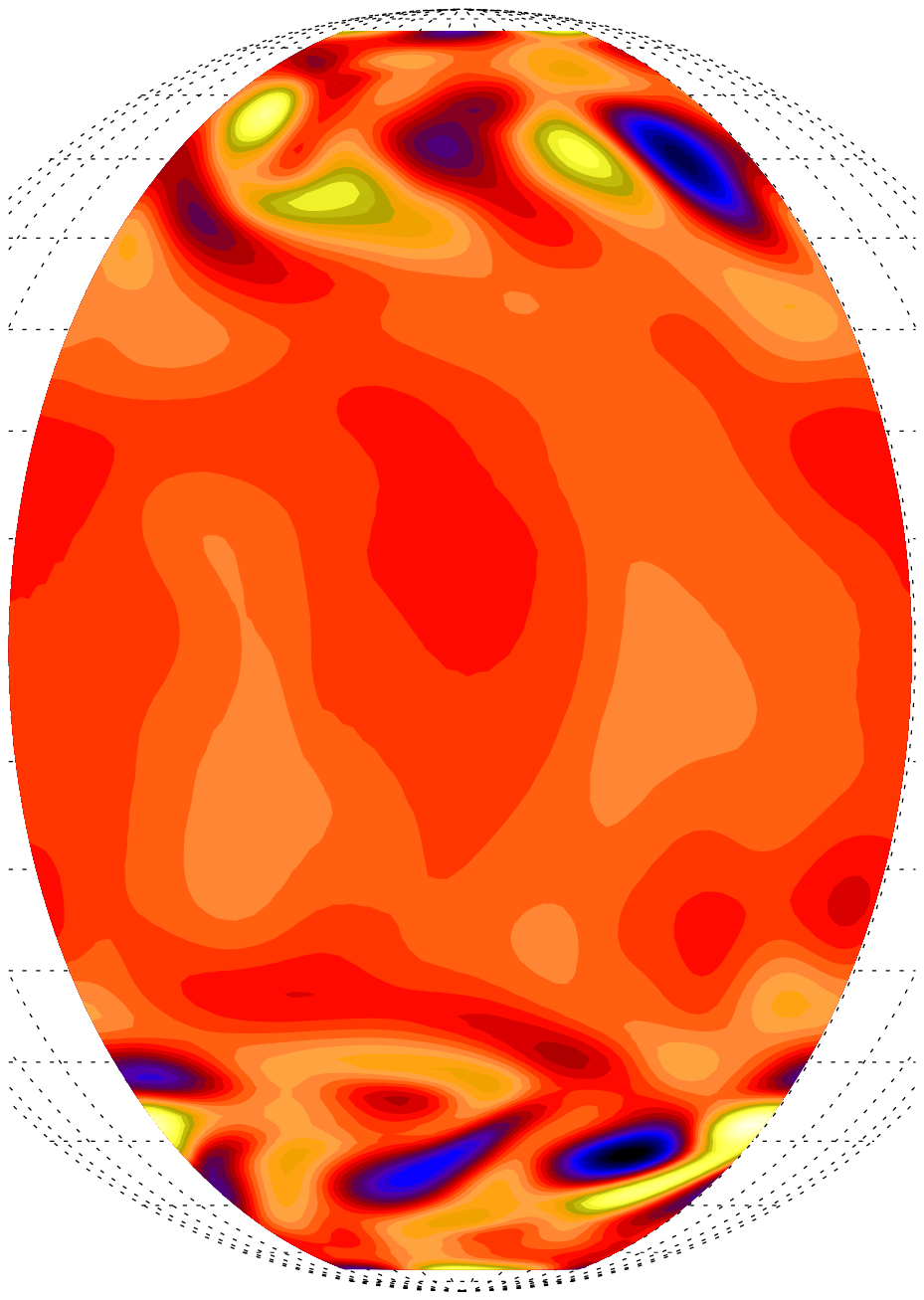}
\end{center}\caption[]{
Radial velocity ($U_r$) above the surface for $r=1.15$, 1.25, $1.35\,\Rsun$
from left to right, for Run~A5.
Dark blue shades represent negative and light yellow positive values.
}
\label{urad2}
\end{figure}

\begin{figure}[t!]
\begin{center}
\includegraphics[width=0.9\textwidth]{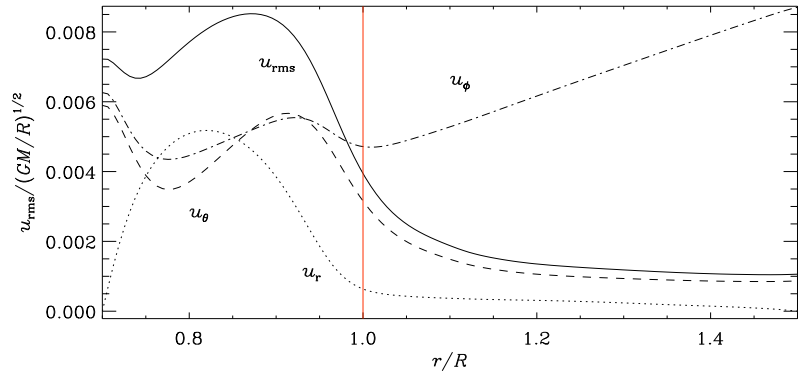}
\end{center}\caption[]{
Root-mean-square values of $U_r$ (dotted), $U_{\theta}$ (dashed),
and $U_{\phi}$ (dash-dotted) as a function of radius for Run~A5.
The solid line shows the radial profile of our
nominal rms velocity, $\urms=\sqrt{3/2(U_r^2+U_{\theta}^2)}$.
The (red) vertical line indicated the surface at ($r=R$).
The values are normalized by $\sqrt{GM/\Rsun}$.
}
\label{urms}
\end{figure}

To determine the degree of overshooting and penetration into the
stably stratified layers above the convection zone, we show in
\Fig{urad2} the radial velocity above the surface at $r=1.15$,
$1.25$, and $1.35\,\Rsun$ for Run~A5.
At low latitudes, there is very little radial penetration
(velocity features are only seen until $r=1.15\,\Rsun$),
while at higher latitudes the radial velocity pattern is
transmitted all the way to $1.35\,\Rsun$.
This is not surprising in view of the Taylor--Proudman theorem, which
states that for rapid rotation (large values of $\Co$) the local angular
velocity of the gas is constant along cylindrical surfaces.

Next, we plot in
\Fig{urms} the rms values of all three velocity components for Run~A5.
The amplitude of the radial velocity component falls off the fastest.
The latitudinal component also falls off with radius, but remains
about three times larger than the radial component.
The longitudinal component, on the other hand, increases with radius
in a way that is compatible with rigid rotation with an angular
velocity that is somewhat larger than the rotation rate of the
frame of reference.

\begin{figure}[t!]
\begin{center}
\includegraphics[width=0.327\textwidth]{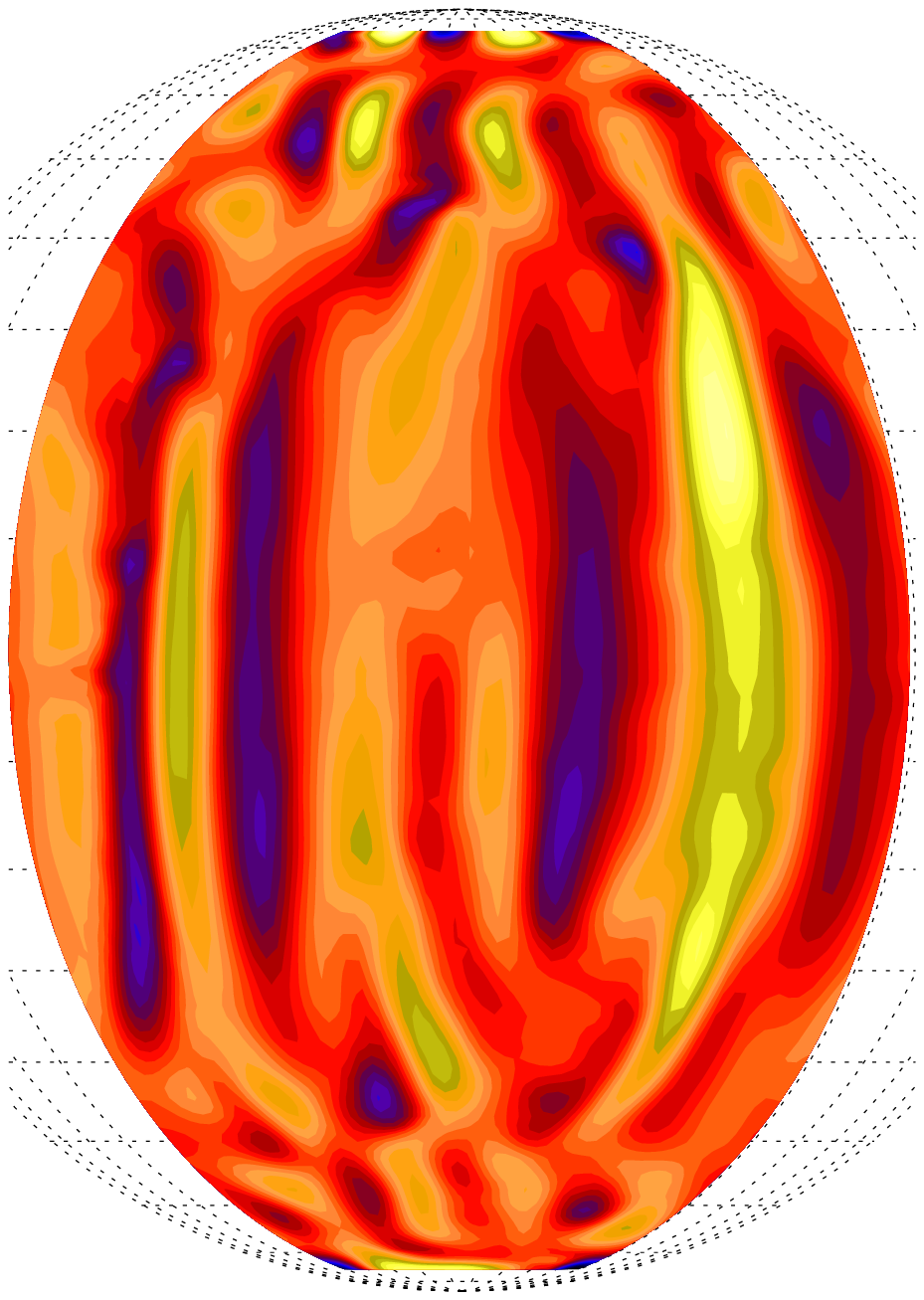}
\includegraphics[width=0.327\textwidth]{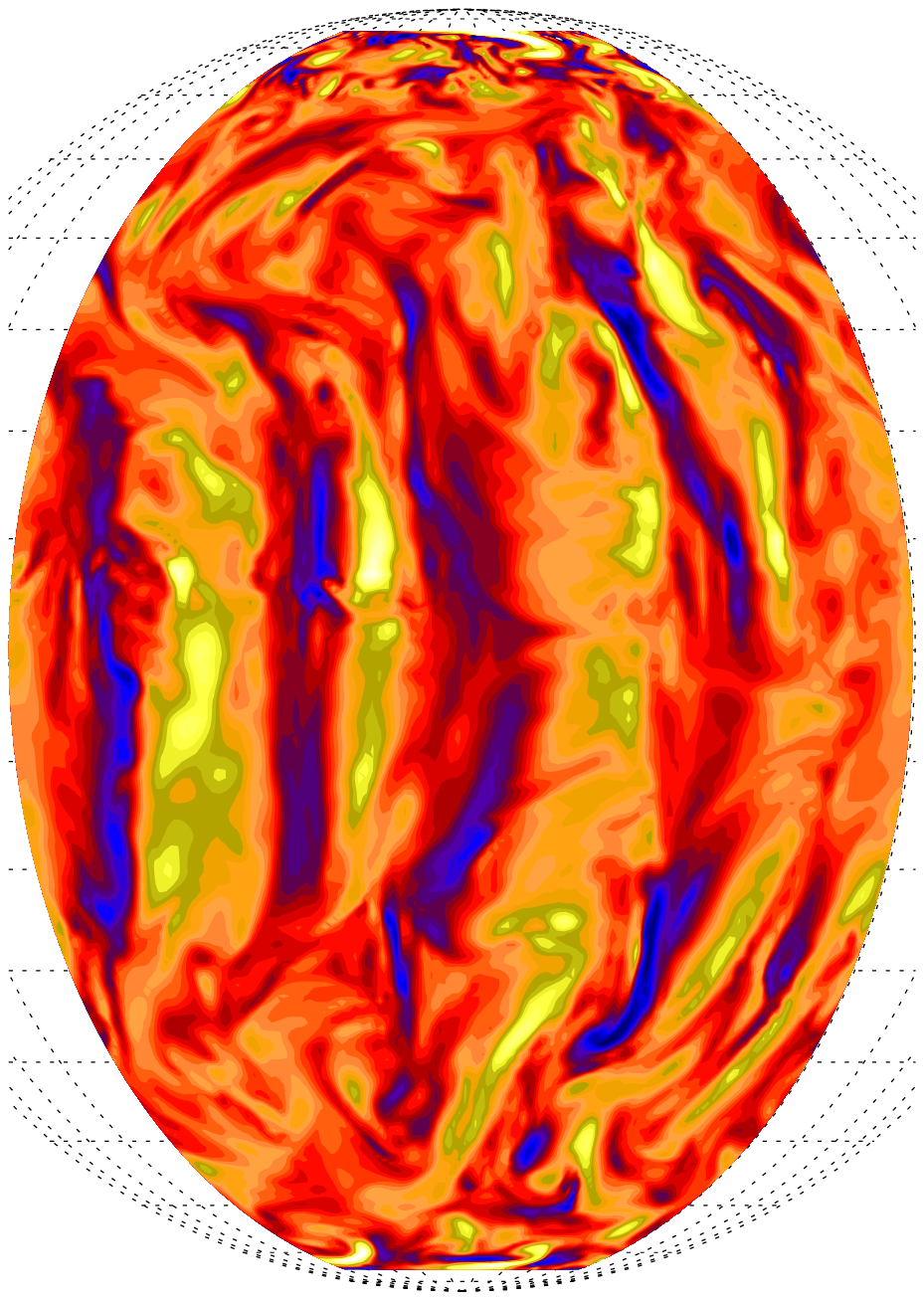}
\includegraphics[width=0.327\textwidth]{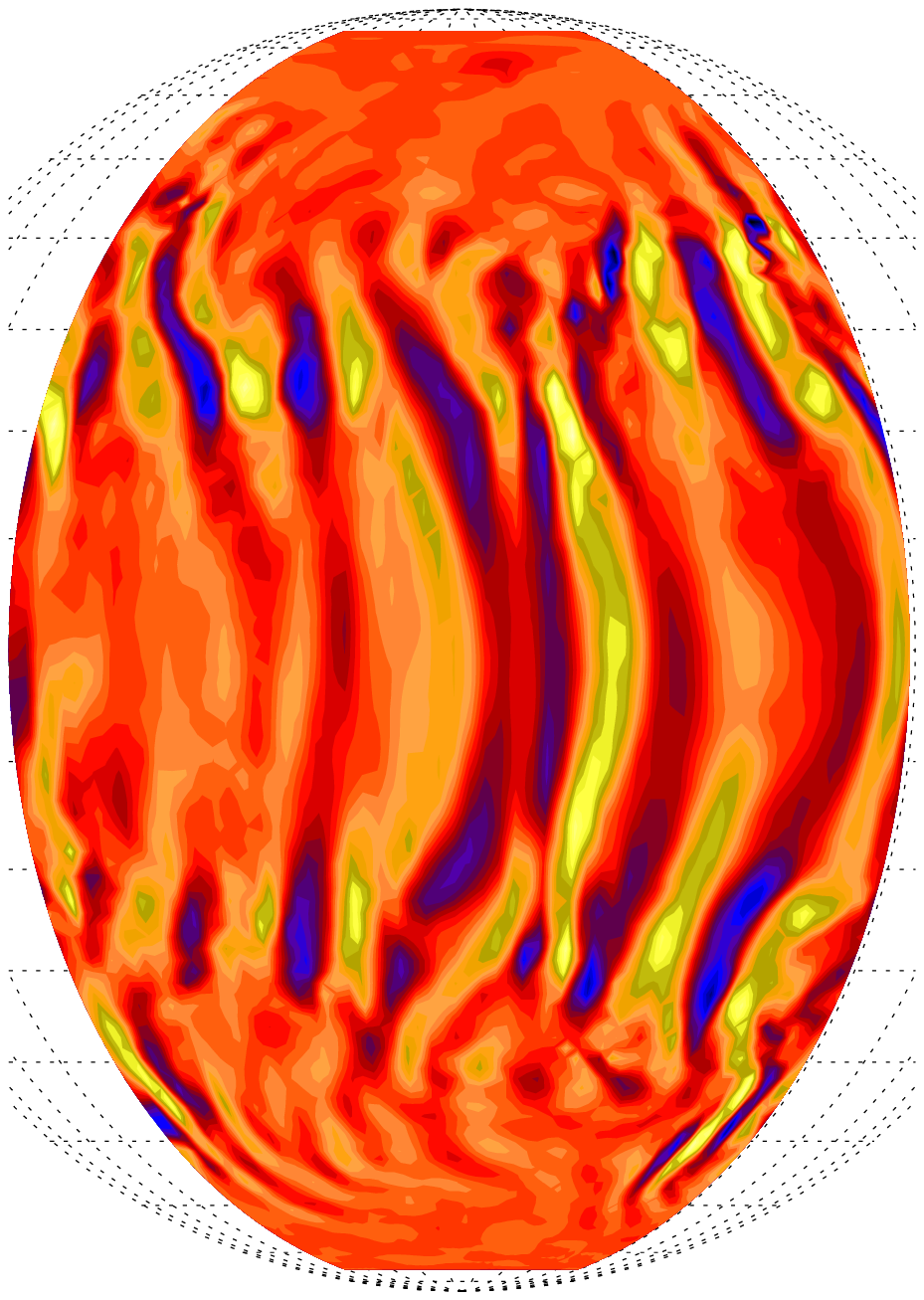}
\end{center}\caption[]{
Radial velocity ($U_r$) beneath the surface ($r=0.89\,\Rsun$) for
Runs~A5, A5a, and Ar1 from left to right.
Dark blue shades represent negative and light yellow positive values.
}
\label{urad}
\end{figure}

The size of the convection cells depends strongly on the strength of rotation
and the degree of density stratification; see also \cite{KMB11}.
We plot the radial velocity $U_r$ at
$r=0.89\,\Rsun$ for Runs~A5, A5a, and Ar1 in \Fig{urad}.  
The Run~A5 has a low fluid Reynolds number and therefore the convection
cells are large; see \Tab{summary}. The flow pattern shows clear
`banana cells' as in previous work with comparable Coriolis parameter,
cf.\ \cite{KKGBC11}. A higher fluid Reynolds number and
higher resolution, as in Run~A5a, allow the velocity field to form more
complex structures. However, the banana cells are still visible.  
If one now looks at a simulation with more rapid rotation (Run~Ar1,
plotted in the right-most panel of \Fig{urad}) with a Coriolis number
of $\Co=50$, the number of banana cells increases and they are more
clearly visible than in Run~A5a.
Note also that the radial velocity is now significantly reduced at
high latitudes inside the inner tangent cylinder.

\begin{figure}[t!]
\begin{center}
\includegraphics[width=0.327\textwidth]{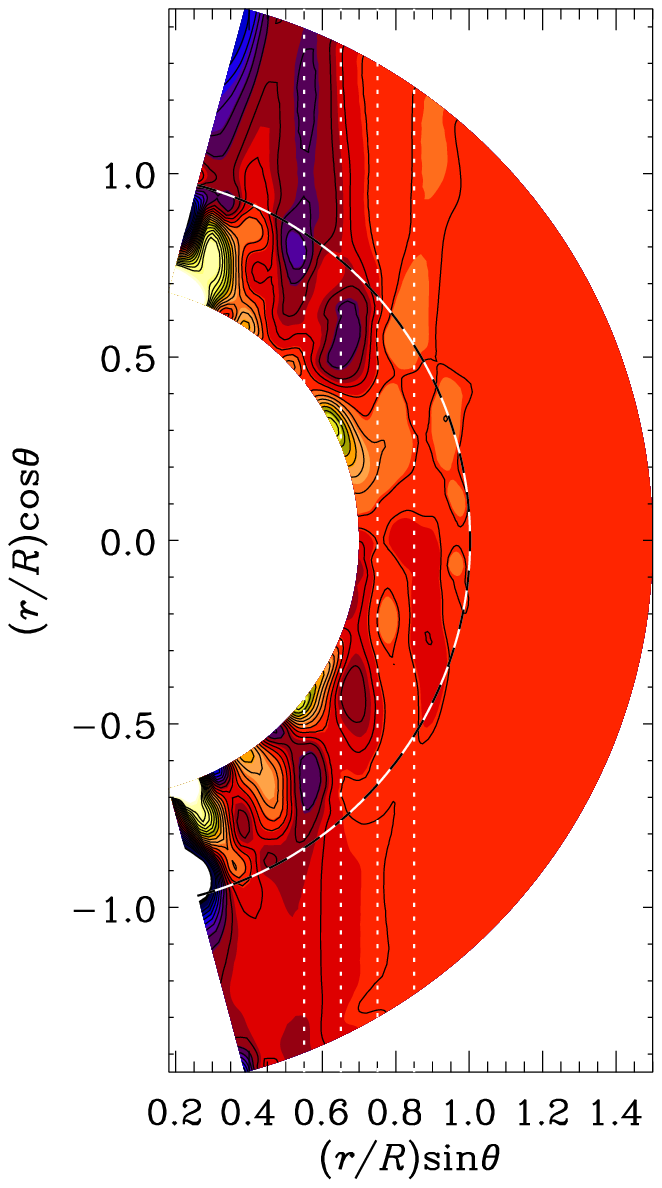}
\includegraphics[width=0.327\textwidth]{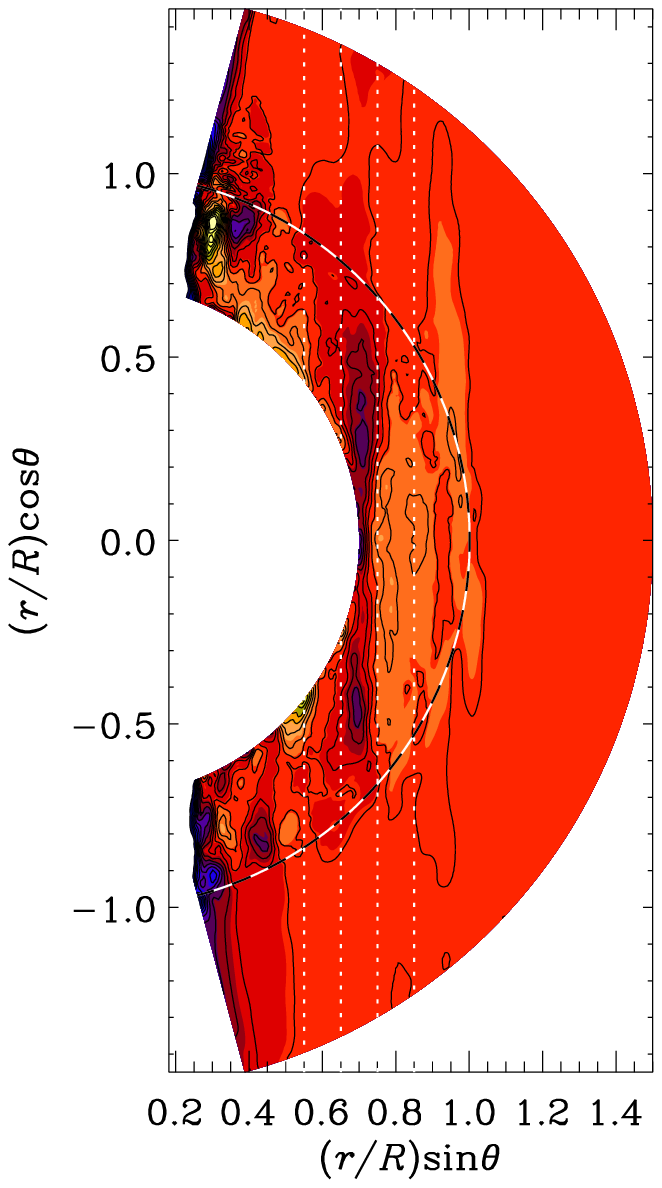}
\includegraphics[width=0.327\textwidth]{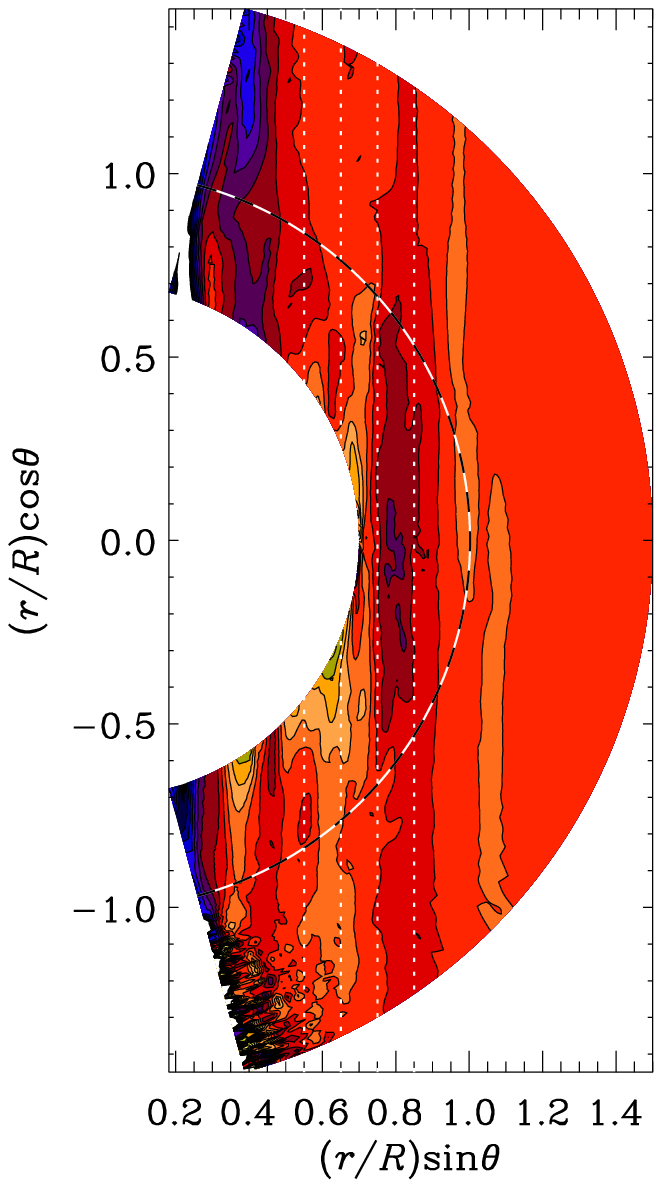}
\end{center}\caption[]{
Differential rotation profiles $\overline{\Omega}(r,\theta)=
\overline{U}_{\phi}/(r\sin{\theta})+\Omega_0$ for
Runs~A5, A5a, and Ar1 from left to right.
Dark blue shades represent low and light yellow high values,
overplotted by the isocontours with solid black lines.
The dotted white lines parallel to the rotation axis are given for
orientation and the dashed line indicates the surface ($r=R$).
}\label{difrot}
\end{figure}

In the Sun, differential rotation is an important element to
produce the magnetic field structures observed at large scales,
exhibiting a cyclic behavior over time, as manifested by the sunspot cycle.
To illustrate the differential rotation profiles generated in the
simulations, we plot the azimuthally averaged angular velocity,
$\overline{\Omega}(r,\theta)=\overline{U}_{\phi}/(r\sin{\theta})+\Omega_0$,
for Runs A5, A5a, and Ar1 in the saturated state of the simulation; see
\Fig{difrot}.  
In the plot, we show isocontours of angular velocity with solid black lines. 
In the convection zone the contours of local angular velocity tend to be
cylindrical, which is likely a consequence of
the absence of a strong latitudinal modulations of
specific entropy \citep{BMT92,KR95,MBT06}.
The coronal part seems to rotate as a solid body outside the outer
tangent cylinder (i.e., for $r\sin\theta>R$),
while inside it some differential rotation occurs also in the coronal part.
In the convection zone between the inner and outer tangent
cylinders, the angular velocity is enhanced relative to that
inside the inner tangent cylinder (see the first and second panels
of \Fig{difrot}), while in the case of extremely rapid rotation
this may actually be reversed.

In the three runs shown in \Fig{difrot} the stratification in the
whole domain is just ${\rhob/\rhot}=40$, which is rather small
compared to the stratification of the Sun (${\rhob/\rhot}\sim 10^{14}$).
It seems, therefore, that the Coriolis force is acting much more strongly in the
coronal part of our simulation than in reality.
In the Sun the Lorentz force plays a more important role in the corona
than in our model.
In the convection zone,
we find quenching of convection due to rapid rotation. 
In Run~A5, where $\Co=7$, the lines of constant
rotation rate are more radial than vertical and show super-rotation,
i.e., the equator rotates faster than the poles.
As expected, this tends to coincide with locations where
the Reynolds stress in the radial direction is negative
\citep[see, e.g.,][]{Rue80}.
However, the convection cells are rather big and have a strong
local influence on $U_{\phi}$ and could in principle lead to
subrotation; see the corresponding discussion in \cite{DSB06}.
Note that the rms velocity in Run~A5
is two times smaller than in Run~A5a, which has higher resolution
and higher fluid and magnetic Reynolds numbers ($\Rey=\Rm=100$). Due
to this, we find clear super-rotation, even though the Coriolis
number is slightly lower ($\Co=4.5$) than what is realized in Run~A5. In the
third case, Run~Ar1, where the rotation is extremely rapid ($\Co=50$),
we also find super-rotation, where the lines of constant rotation rate are
almost all vertical.  In comparable work \citep{KKGBC11,KMB11},
super-rotation has been found, when the Coriolis number
was larger than 4. 
This is similar to our results including a coronal part.  
In addition, there is a minimum of the rotation
rate at mid-latitudes and a polar vortex at high latitudes.
Rotation profiles, which show a comparable behavior, have been found by
several groups \citep{METCGG00,EMT00,KKGBC11,KMB11}.
The region with the higher rotation rate near the equator is limited to the
upper convection zone and can even penetrate into the coronal part.
In Run~Ar1 the velocity damping described in \Sec{damp} is used. 
By comparing the right-most panel of \Fig{difrot}, with damping, to the
left-most panels, without it, we conclude that the damping does 
not make much of a difference to the coronal velocity structures.

\begin{figure}[t!]
\begin{center}
\includegraphics[width=0.327\textwidth]{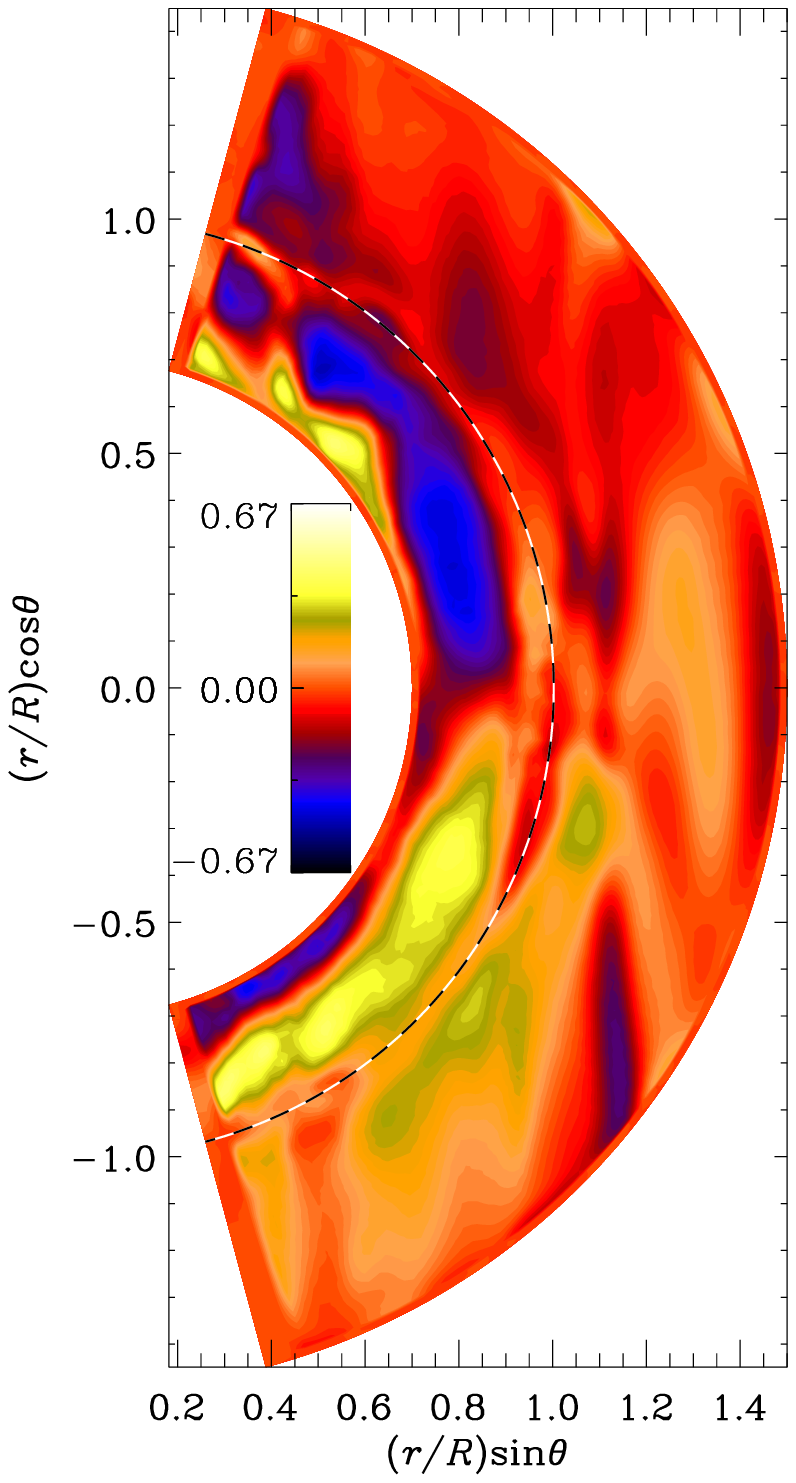}
\includegraphics[width=0.327\textwidth]{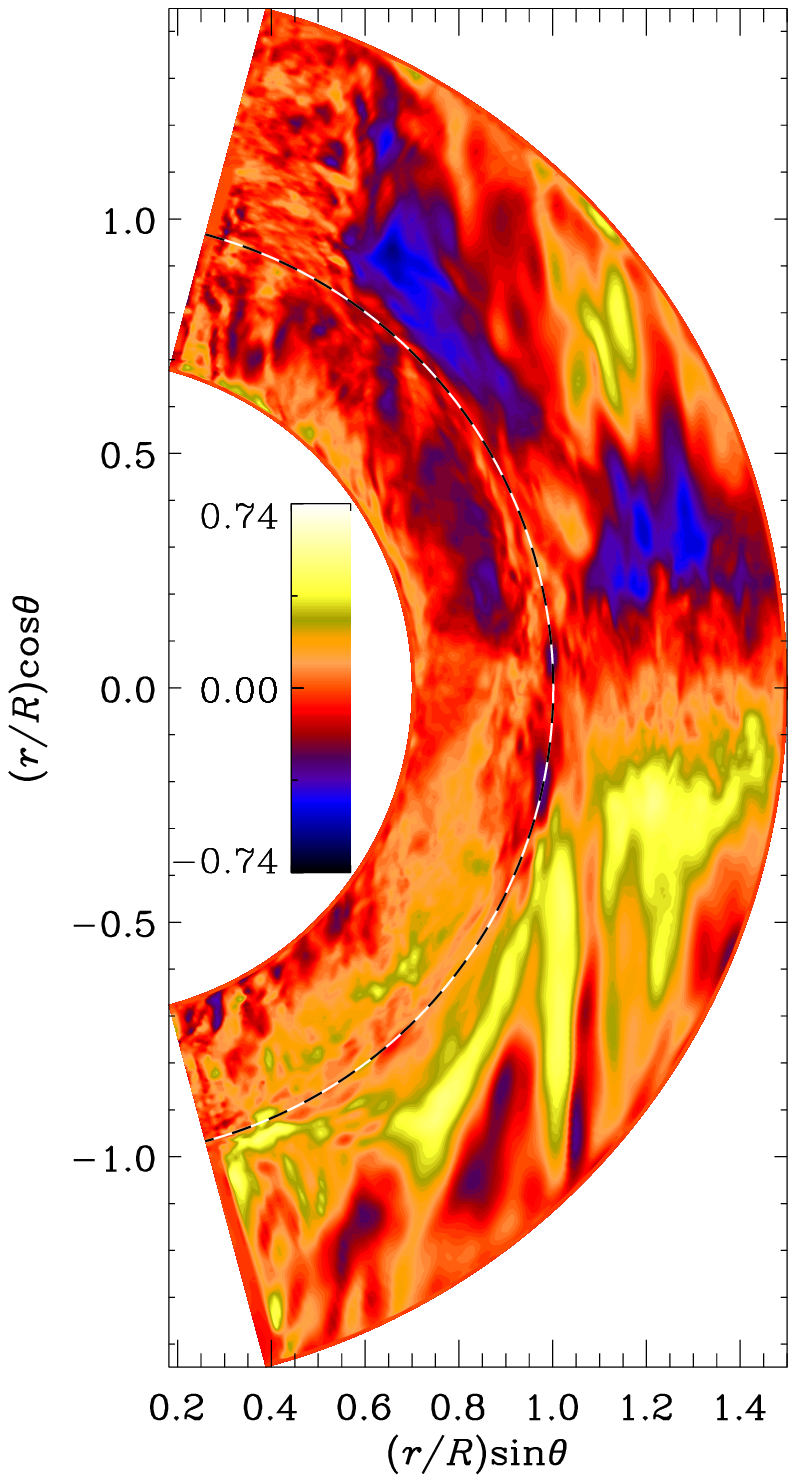}
\includegraphics[width=0.327\textwidth]{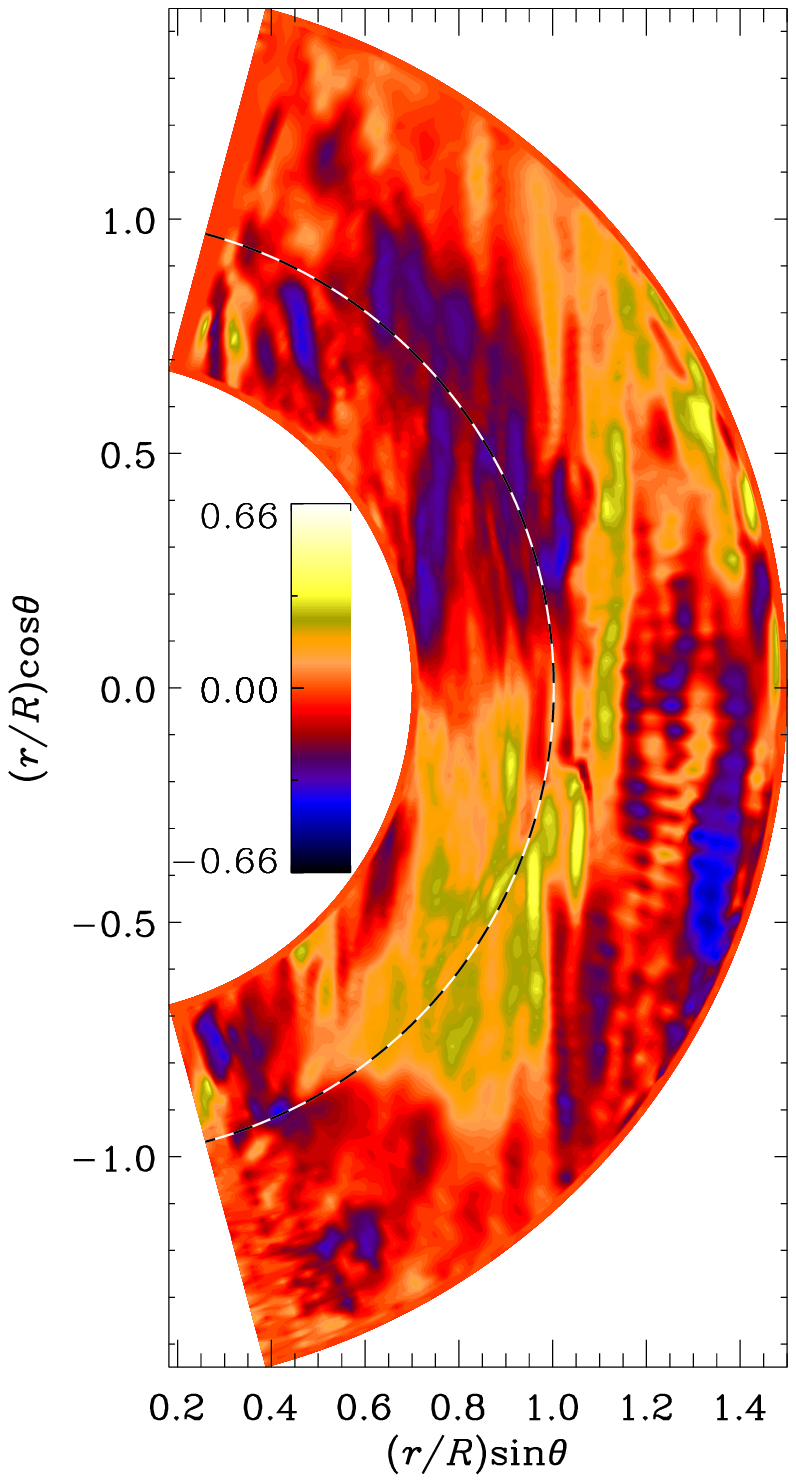}
\end{center}\caption[]{
Relative helicity $h_{\rm rel}(r,t)=\overline{\oo\cdot\uu}/\omega_{\rm rms}\urms$ plotted for
Runs~A5, A5a, and Ar1 from left to right.
Dark blue shades represent negative and light yellow positive values.
The dashed line indicates the surface ($r=R$).
}
\label{relhel}
\end{figure}

Simulations with randomly forced turbulence (WB,WBM) have
shown that the relative kinetic helicity $h_{\rm rel}$ has a strong
influence both on the generation of large-scale magnetic fields and the ejection
events.
In WB and WBM, values of $h_{\rm rel}$ of order unity were achieved by
using a forcing function with purely helical plane waves. 
In the convection runs presented here, however, values of
large relative helicity, $h_{\rm rel}=0.5$,
are obtained (for Run~A5), at least at certain radii.  
In \Fig{relhel}, we present contour plots of azimuthally averaged
relative helicity in the meridional plane for Runs~A5, A5a, and Ar1. 
All three show the typical sign rule of kinetic helicity under the
influence of rotation, i.e.\ the northern hemisphere has
predominantly a negative sign and the southern a positive one.  
Close to the bottom of the convection zone, the sign changes, which
has earlier been reported by several authors both in Cartesian
\citep[e.g.][]{BNPST90,OSB01}
and spherical geometries \citep[e.g.][]{METCGG00,KKBMT10}.
Only in Run~Ar1 with rapid rotation, the behavior is not that clear.
The relative helicity is no longer confined to the convection zone,
but significant values occur also in the coronal region. The sign rule
still holds within the convection zone, while a more complicated
sign behavior is visible in the coronal part. 
The maximal values of the azimuthally averaged helicity are around
$h_{\rm rel}=0.3$, occurring close to the surface.
In Run~A5a, the maximum value is slightly higher and
is located in the middle of the convection zone,
although relatively high values are present in the coronal part as well.
It is not yet completely clear how high values of relative kinetic
helicity can be achieved; strong rotation tends to suppress it,
whereas strong stratification increases it.
Its exact role in generating coronal ejections is yet unclear.

\subsection{Convective dynamo}

\begin{figure}[t!]
\begin{center}
\includegraphics[width=0.327\columnwidth]{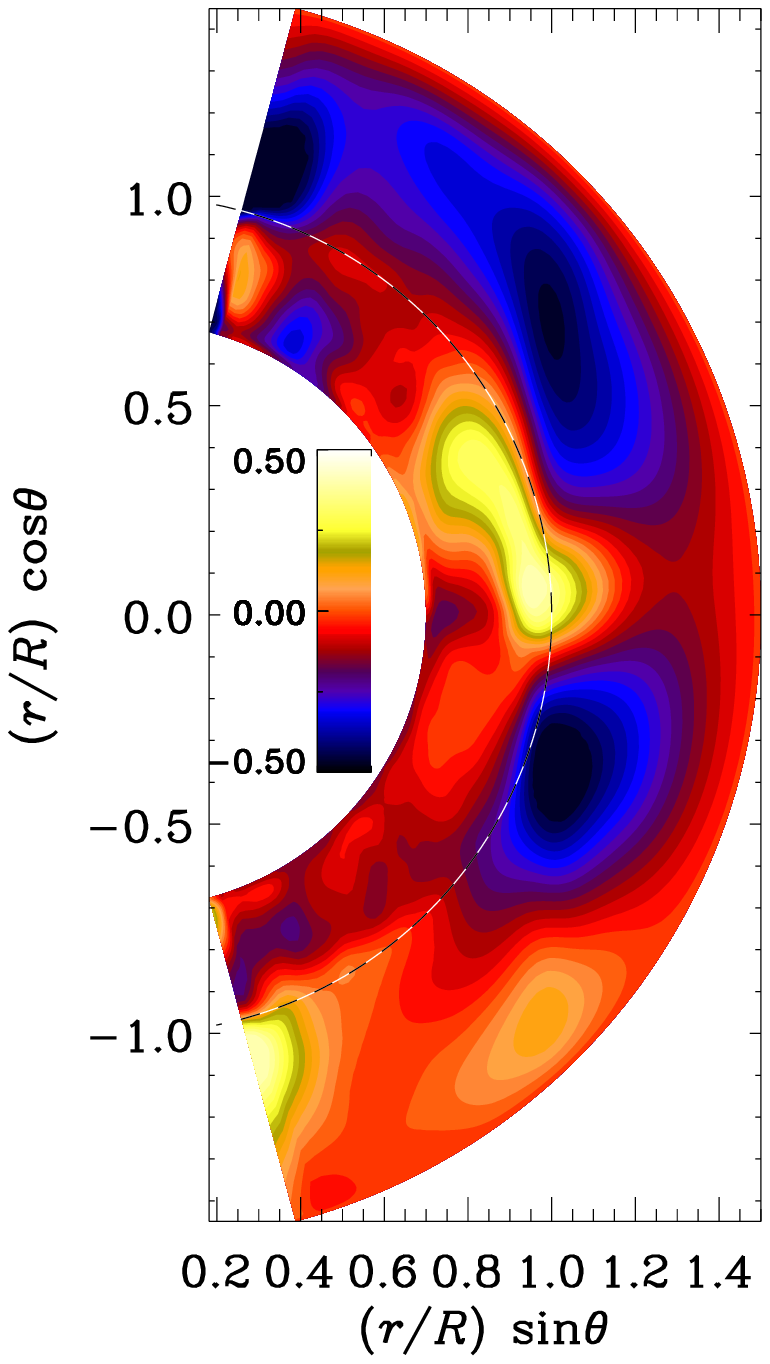}
\includegraphics[width=0.327\columnwidth]{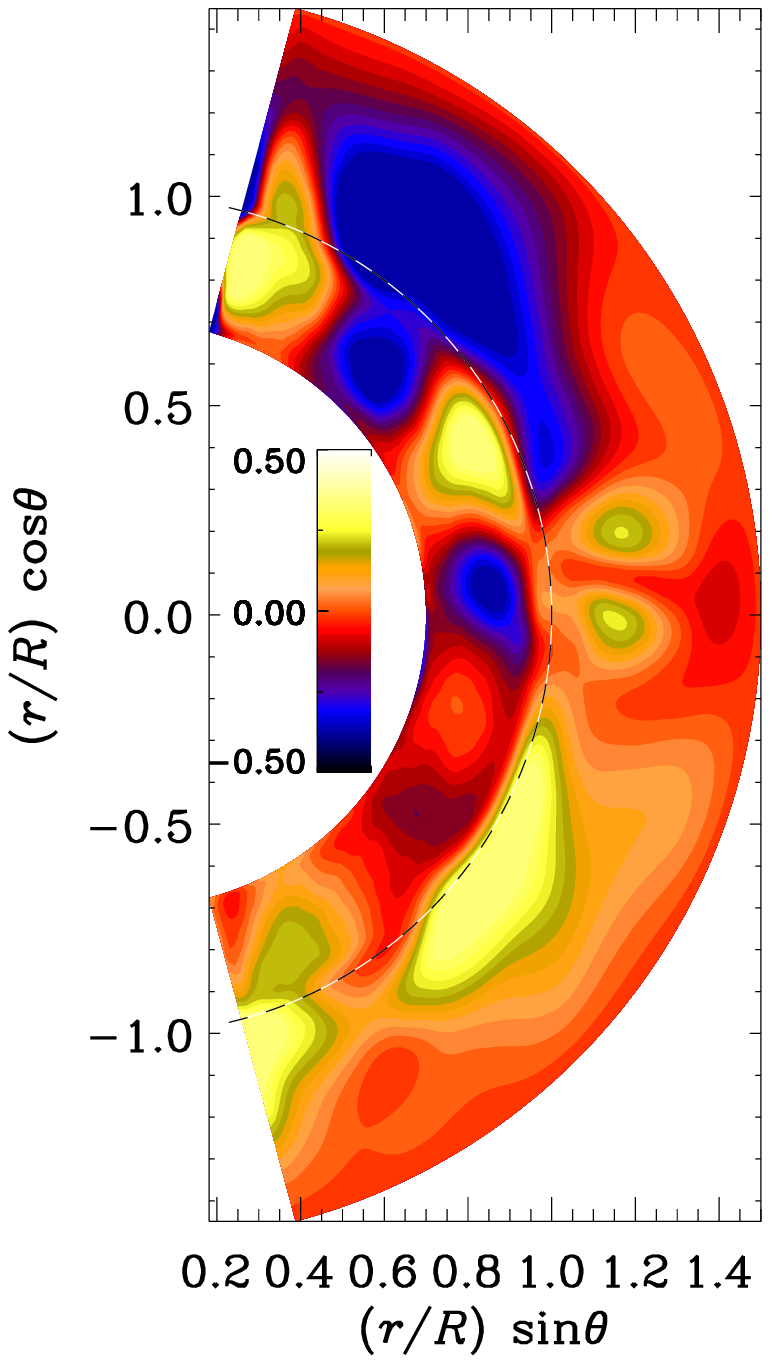}
\includegraphics[width=0.327\columnwidth]{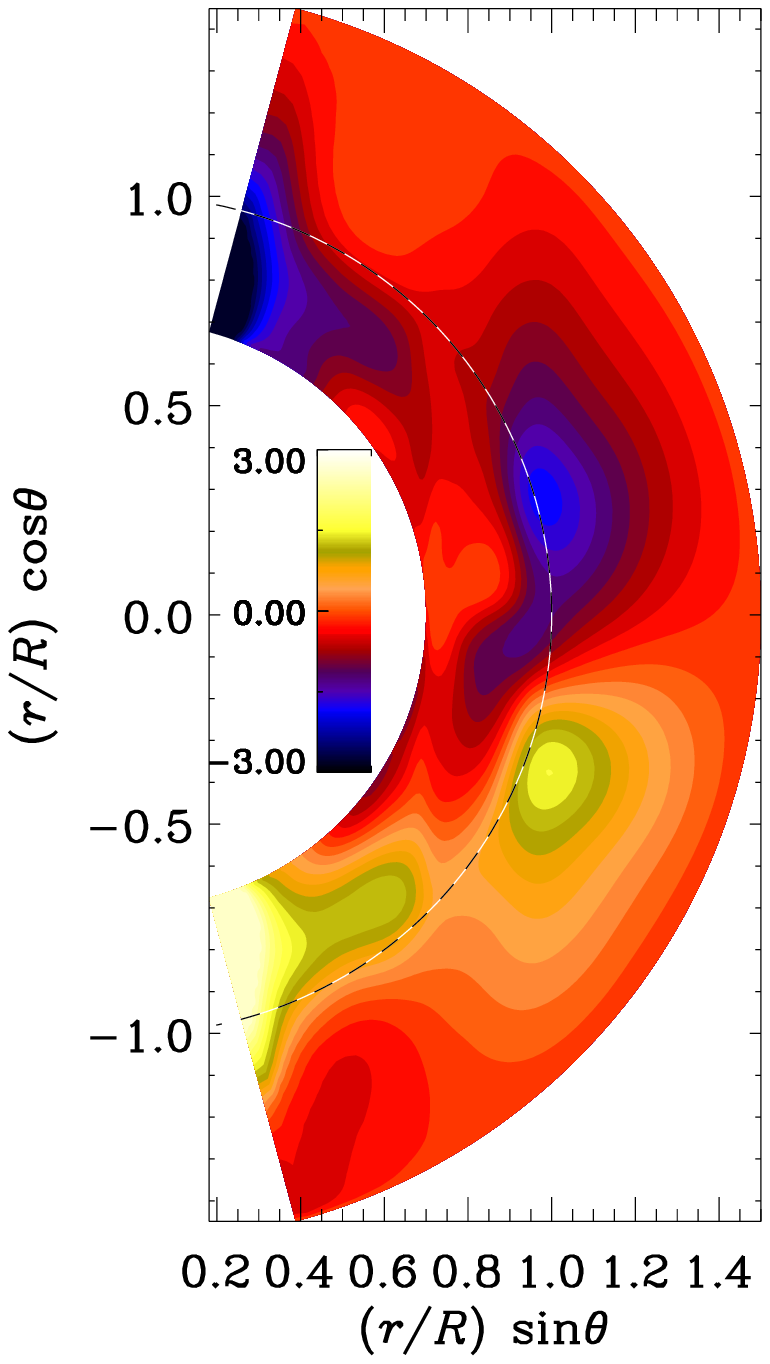}
\end{center}\caption[]{
Time-averaged $\meanB_{\phi}$ for Run~A5, Ar5a, and Ar1 from left to right.
Dark blue shades represent negative and light yellow positive values.
The magnetic field is normalized by the equipartition value.
The dashed line indicates the surface ($r=R$).
}\label{bphi}
\end{figure}

The convective motions generate a large-scale magnetic field due to
dynamo action.
The magnetic field grows first exponentially and begins then to affect the
velocity field.
The effects of this backreaction can be subtle in that we found
a 6\% enhancement of the rms velocity after saturation.
The growth of the magnetic field saturates after around 200 to 1000
turnover times, depending on the Coriolis and Reynolds numbers.  
In the runs in \Tab{summary}, we obtain different dynamo
solutions for the saturated field. 

\begin{figure}[t!]
\begin{center}
\includegraphics[width=0.49\columnwidth]{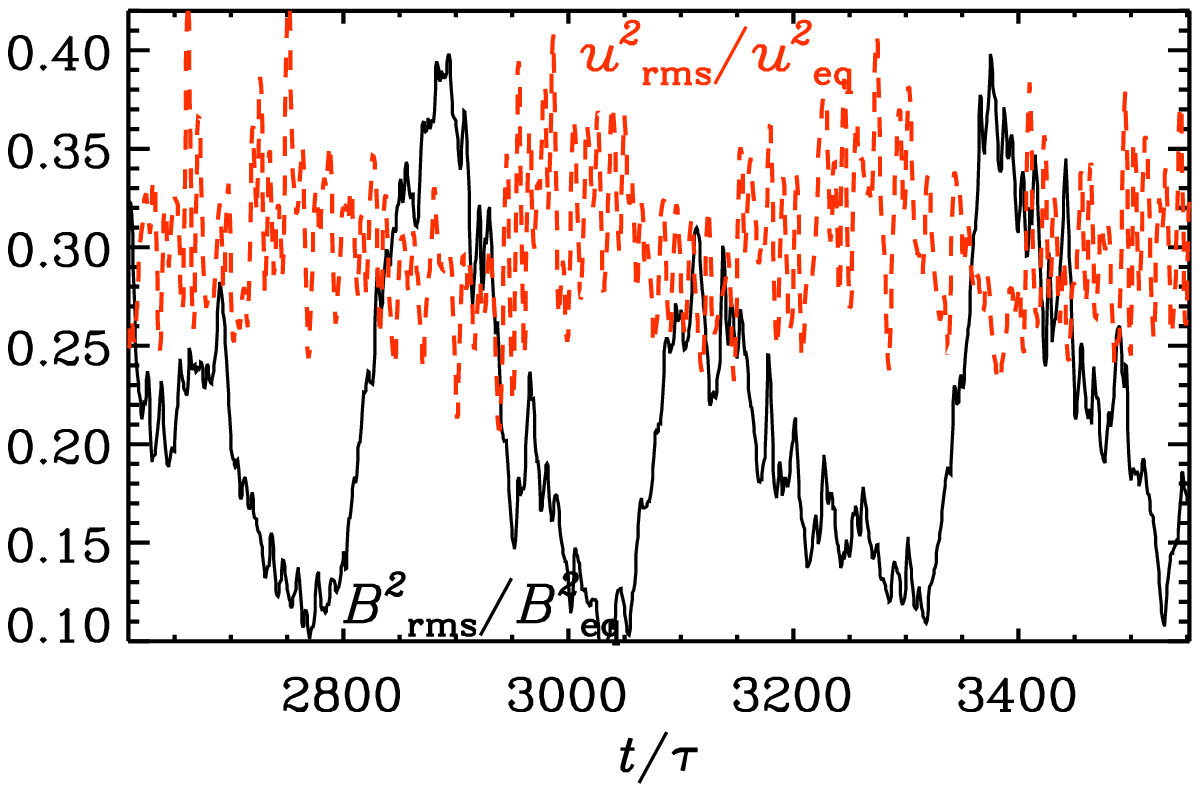}
\includegraphics[width=0.49\columnwidth]{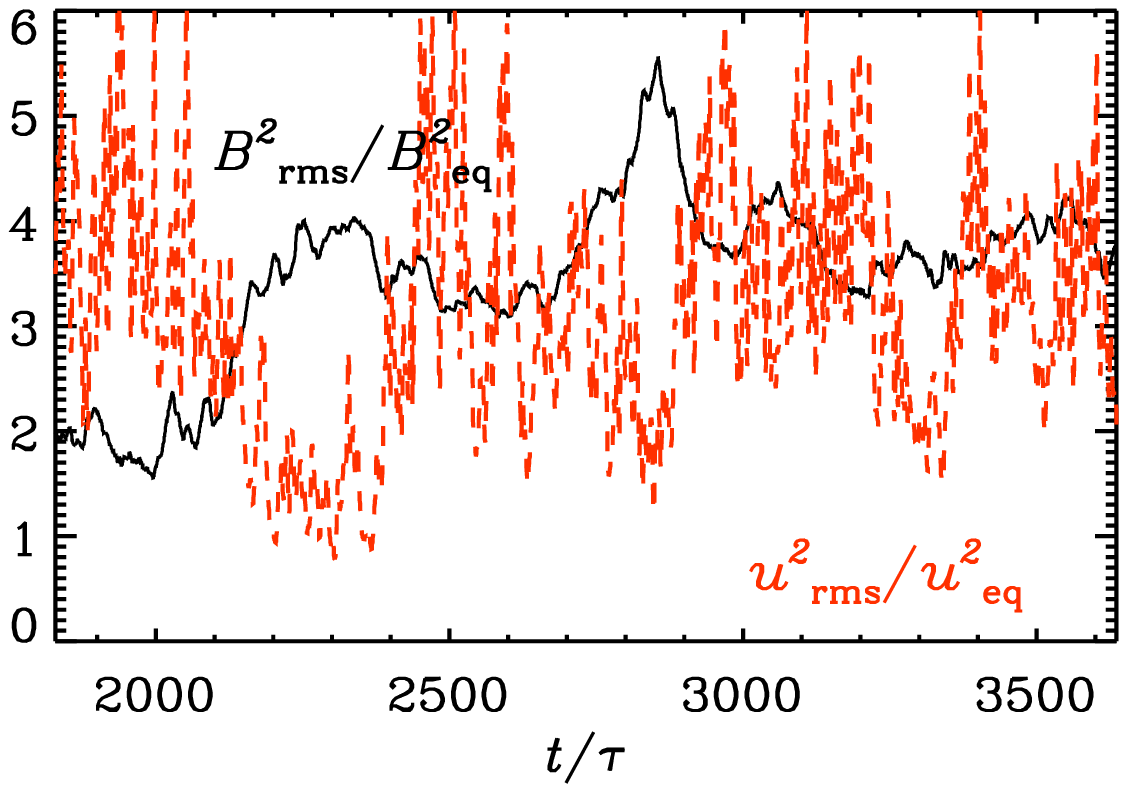}
\end{center}\caption[]{
Phase relation of the magnetic field ($\Brms^2/\Beq^2$, solid black 
lines) and the velocity field ($\urms(t)^2/\bra{\urms^2}_t$, dashed 
red lines) in the convection zone for Runs~A5 (left panel)
and Ar1 (right).
The velocity has been multiplied by a factor of 0.3
(left panel) and 3 (right), respectively, and smoothed over five
neighboring data points.
}
\label{bu}
\end{figure}

\begin{figure}[t!]
\begin{center}
\includegraphics[width=0.8\columnwidth]{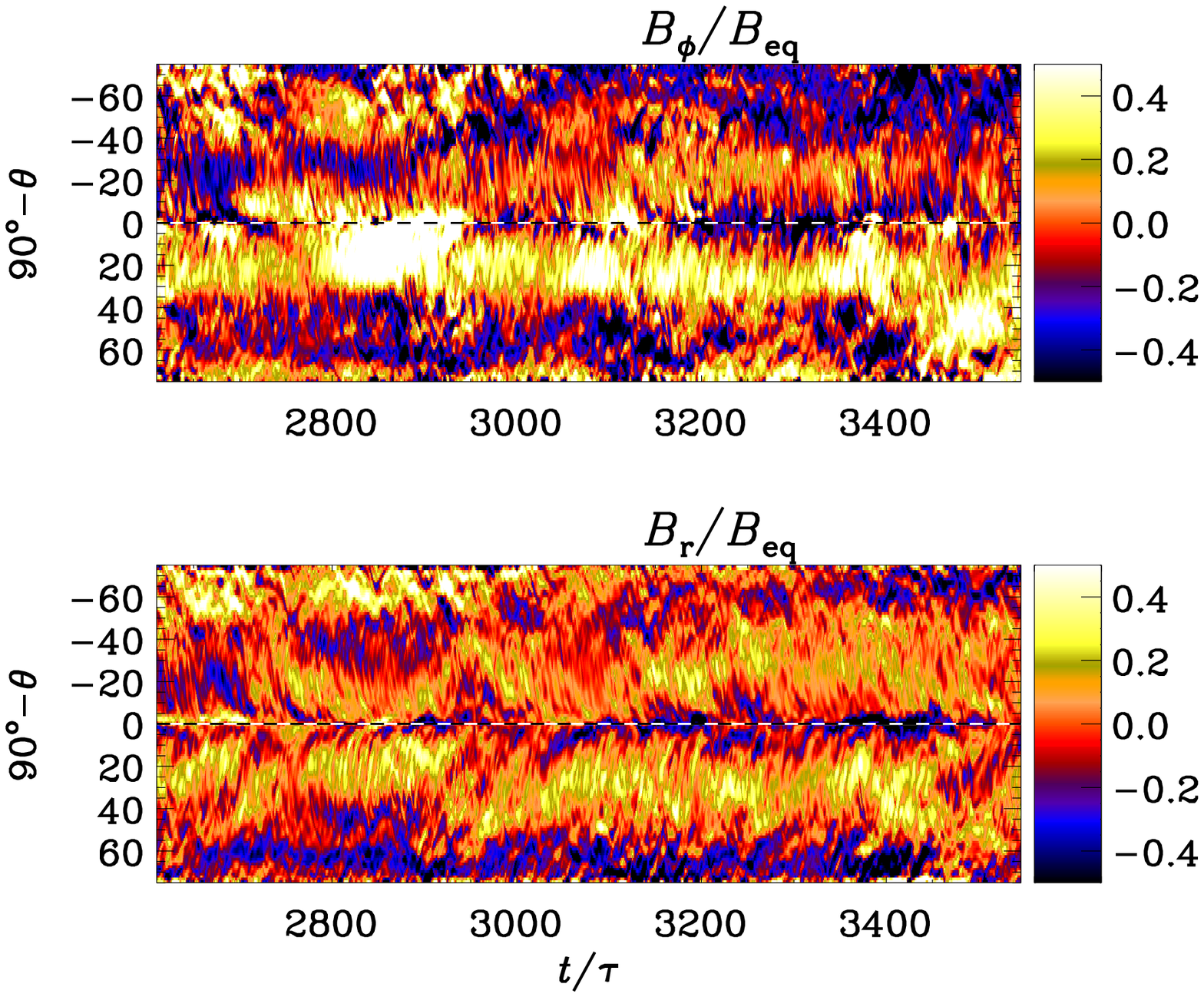}
\end{center}\caption[]{
Variation of $\meanB_{\phi}$ and $\meanB_{r}$ 
in the convection zone at $r=0.9\Rsun$ for Run~A5.
Dark blue shades represent negative and light yellow positive values.
The dashed horizontal lines show the location of the equator at
$\theta=\pi/2$.
The magnetic field is normalized by the equipartition value.
}
\label{but1}
\end{figure}

In \Fig{bphi} we show the time-averaged azimuthal magnetic field
$\meanB_{\phi}$ for Run~A5, A5a, and Ar1.
Note that the $\phi$ component of the magnetic field is also strong in the
coronal part and roughly antisymmetric about the equator.
Furthermore we find an oscillation of the volume-averaged rms
magnetic field in the convection zone;
see the left-hand panel of \Fig{bu} for Run~A5.
The growth tends to be steeper than the decline, the period
being around $t/\tau=220$.  
The field reaches a maximum of
60\% of the equipartition field strength, $\Beq$, which is comparable
to the values obtained in the forced turbulence counterparts both in
Cartesian and spherical coordinates (WB,WBM).  Comparing this
with the change of the kinetic energy, plotted as fluctuations of the
rms velocity squared, we find an anti-correlation with respect to the
magnetic field oscillation.  
The magnetic field is high (low), when
the velocity is low (high).
In the work by \cite{Br05}, the authors interpret this behavior as
the interplay of the magnetic backreaction and
the dynamo effect of the differential rotation.
Due to the Lorentz force a higher magnetic field strength leads to
quenching of the differential rotation.
An increased magnetic field quenches the Reynolds stress and thus
lowers the differential rotation, which limits the magnetic field.
A weak magnetic field leads to stronger differential rotation.
Similar behavior has been observed also in previous forcing
simulations (WBM).
This behavior is not seen as clearly in the
large-scale magnetic field which shows variations in strength, but not in 
sign.  
As shown in \Fig{but1} for Run~A5,
the $\meanB_{\phi}$ and $\meanB_{r}$ have local maxima in time and
in latitude, but the overall structure is nearly constant in time.
Even though the large-scale field structure is stationary,
the small-scale structures show an equatorward migration near the equator.
The reason for this is unclear, but meridional circulation does not
seem to play a role here.
\begin{figure}[t!]
\begin{center}
\includegraphics[width=0.8\columnwidth]{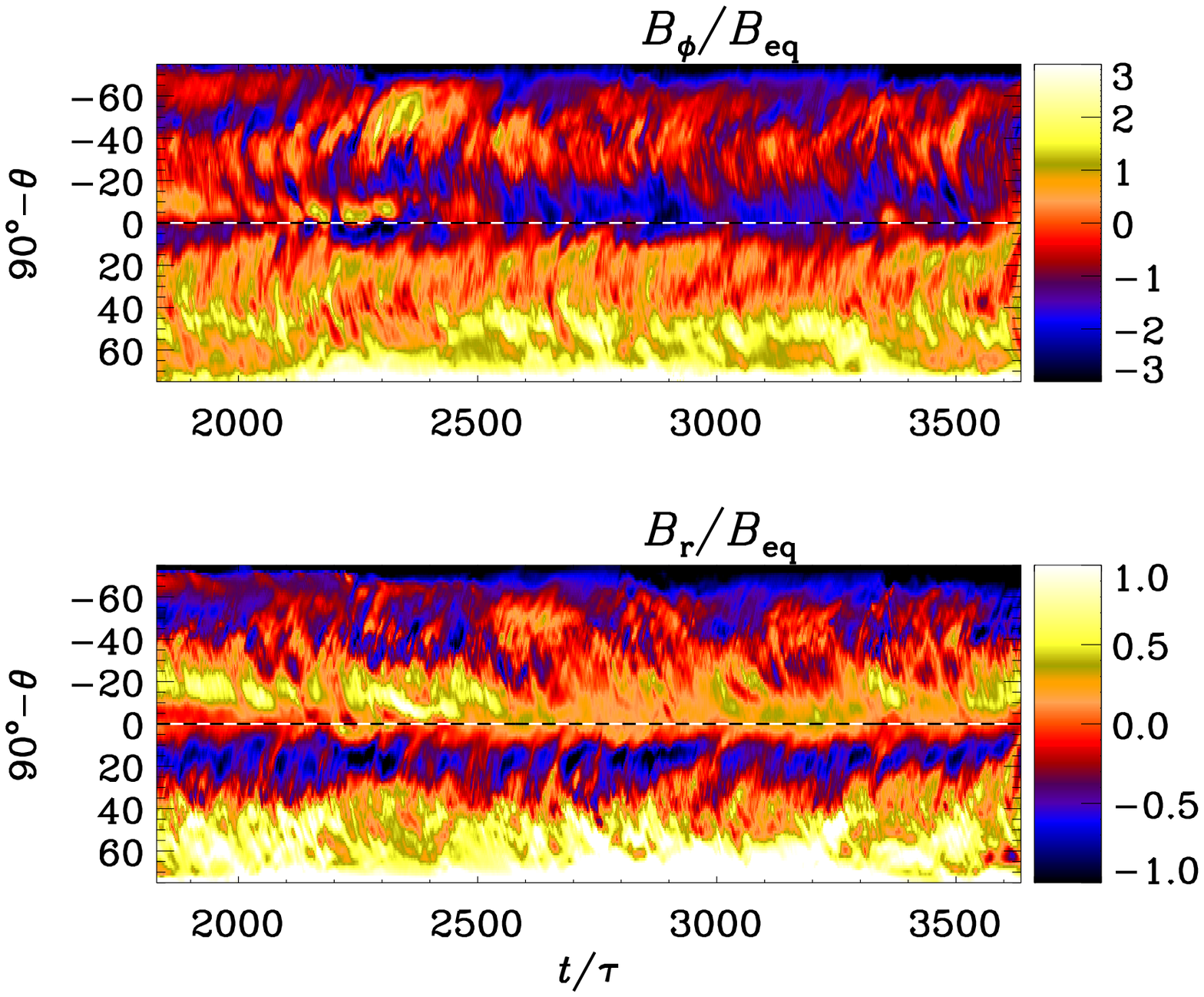}
\end{center}\caption[]{
Variation of $\meanB_{\phi}$ and $\meanB_{r}$ 
in the convection zone at $r=0.9\Rsun$ for Run~Ar1.
Dark blue shades represent negative and light yellow positive values.
The dashed horizontal lines show the location of the equator at
$\theta=\pi/2$.
The magnetic field is normalized by the equipartition value.
}
\label{but2}
\end{figure}
\begin{figure}[t!]
\begin{center}
\includegraphics[width=0.49\columnwidth]{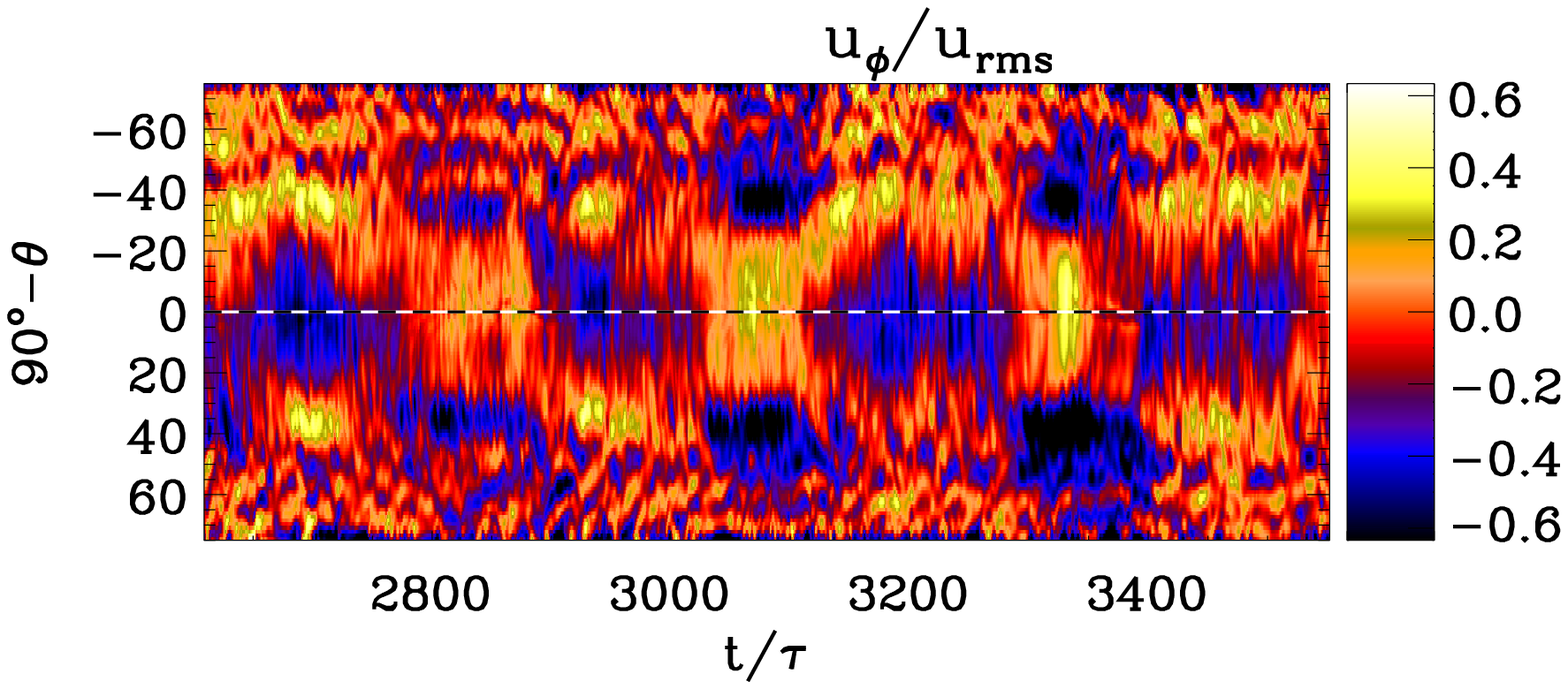}
\includegraphics[width=0.49\columnwidth]{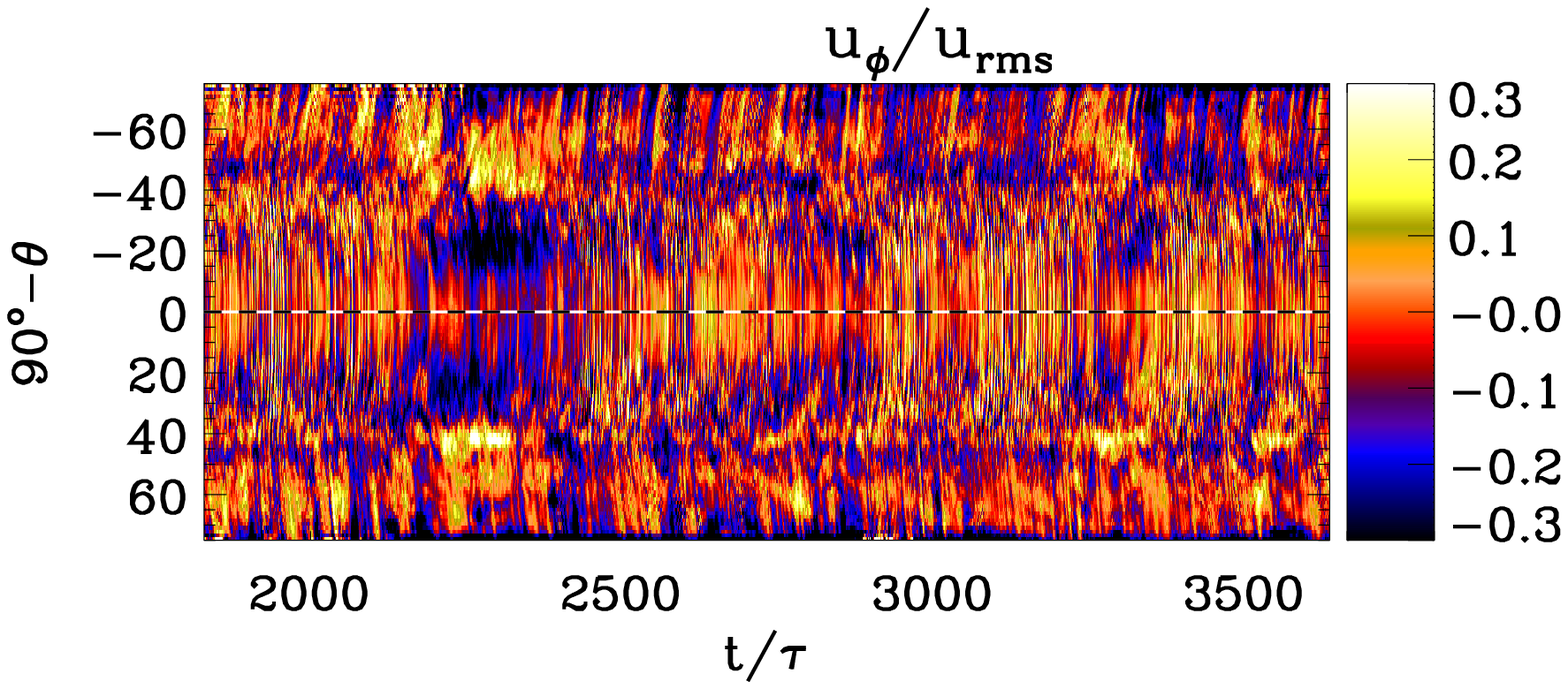}
\end{center}\caption[]{
Variation of $\overline{U}_{\phi}$ in the convection zone at
$r=0.9\Rsun$
for Run~A5 (left panel) and Run~Ar1 (right panel).
Dark blue shades represent negative and light yellow positive values.
The dashed horizontal lines show the location of the equator at
$\theta=\pi/2$.
The velocity is normalized by the mean rms velocity in the
convection zone.
}
\label{uphi}
\end{figure}

In Run~Ar1, the magnetic field reaches up to 5.5
times the equipartition value, but does not show a periodic oscillation;
see the right hand panel of \Fig{bu}.
In comparable work \citep{KKBMT10}, similar values for the field strength
were found.
However, the rms velocity is also quenched, when the magnetic field is
high.
Looking at $\meanB_{\phi}$ and $\meanB_{r}$, plotted over time and
latitude in \Fig{but2}, the large-scale magnetic field is similar to
Run~A5, which is constant in time without any oscillation.
In the recent work by \cite{KMB12}, the authors found an oscillatory
behavior of $\meanB_{\phi}$ and $\meanB_{r}$ including equatorward
migration for latitudes below $60^\circ$, which is the first
time that such a result is obtained from direct numerical convection
simulation of rotating convection.

The azimuthal velocity $\overline{U}_{\phi}$ versus time and latitude 
(\Fig{uphi}) shows minima at the same times as the maxima 
of the magnetic field occur.
In Run~A5a, the
occurrence of strong magnetic fields suppresses the differential rotation.
The pattern of the azimuthal velocity is symmetric about the equator and
shows an oscillatory behavior, which is not that clear in the
large-scale magnetic field.
In the $\overline{U}_{\phi}$ plot in \Fig{uphi} of Run~Ar1, we find just one localized
minimum, which coincides with the low values of
$\urms(t)^2/\bra{\urms^2}_t$ between $t/\tau=2100$ and $2400$.

\subsection{Coronal ejections}

\begin{figure*}[t!]
\begin{center}
\includegraphics[width=0.327\textwidth]{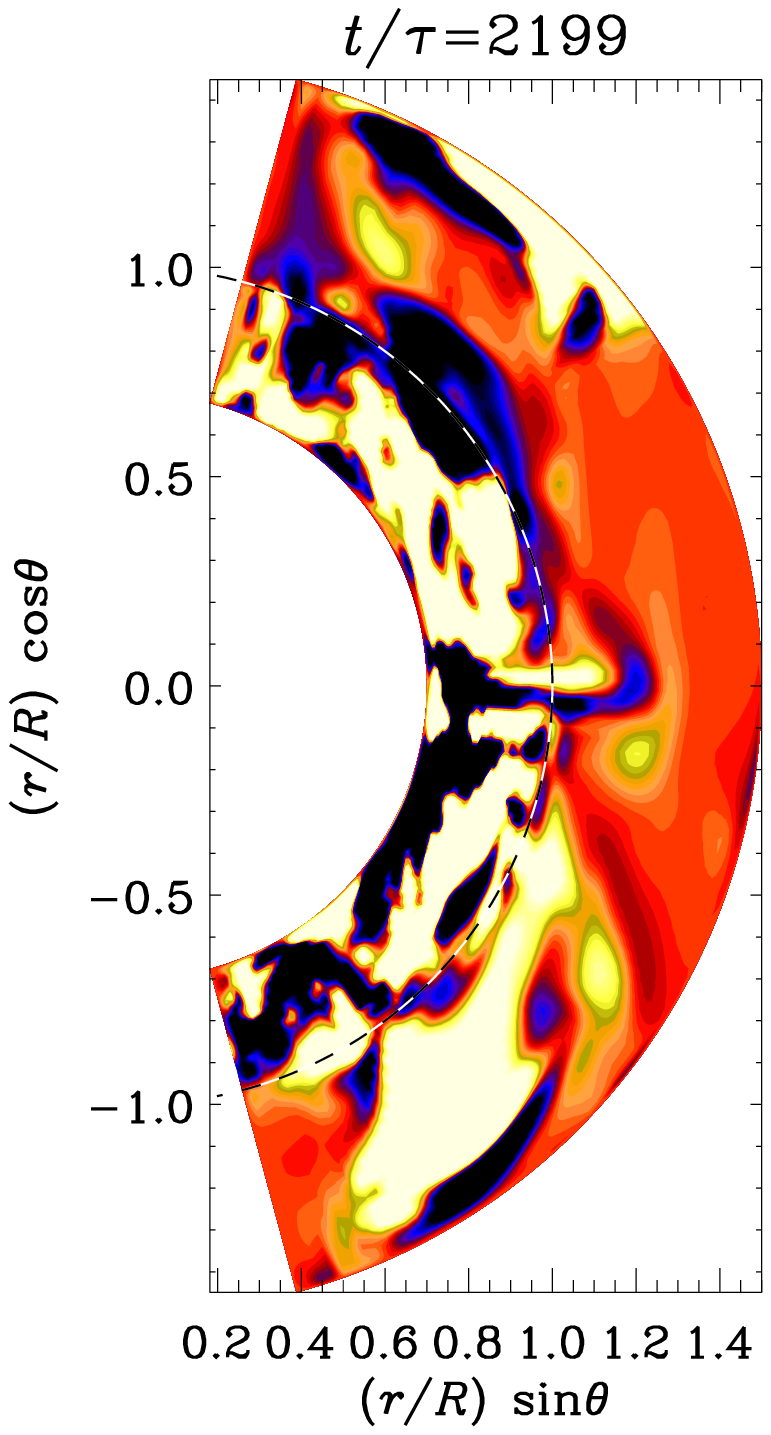}
\includegraphics[width=0.327\textwidth]{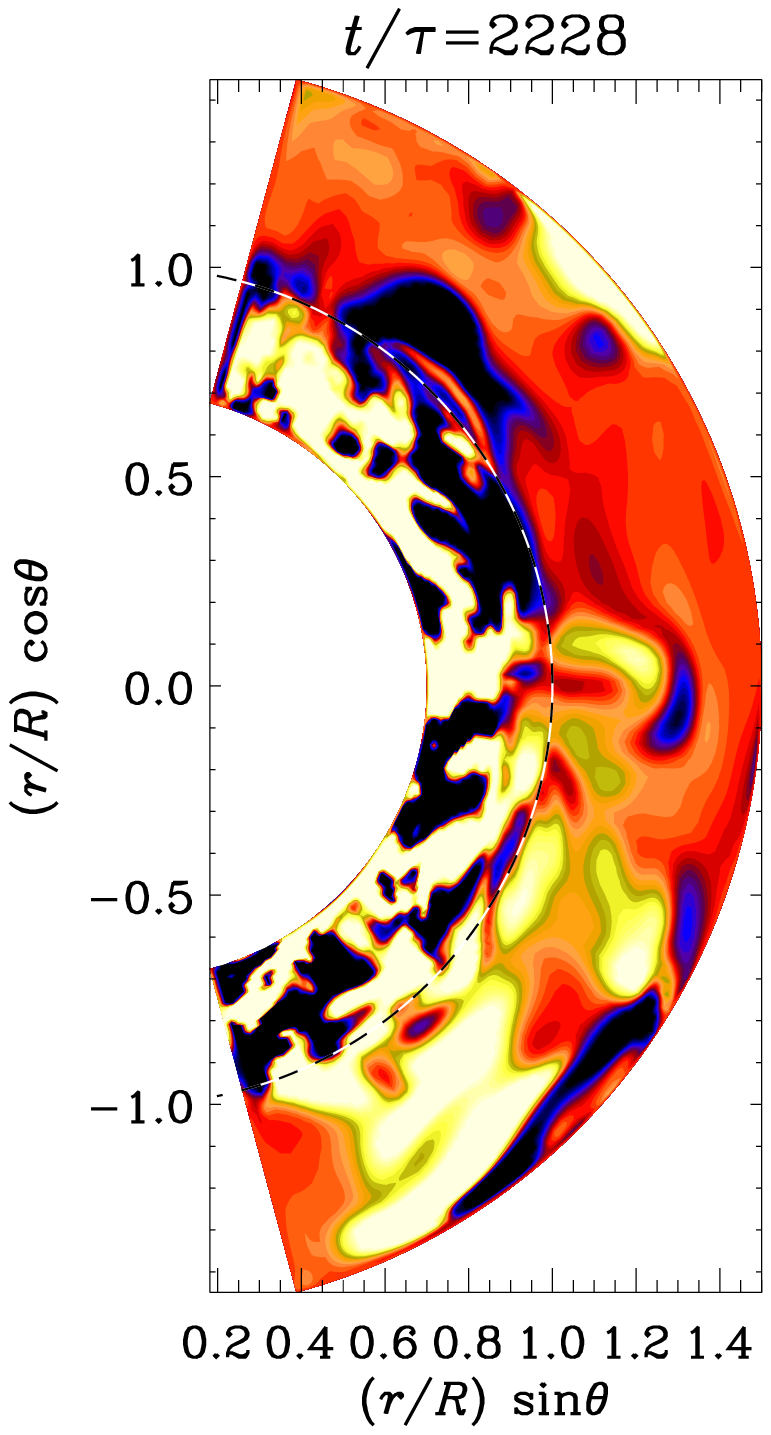}
\includegraphics[width=0.327\textwidth]{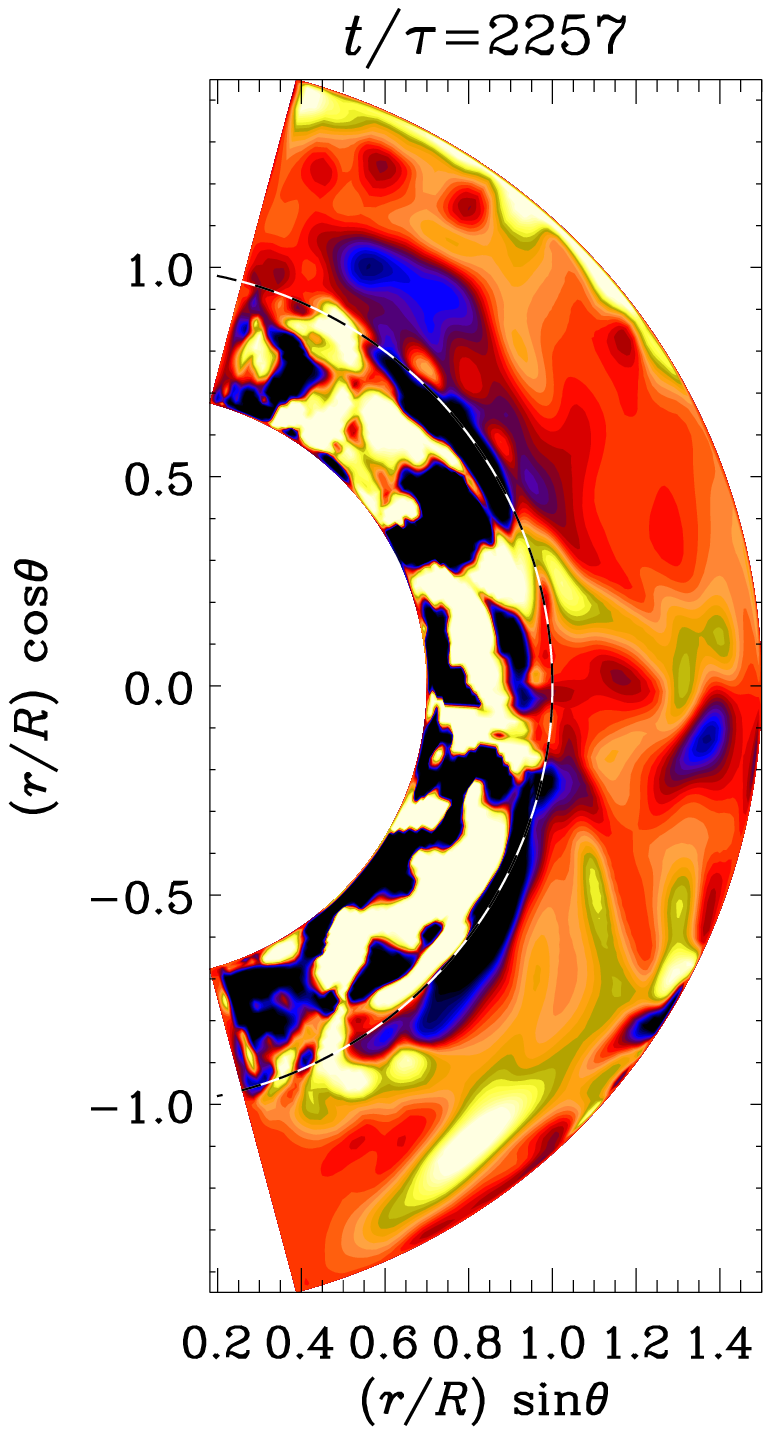}
\end{center}\caption[]{
Time series of a coronal ejection near the equator ($\theta=\pi/2$),
taken from Run~A5.
The normalized current helicity,
$\mu_0 R\,\overline{\JJ\cdot\BB}/\bra{\overline{\BB^2}}_t$, is shown in a
color-scale representation from different times; dark blue represents
negative and light yellow positive values.
The dashed horizontal lines show the location of the surface at
$r=\Rsun$.
}
\label{jb}
\end{figure*}

\begin{figure*}[h!]
\begin{center}
\includegraphics[width=0.331\textwidth]{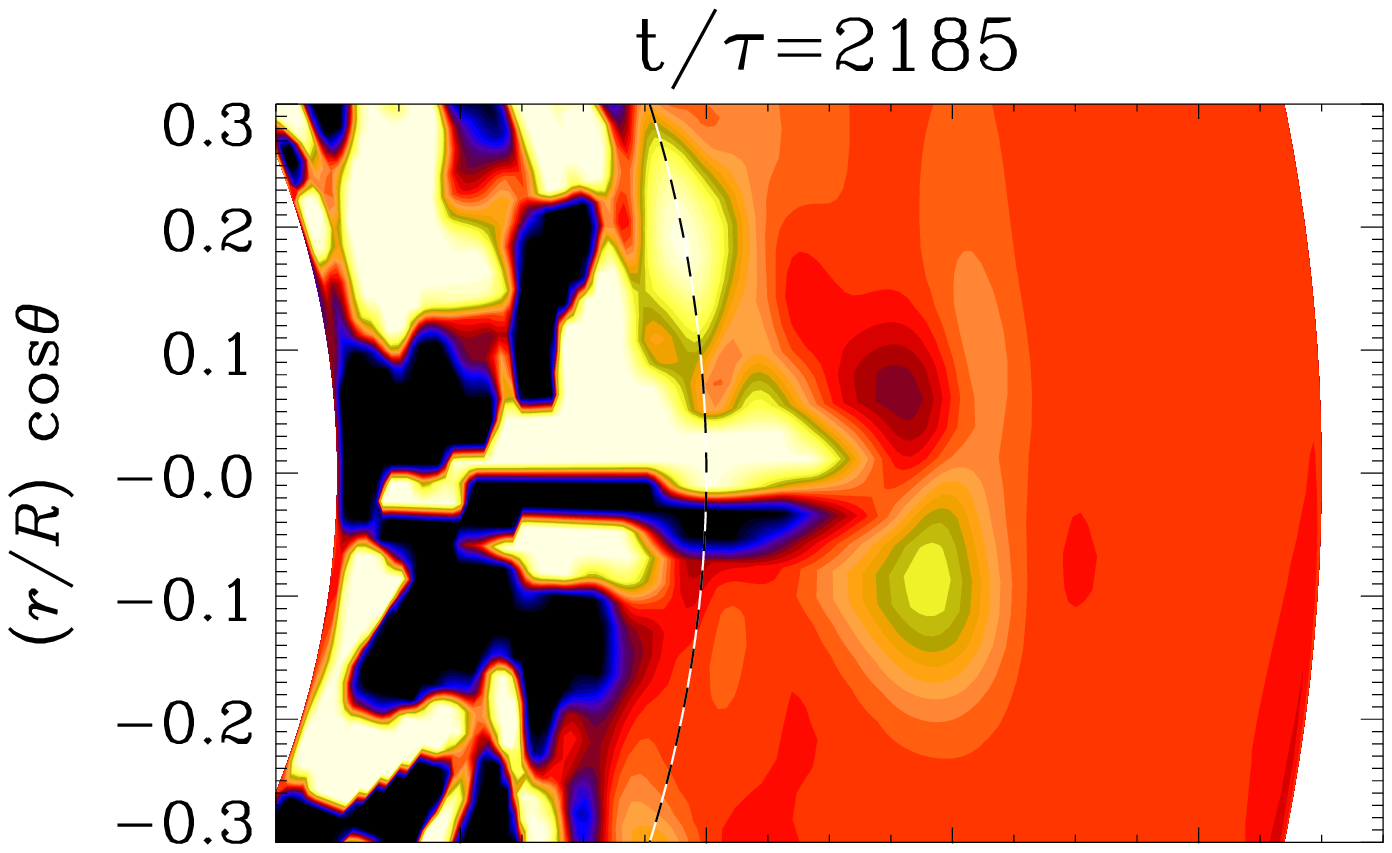}
\includegraphics[width=0.269\textwidth]{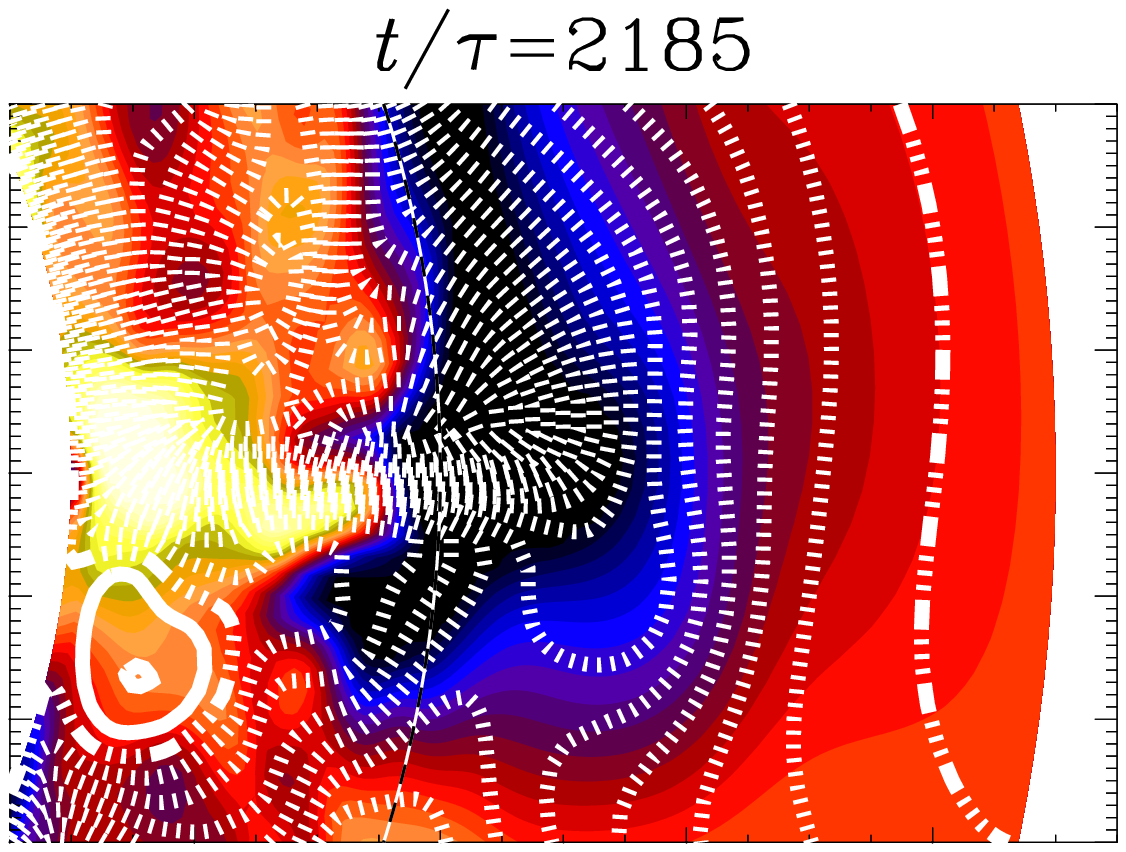}
\includegraphics[width=0.269\textwidth]{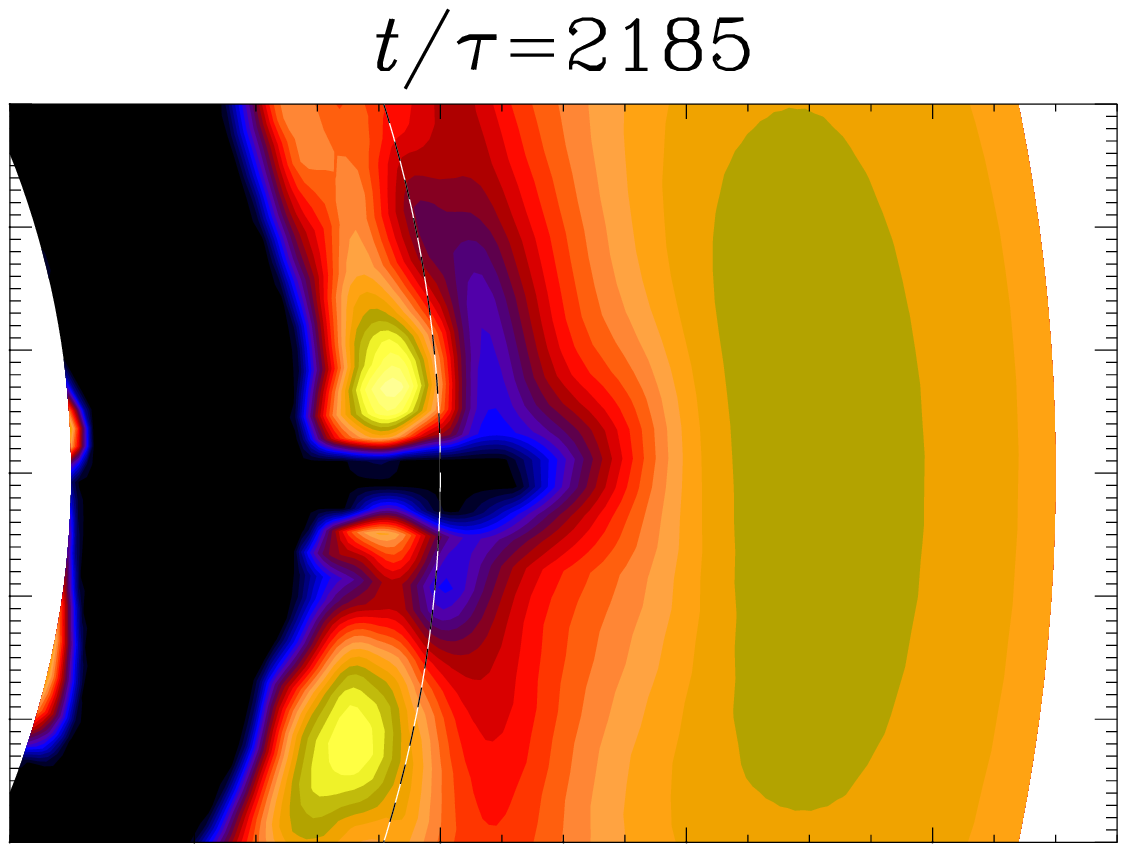}
\includegraphics[width=0.331\textwidth]{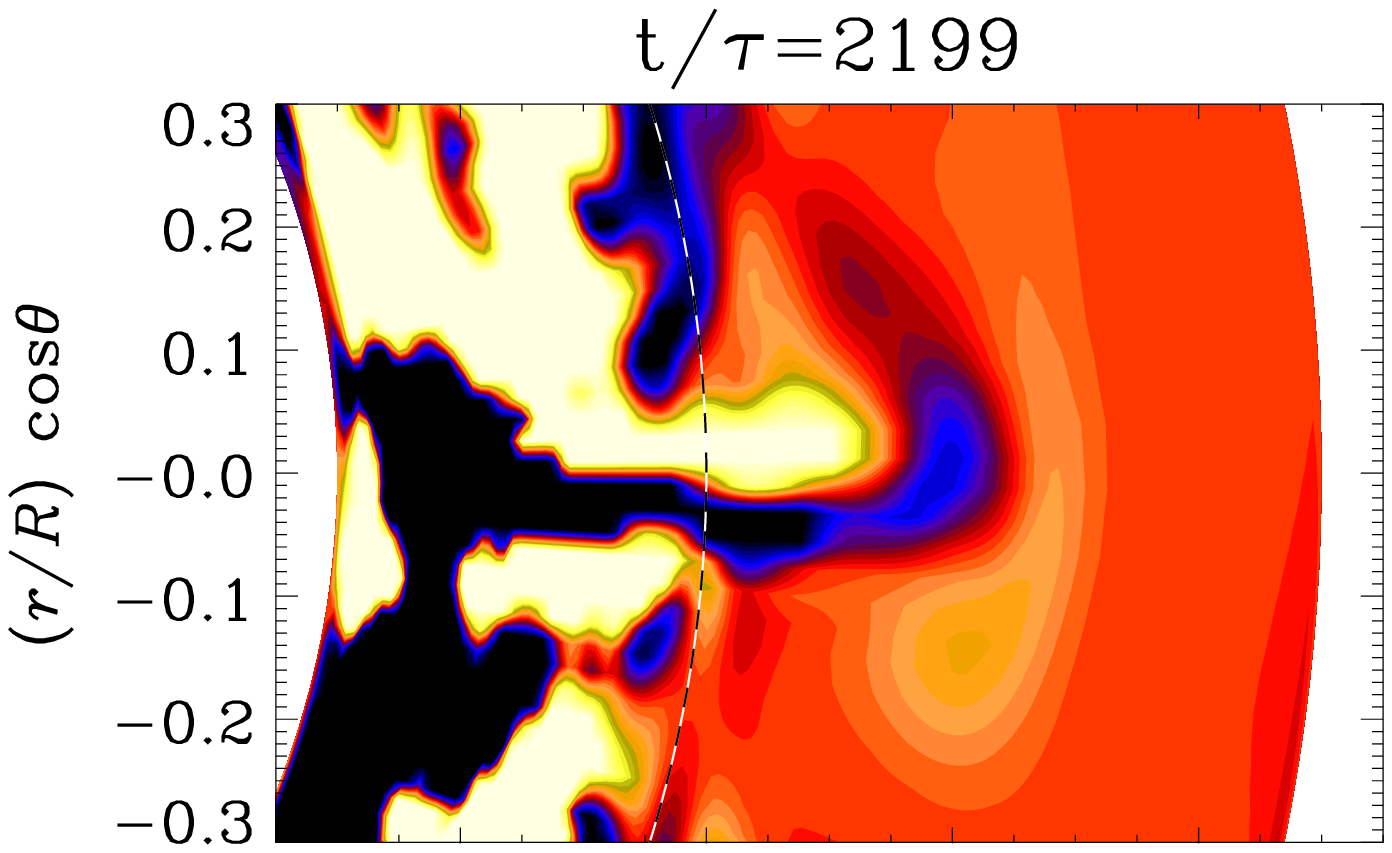}
\includegraphics[width=0.269\textwidth]{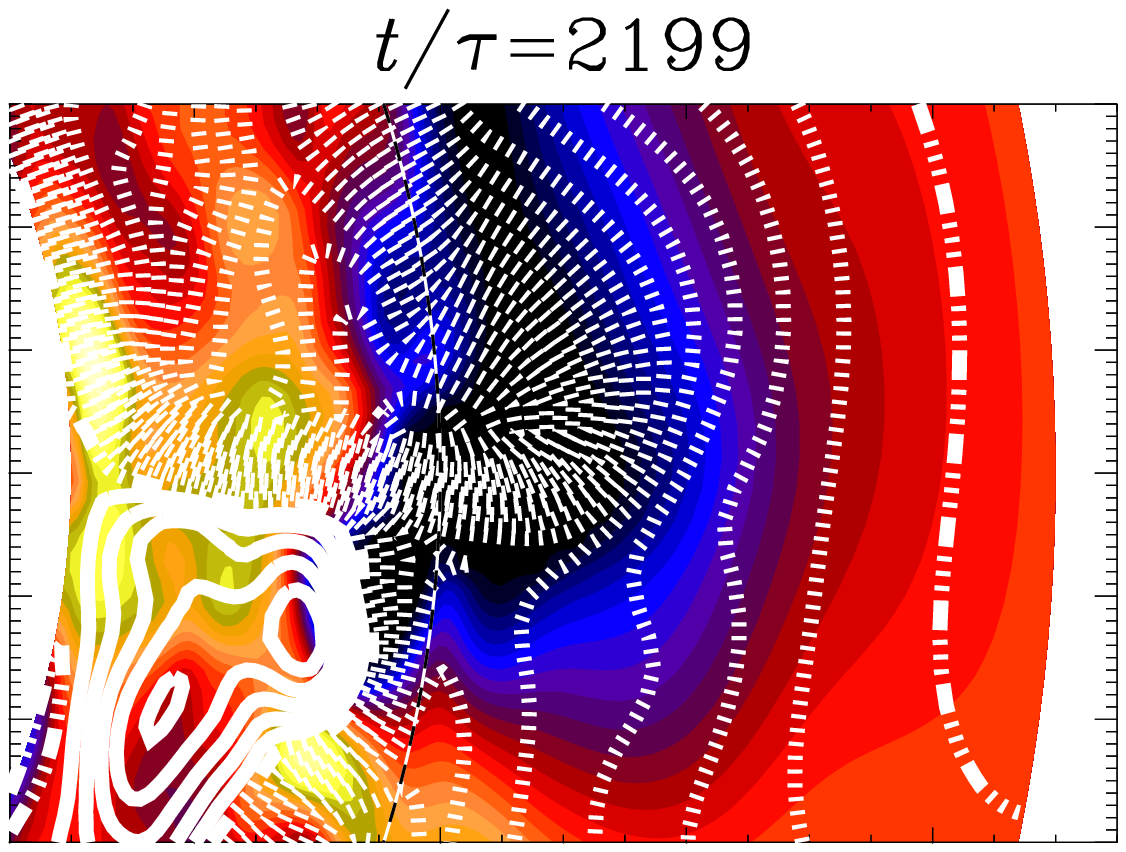}
\includegraphics[width=0.269\textwidth]{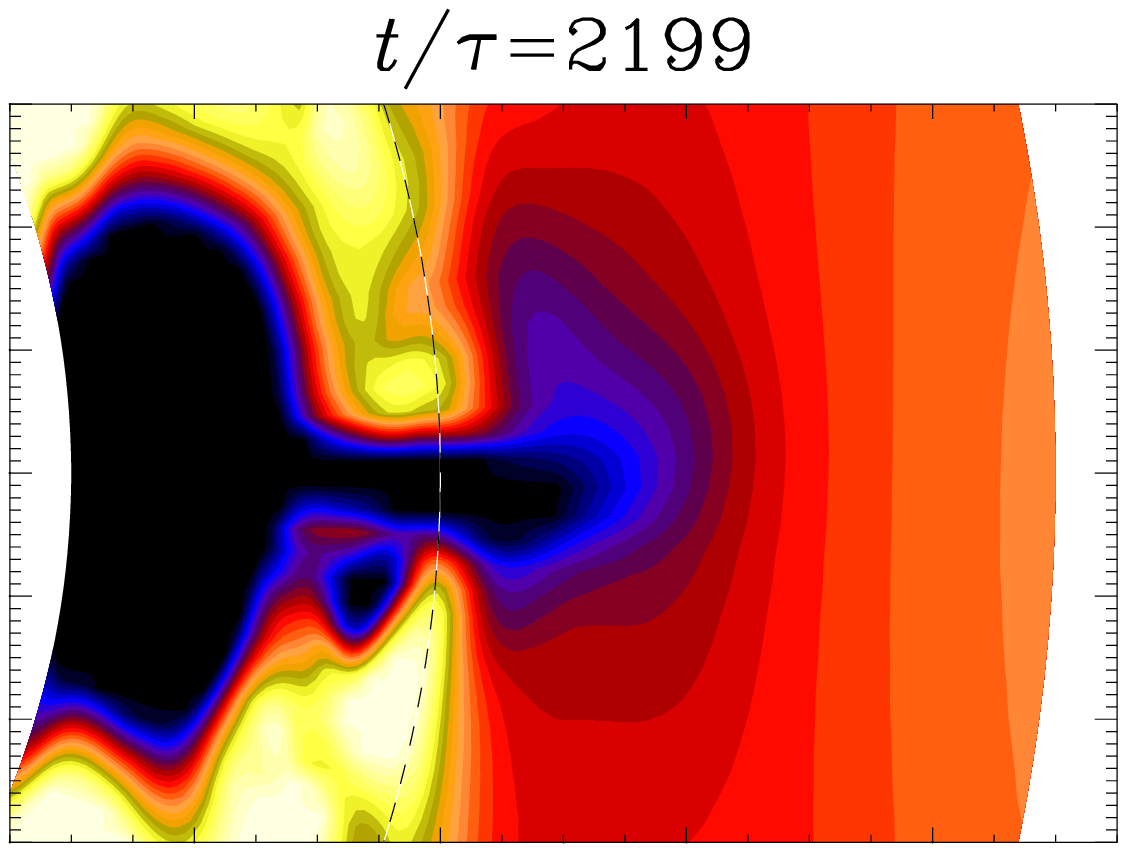}
\includegraphics[width=0.331\textwidth]{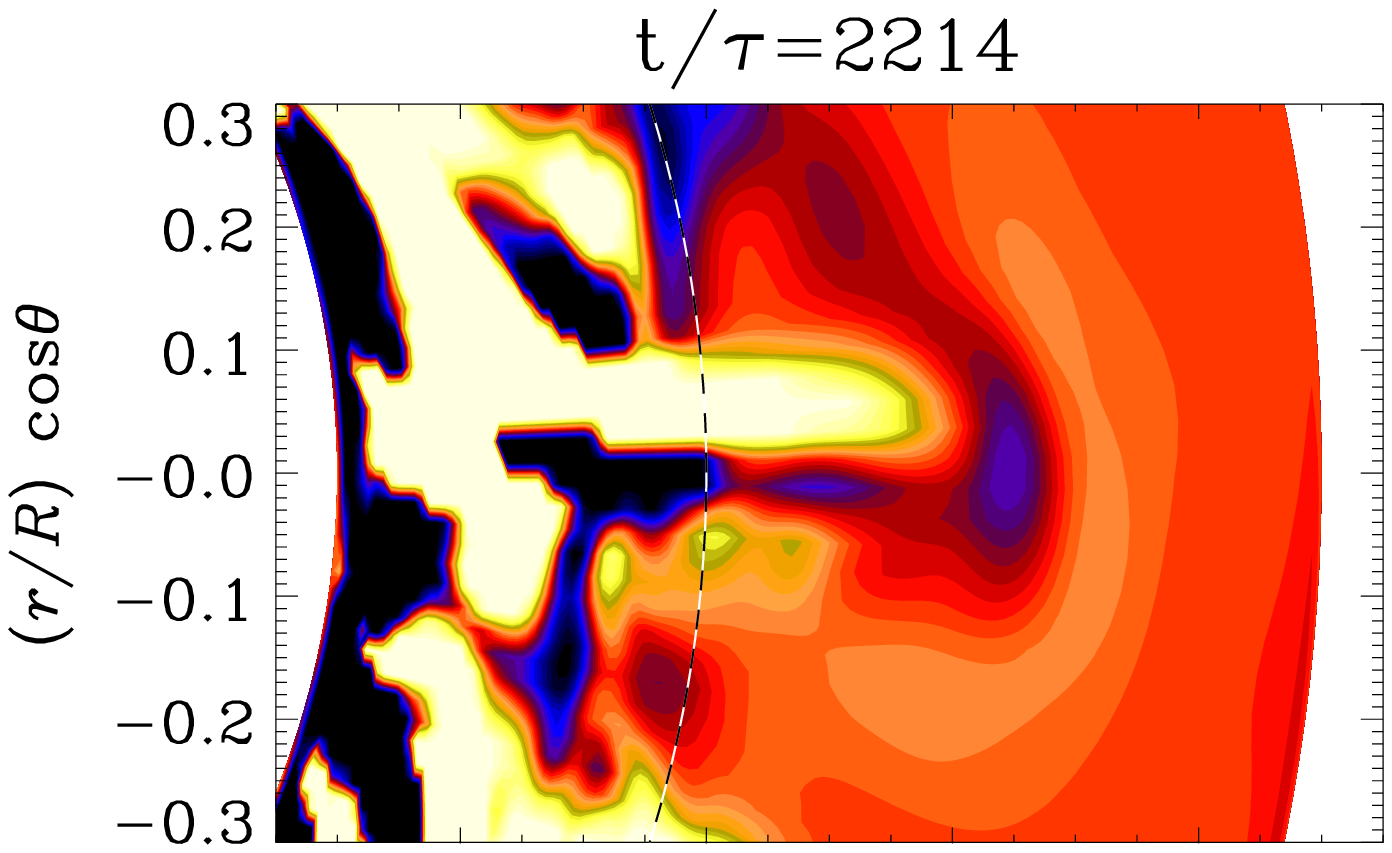}
\includegraphics[width=0.269\textwidth]{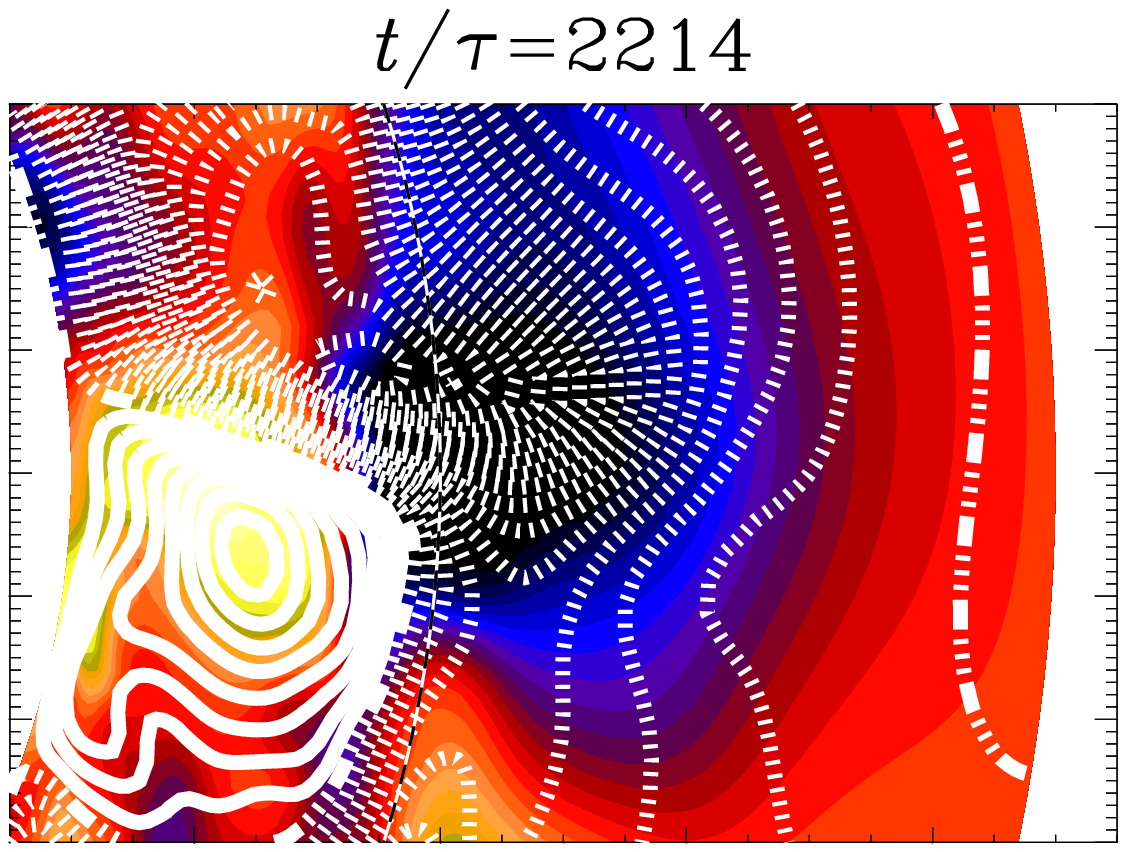}
\includegraphics[width=0.269\textwidth]{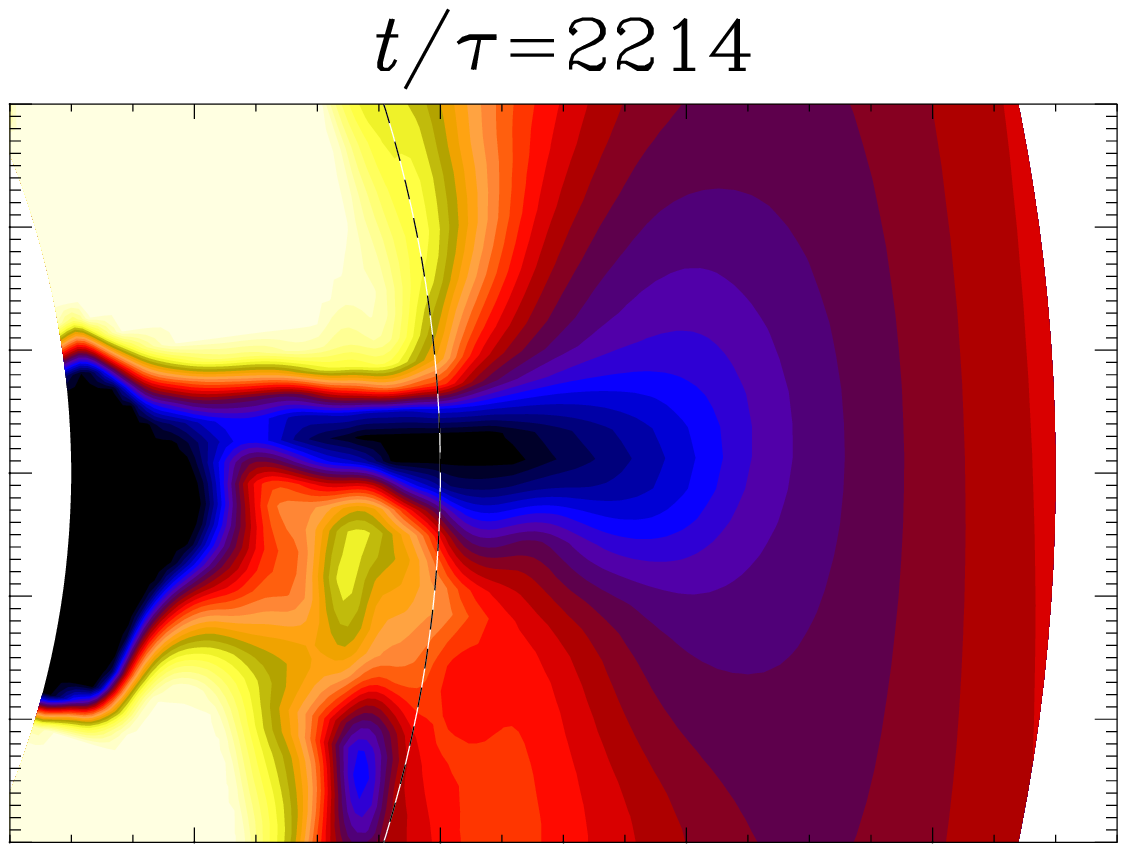}
\includegraphics[width=0.331\textwidth]{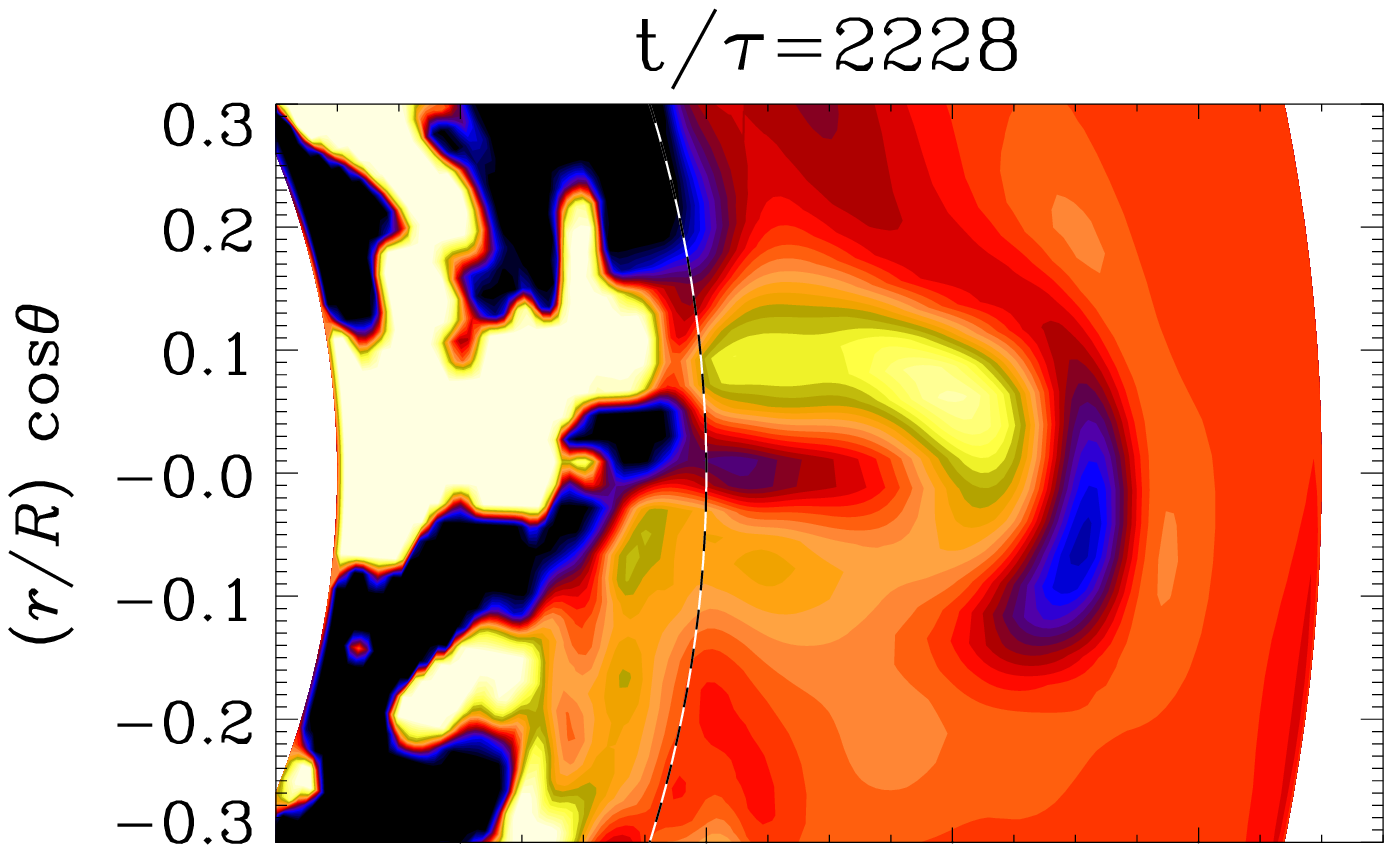}
\includegraphics[width=0.269\textwidth]{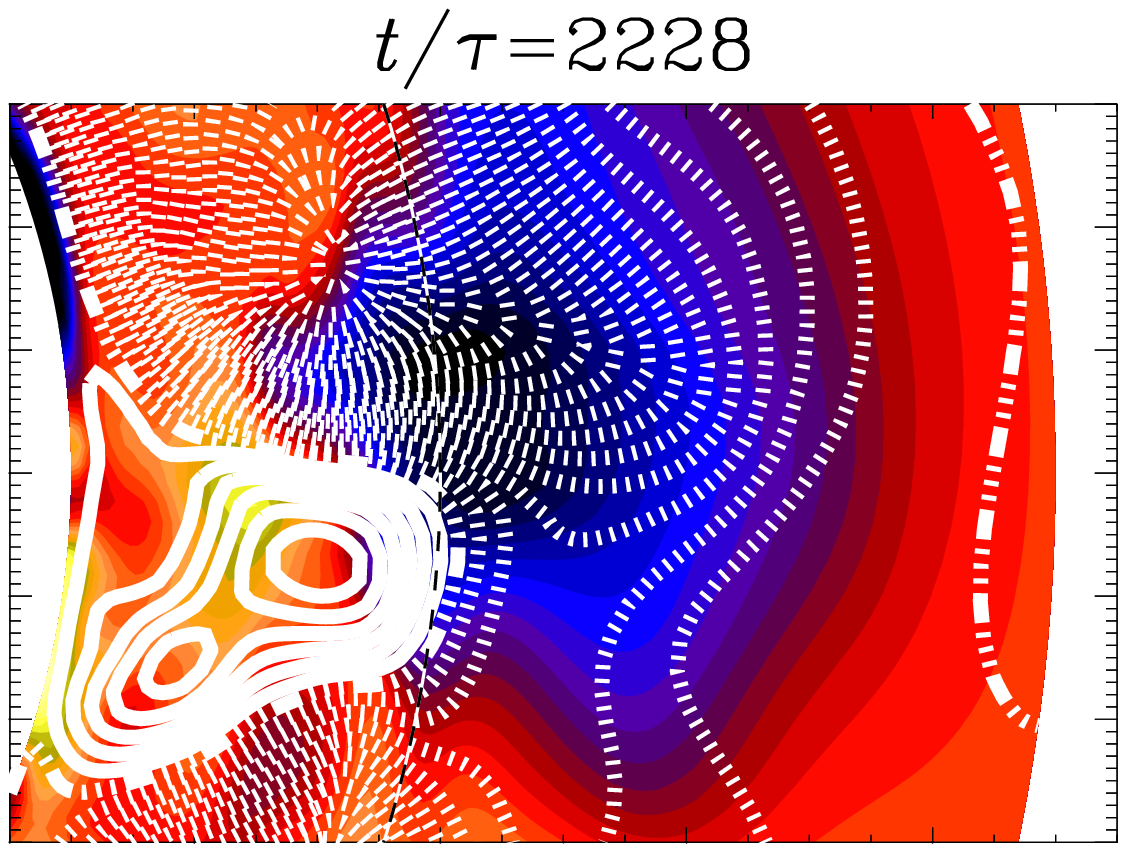}
\includegraphics[width=0.269\textwidth]{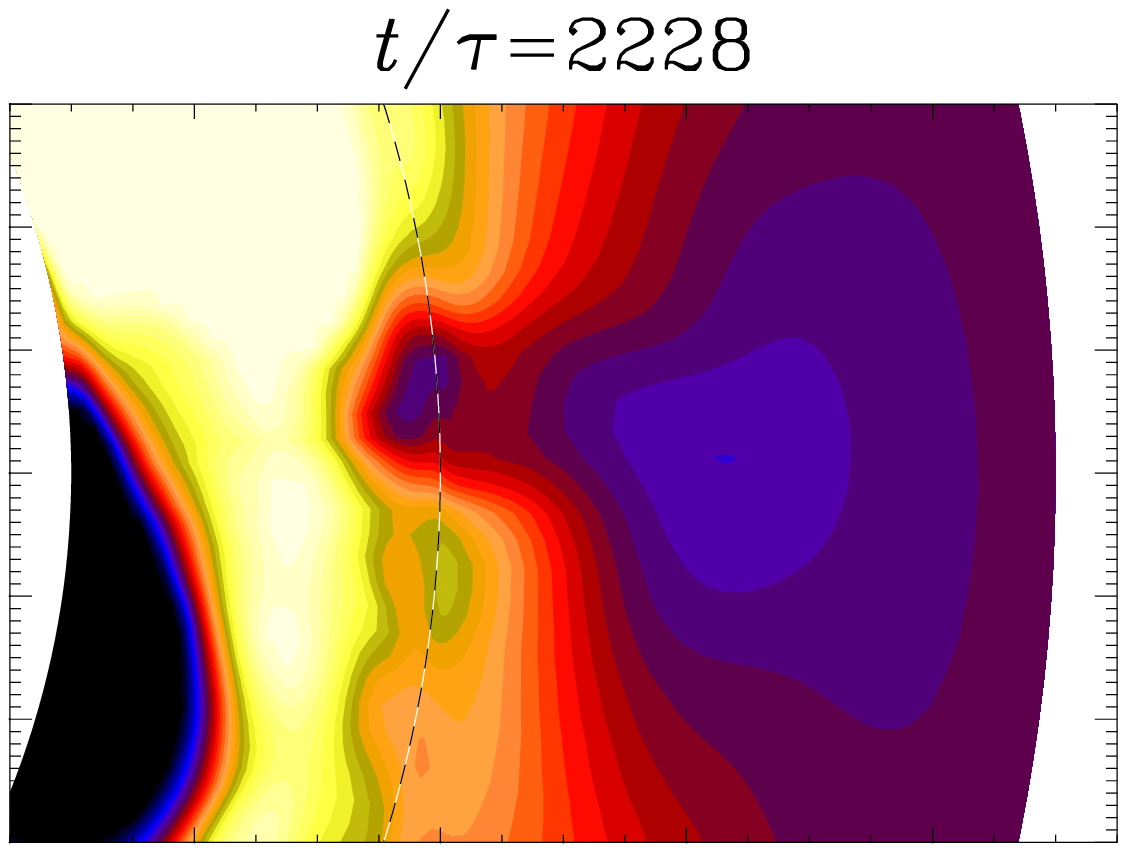}
\includegraphics[width=0.331\textwidth]{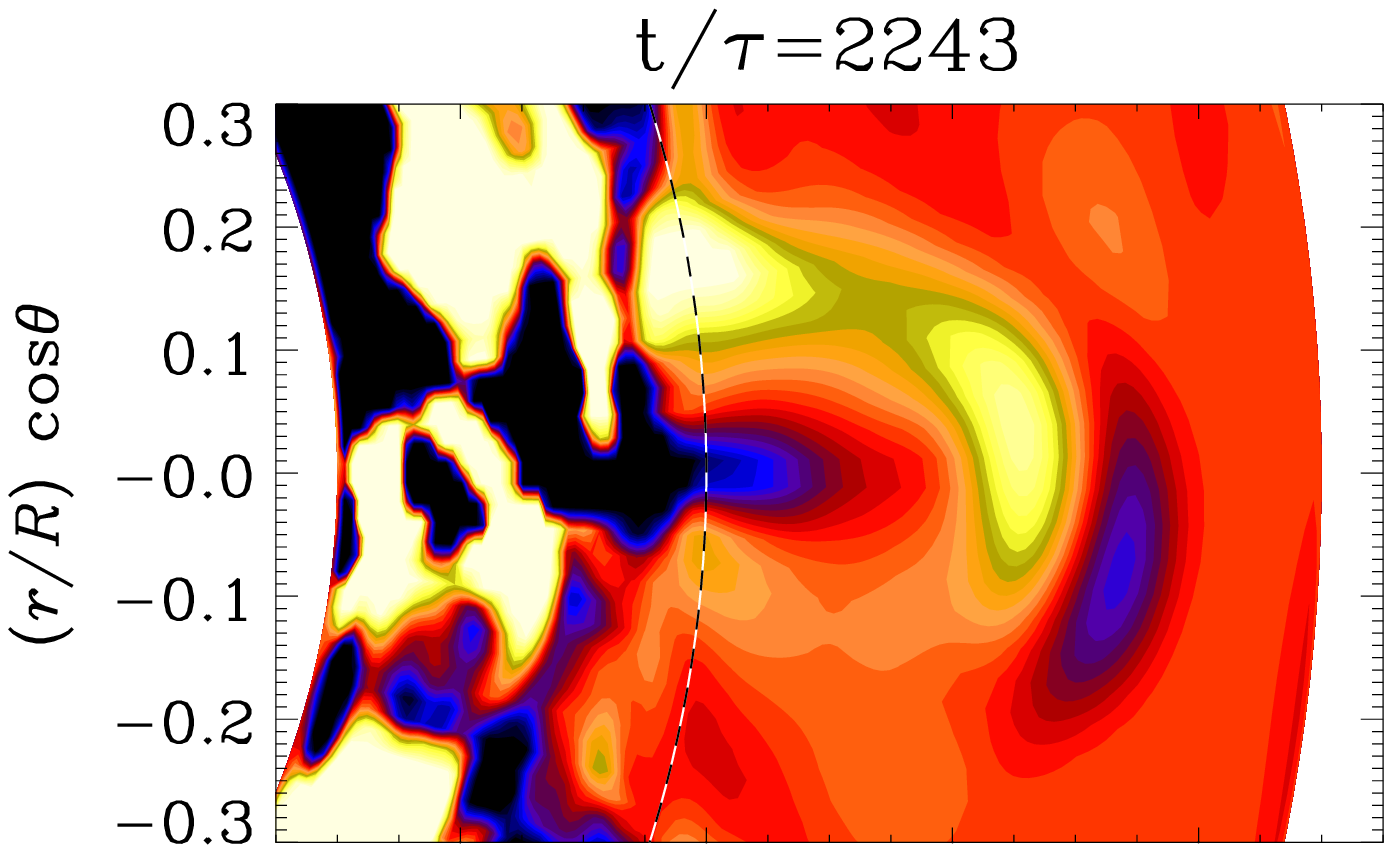}
\includegraphics[width=0.269\textwidth]{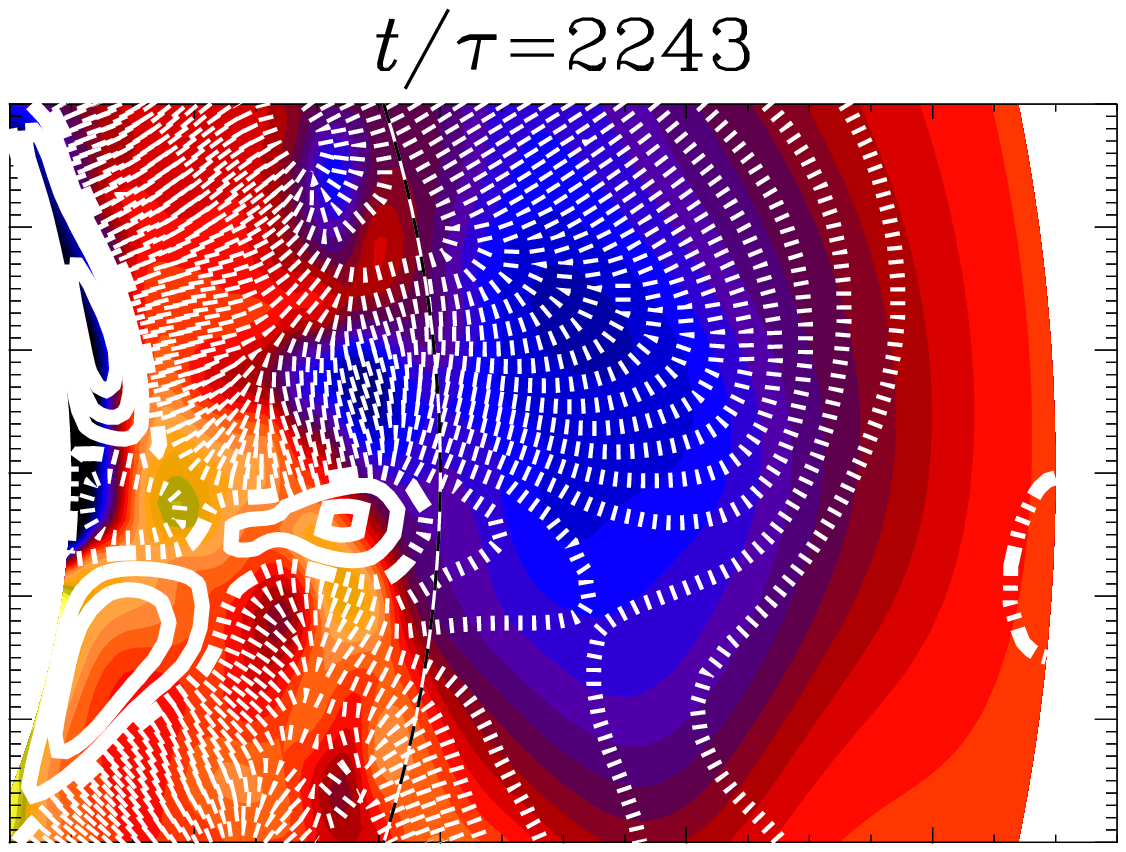}
\includegraphics[width=0.269\textwidth]{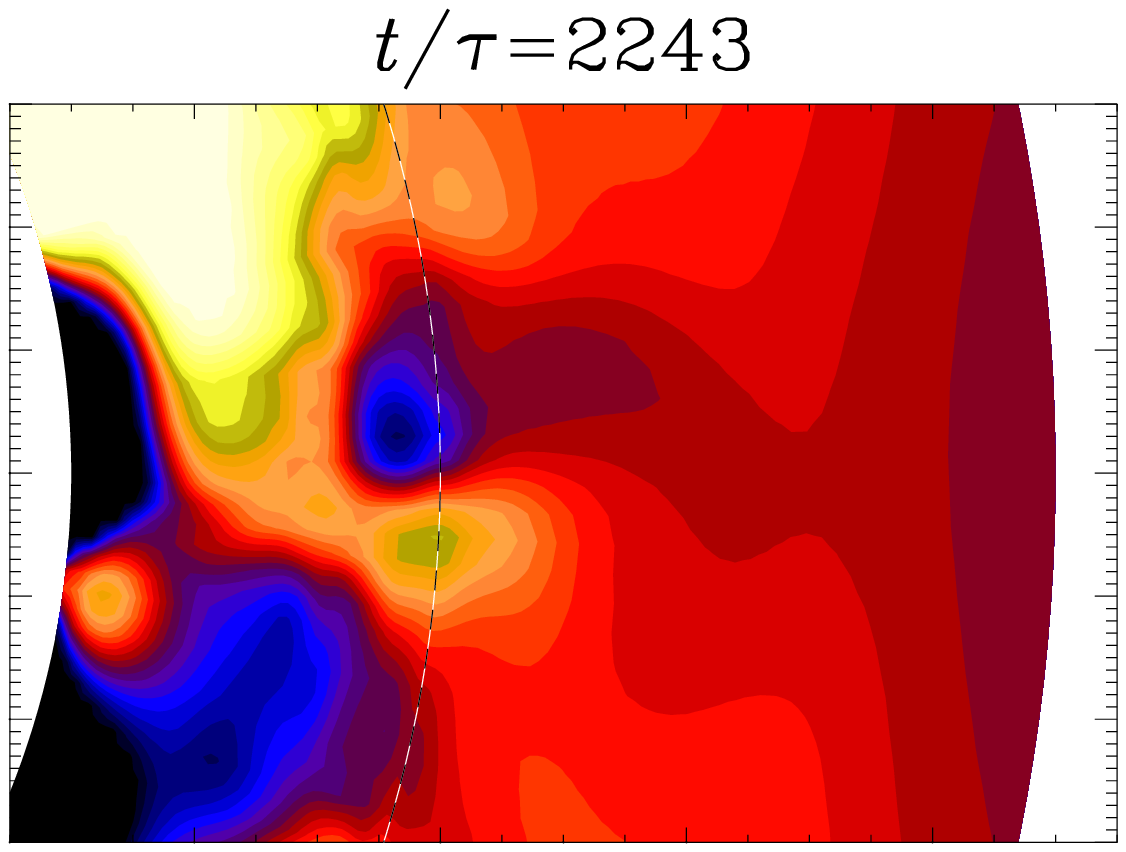}
\includegraphics[width=0.331\textwidth]{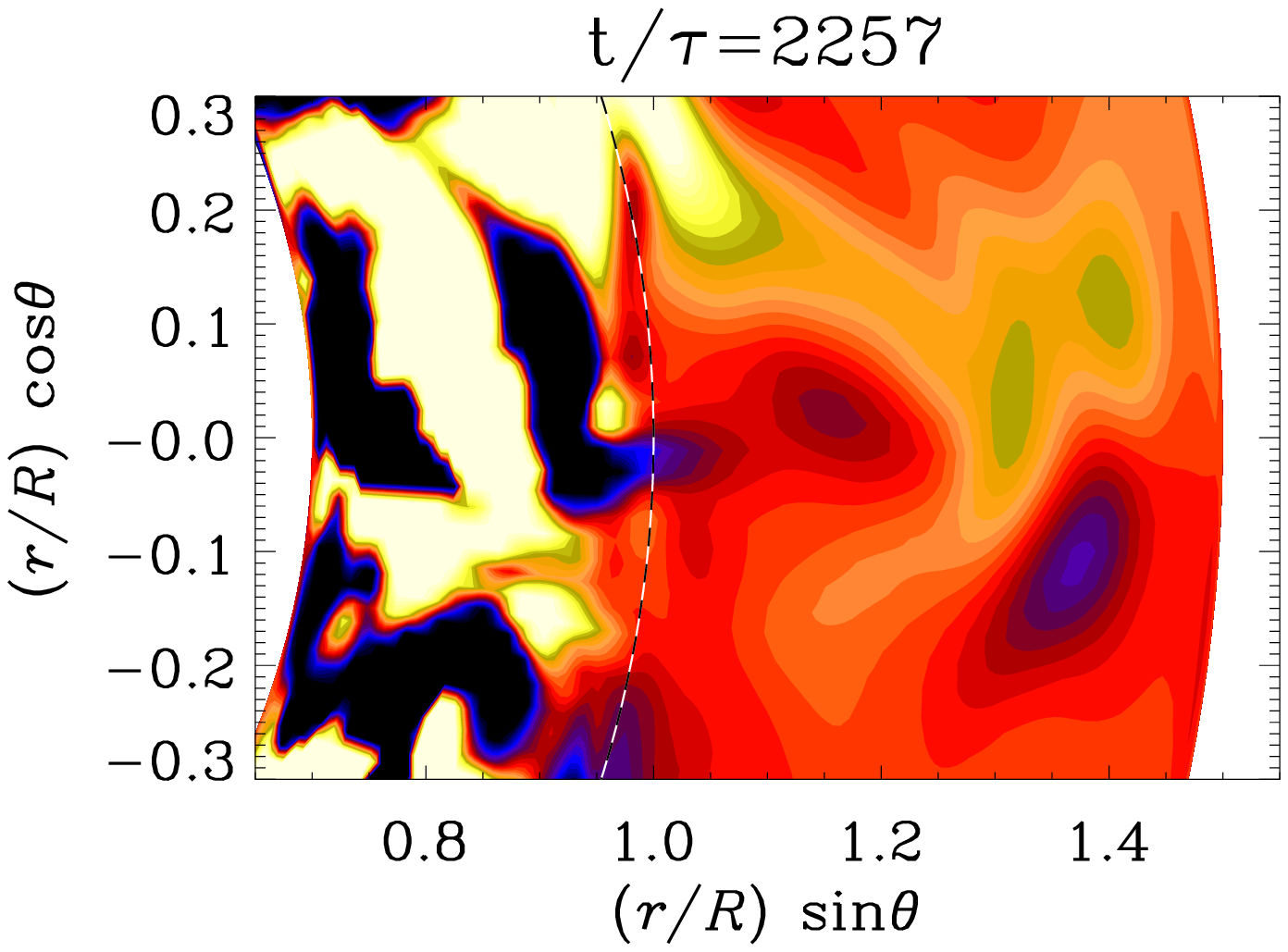}
\includegraphics[width=0.269\textwidth]{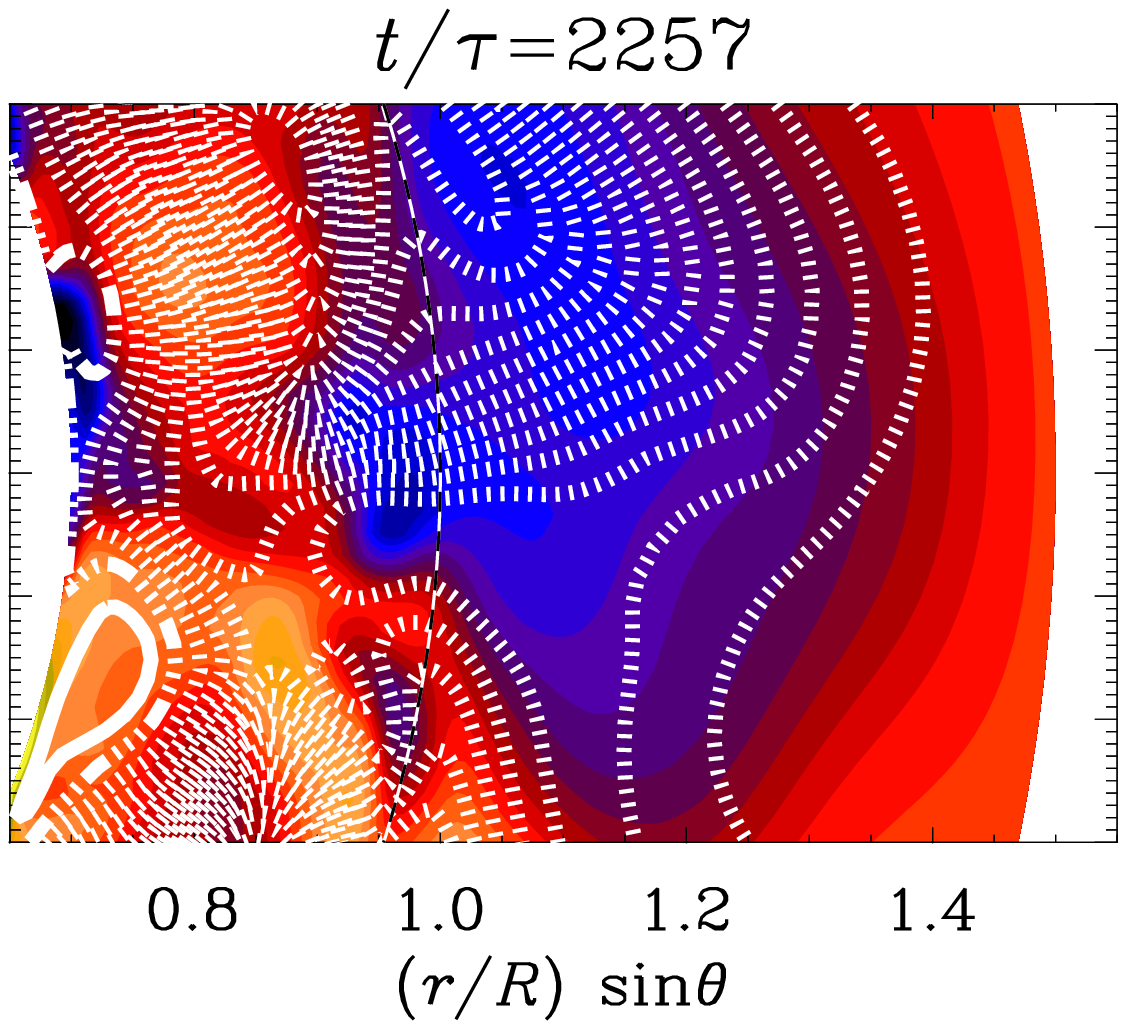}
\includegraphics[width=0.269\textwidth]{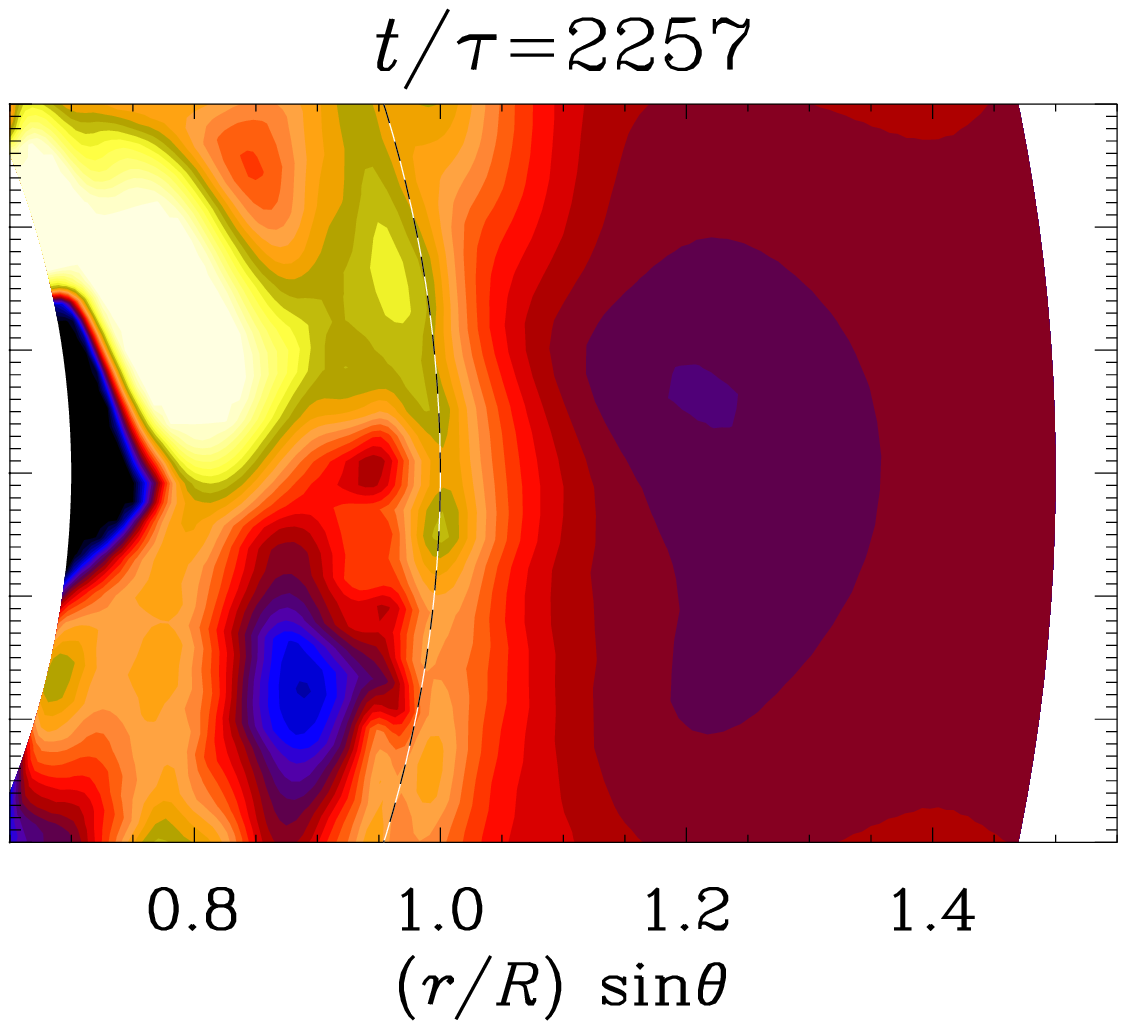}
\end{center}\caption[]{
Time series of a coronal ejection zoomed into the region of the ejection
near the equator ($\theta=\pi/2$),
taken from Run~A5.
The dashed horizontal lines show the location of the surface at
$r=\Rsun$.
{\it Left column}: normalized current helicity,
$\mu_0 R\,\overline{\JJ\cdot\BB}/\bra{\overline{\BB^2}}_t$.
{\it Middle column}: magnetic field, contours of
$r\sin{\theta}\meanA_\phi$ are shown together
with a color-scale representation of $\meanB_\phi$.
The contours of $r\sin{\theta}\meanA_\phi$ correspond to
field lines of $\overline{\BB}$ in the $r$,$\theta$ plane, where
solid lines represent clockwise magnetic field lines and the dashed ones
counter-clockwise.
{\it Right column:} density fluctuations
$\Delta\overline{\rho(t)}=\overline{\rho(t)}-\bra{\overline{\rho}}_t$.
For all plots, the color-scale represents negative as dark blue and
positive as light yellow.
}
\label{eje}
\end{figure*}

\begin{figure*}[t!]
\begin{center}
\includegraphics[width=0.6\textwidth]{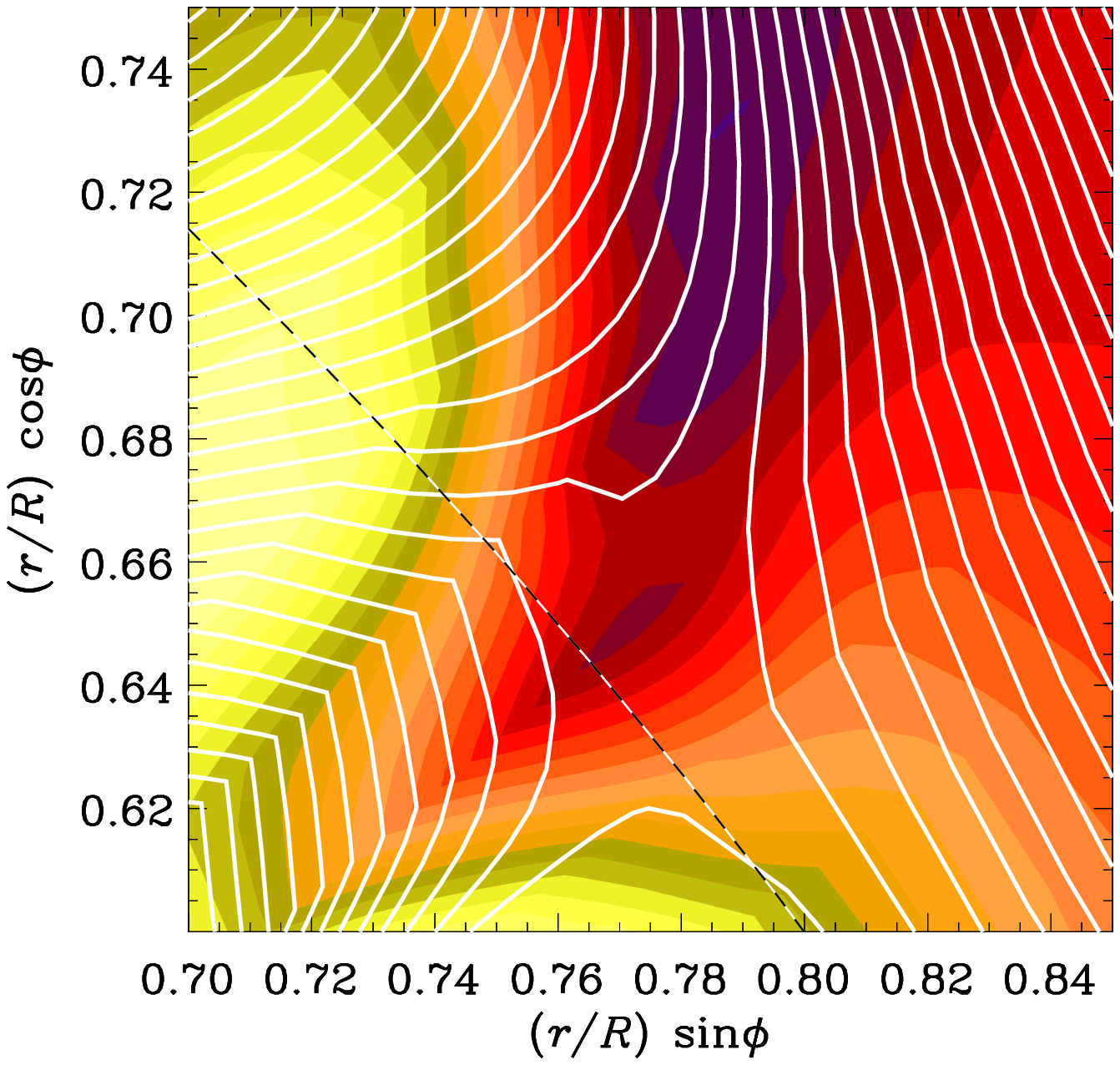}
\end{center}\caption[]{
{\sf X}-point-like structure in the $r$,$\phi$ plane at the equator
($\theta=\pi/2$) at $t/\tau$=2204 zoomed into the ejection region,
taken from Run~A5.
Contours of $rA_{\theta}$ are shown together
with a color-scale representation of $B_{\theta}$; dark
blue stands for negative and light yellow for positive values.
The contours of $rA_{\theta}$ correspond to
field lines of $\BB$ in the $r$,$\phi$ plane.
The dashed horizontal lines show the location of the surface at
$r=\Rsun$.
}
\label{eje_phi}
\end{figure*}

In the runs that we have been performed so far, and of which only three
have been discussed in this paper, only a small fraction of events
can be identified with actual coronal ejections similar to the ones
seen in WB and WBM.
Especially the Runs~A5 and Ar1 show some clear
ejection events. 
There the magnetic field emerges out of the
convection zone and is ejected as an isolated structure.  In \Fig{jb}
we have plotted the normalized current helicity, $\mu_0
R\,\overline{\JJ\cdot\BB}/\bra{\overline{\BB^2}}_t$, as a time series
for Run~A5.
At small scales, the current helicity density, $\JJ\cdot\BB$, is a good proxy
for magnetic helicity density, $\AAA\cdot\BB$, and is, as opposed to
the latter, gauge invariant.
In addition, the
current helicity can be an indicator of helical magnetic structures,
which are believed to be present in coronal mass ejections
\citep{LO94,LO01,PVSK00,RAK02,TKT11}.
Close to the equator a bipolar structure emerges through the surface.
The inner bulk has a positive current helicity, in \Fig{jb} represented
by a yellow color, and it pushes an arc with negative current helicity
ahead of it; see \Fig{eje}.
Such bipolar ejections have been identified in earlier work (WBM)
and compared with the `three-part structure' of coronal mass ejection,
which is described in \cite{LO96}.
The three parts consist of a prominence, which is similar to the
bulk seen in our simulations, a front with an arc shaped structure
corresponding to our arc, and a cavity between these two features.
A bipolarity of twisted magnetic field has also been seen in observed
magnetic clouds by \cite{LLLK11}.
Even though the domain of the simulation is larger in the $\theta$
direction than in WBM, the ejections are much smaller, which is
actually closer to the CMEs observed on the Sun.
In the work of WBM the ejections have a size that corresponds to about 500 Mm,
whereas in this work they seems to have a size corresponding to around 100 Mm
if scaled to the solar radius.
The ejections seem to expand slightly, but no significant expansion
rate can be measured using this resolution.
Comparing with the forced turbulence runs,
the difference in size is mostly due to the more complex and
fluctuating magnetic field in convection runs.
In the sequence of images of \Fig{eje}, an ejection near the equator
reaches the outer boundary and leaves the domain.
To investigate the mechanism driving the ejection, we look at the dynamics
of the magnetic field in \Fig{eje}, where field lines of the
azimuthally averaged mean field are
shown as contours of $r\sin{\theta}\meanA_\phi$, and colors represent
$\meanB_\phi$ together with the density fluctuations and current helicity.
During the ejection, one may notice a strong
concentration of magnetic field lines that are directed radially outwards.
This concentration appears first beneath the surface and then emerges
below the current helicity structure and follows it up into the
coronal part.
Investigating the direction of field lines of the mean field in the
time series in \Fig{eje}, an {\sf X}-point can be found.
In the first panel, at $r=1.07\,\Rsun$ and $\theta=\pi/2+0.1$, the
magnetic field lines form a junction-like shape.
The dotted line represents a counter-clockwise oriented field loop, so
at the two corners of the junction there are field lines with
opposite signs.
After around 14 turnover times this ``junction'' has reconnected at the
same position as where the ejection is detected.
It appears that these two events are related to each other.
Looking at the magnetic field line in the $r$,$\phi$ plane, which here
is {\em not} averaged over the perpendicular direction
(\Fig{eje_phi}),
we identify a structure which has a shape similar 
to an {\sf X}-point.

\begin{figure*}[t!]
\begin{center}
\includegraphics[width=0.275\textwidth]{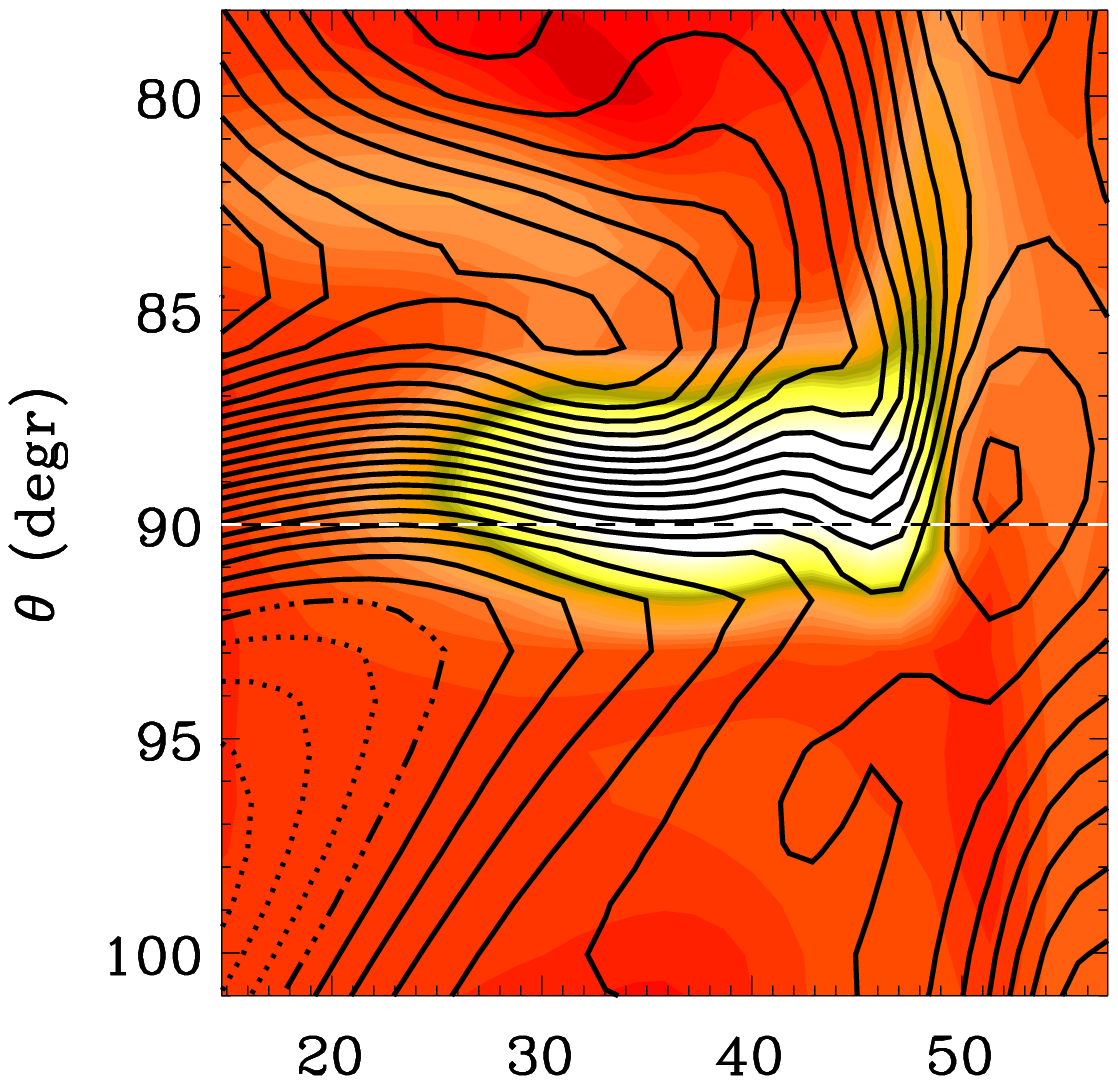}
\includegraphics[width=0.275\textwidth]{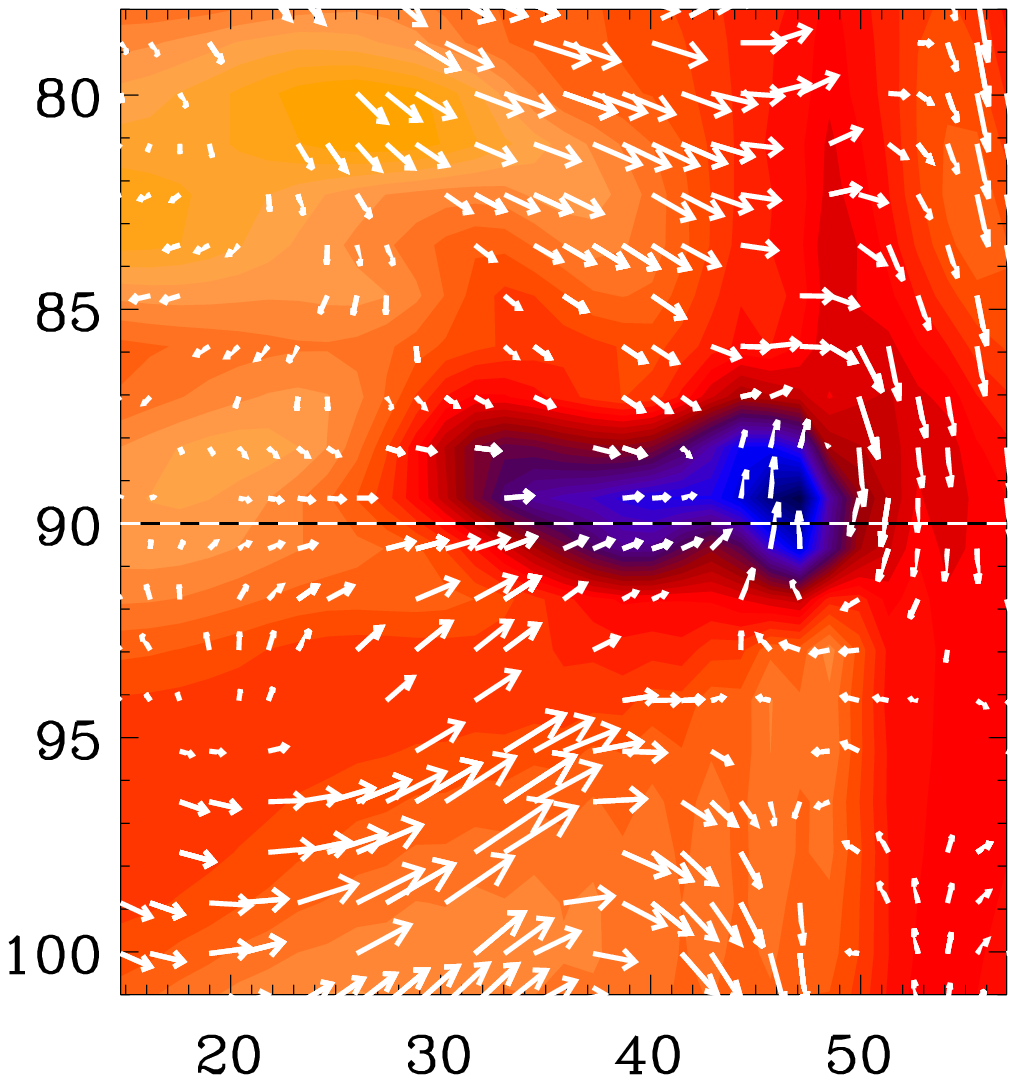}
\includegraphics[width=0.275\textwidth]{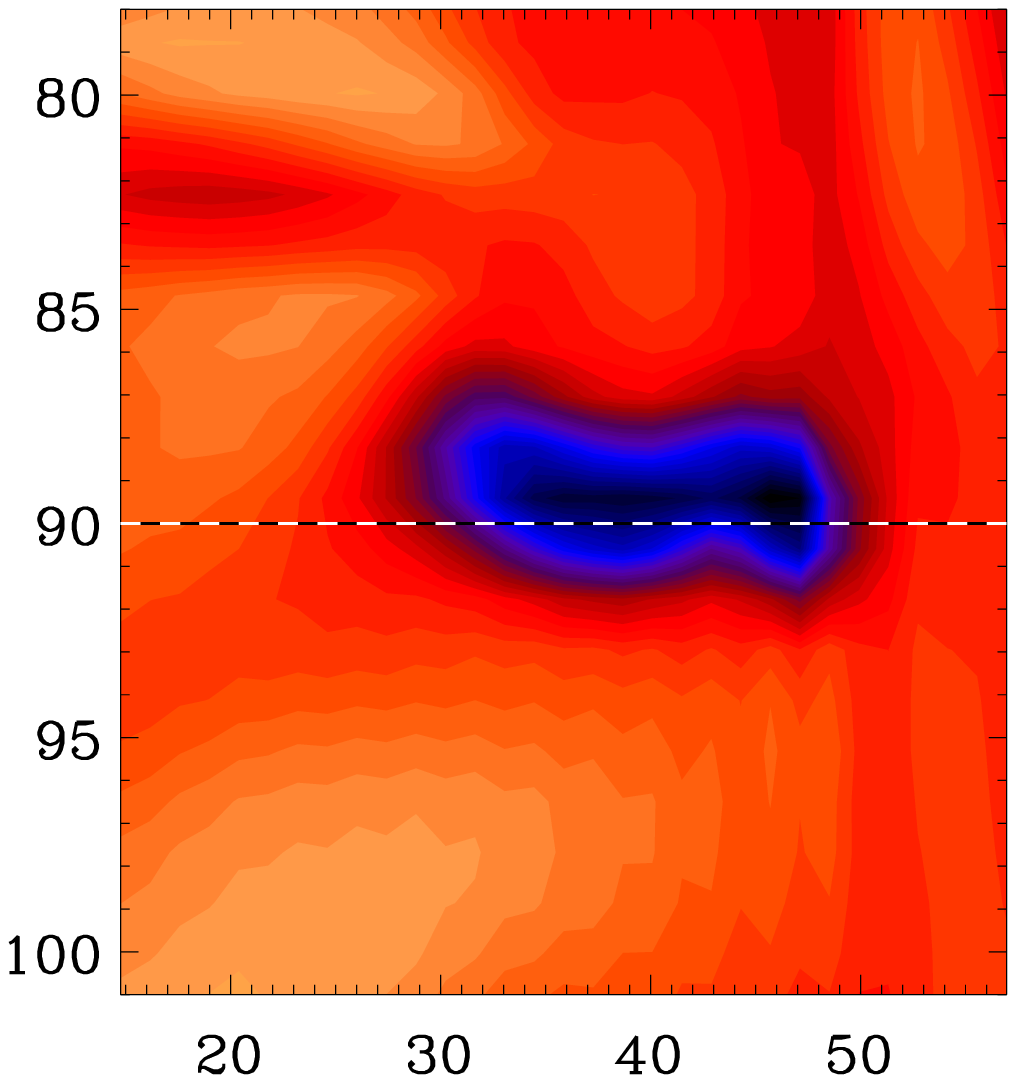}
\includegraphics[width=0.275\textwidth]{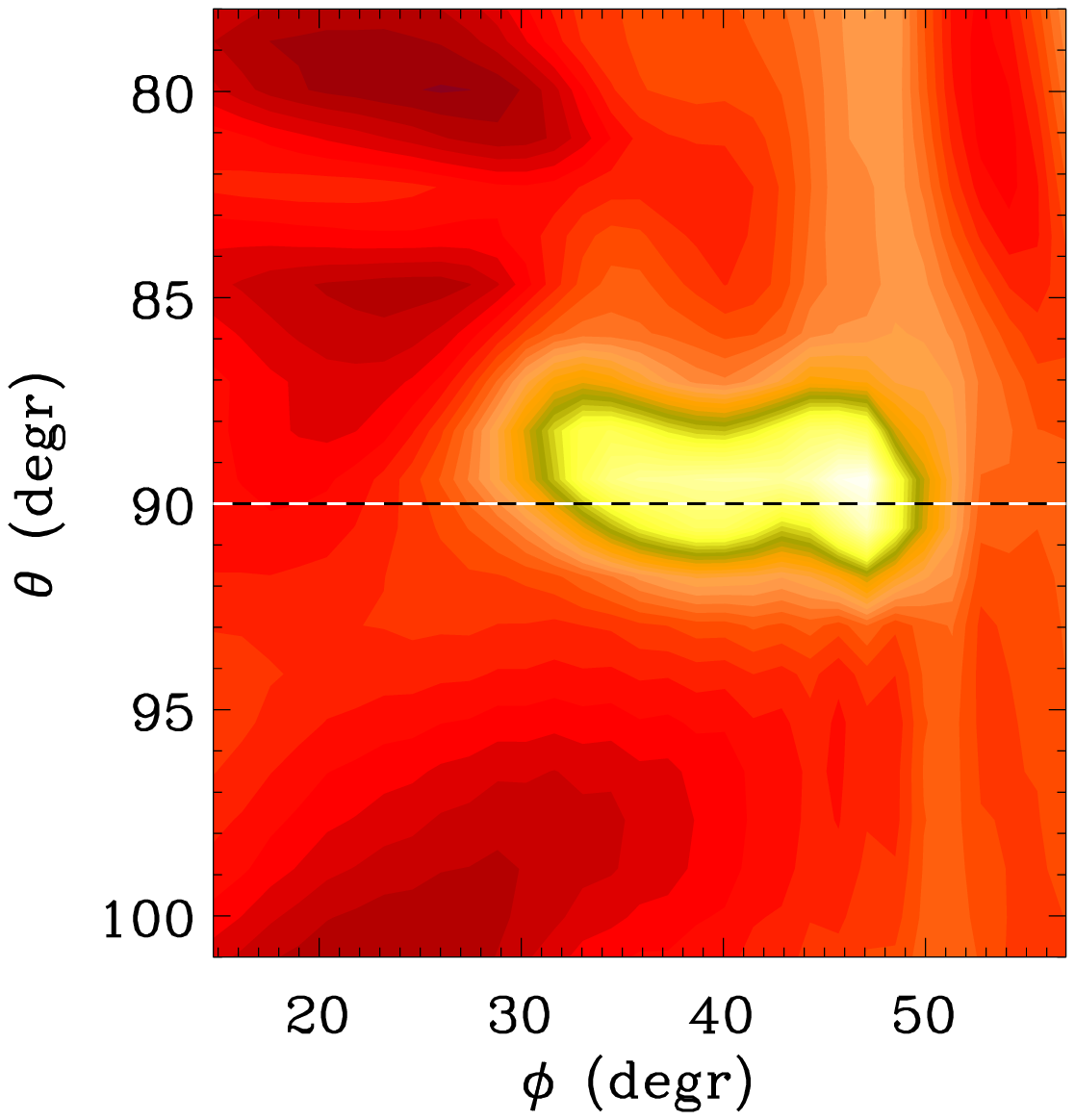}
\includegraphics[width=0.275\textwidth]{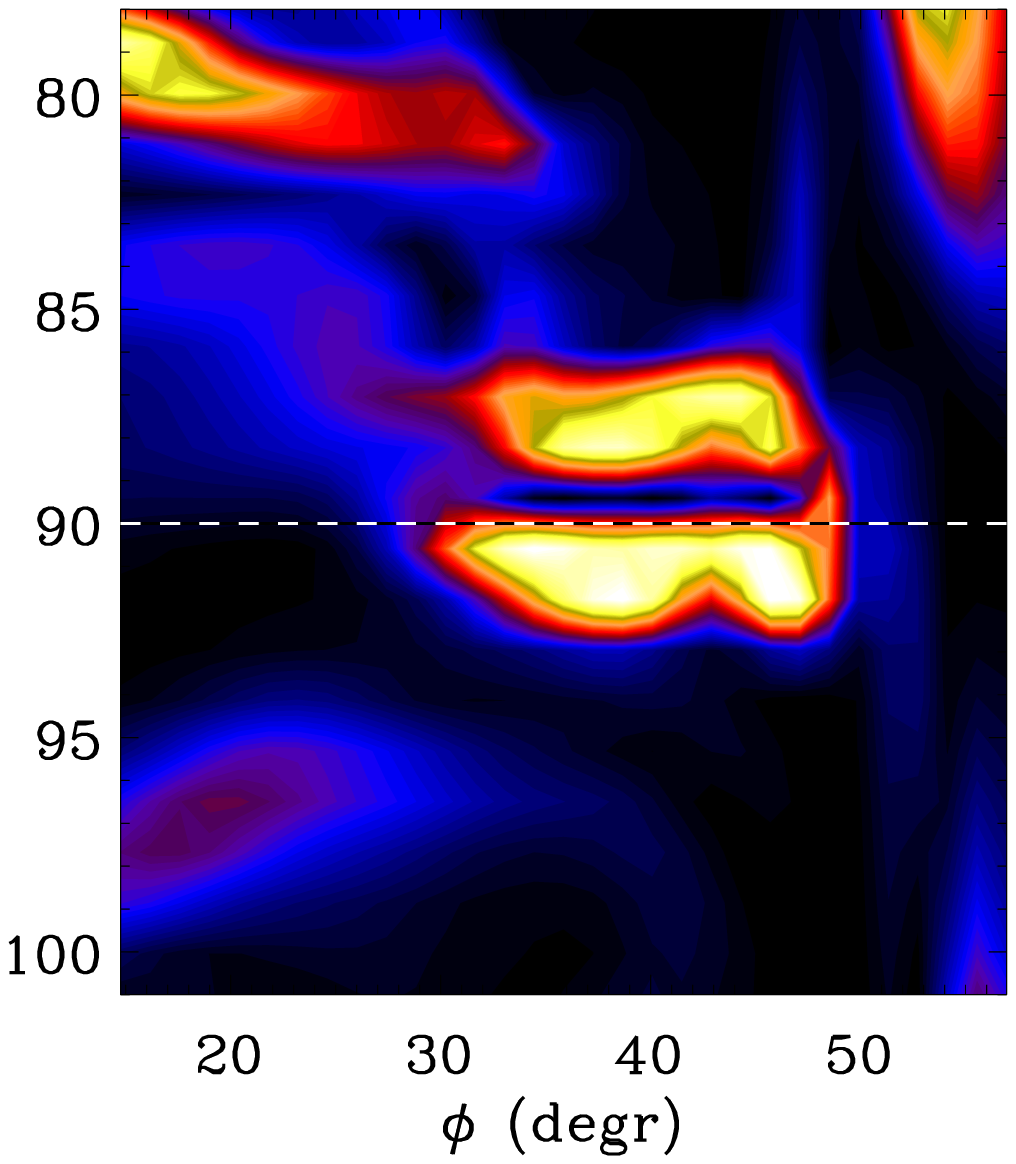}
\includegraphics[width=0.275\textwidth]{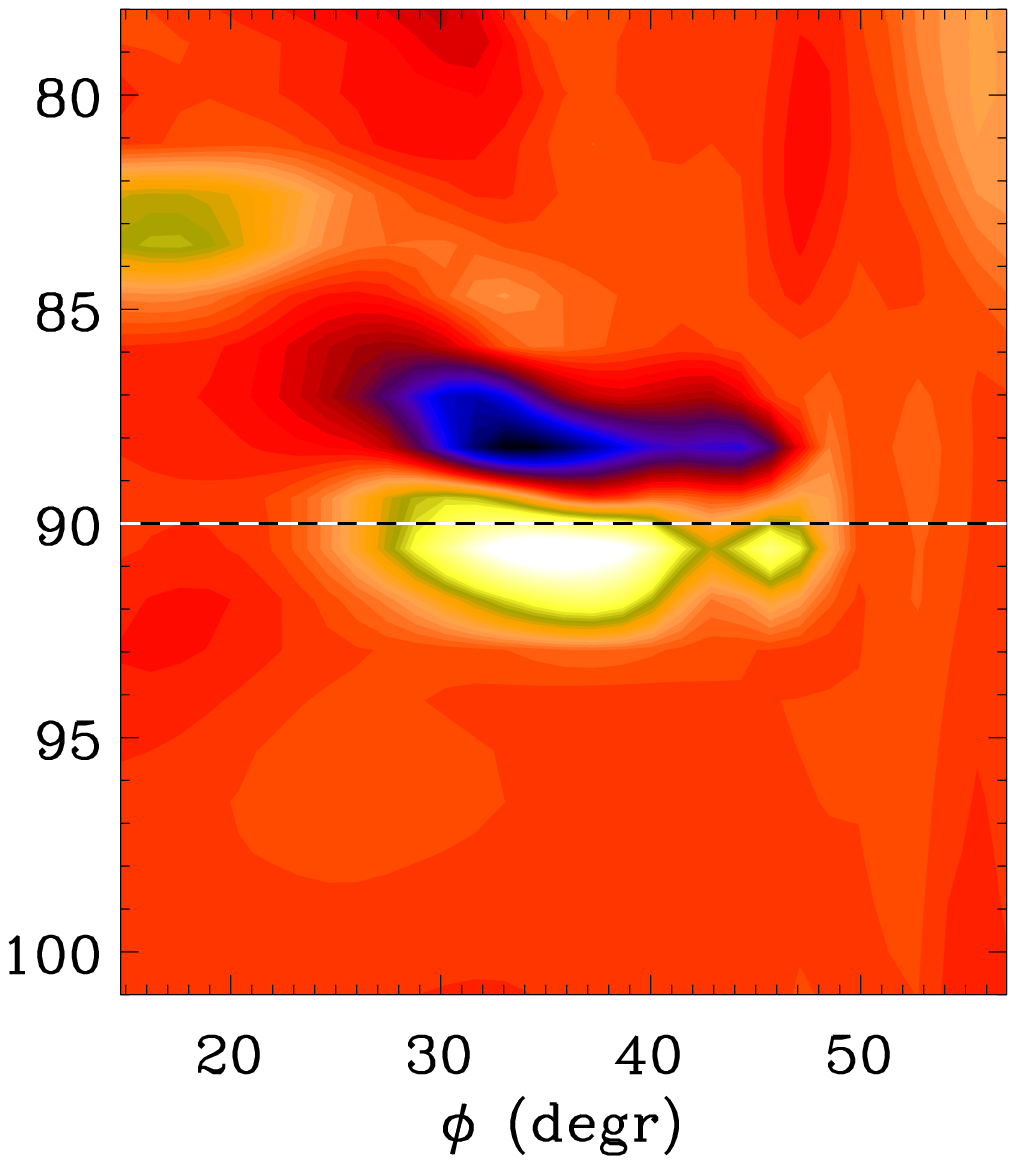}
\end{center}\caption[]{
Different properties of the ejection in the $\theta$,$\phi$ plane at
the surface ($r=\Rsun$) at $t/\tau=2204$,
taken from Run~A5.
{\it Upper row, left panel:} Contours of $A_r$ are shown together
with a color-scale representation of $B_r$; dark
blue stands for negative and light yellow for positive values.
The contours of $A_r$ correspond to field lines of that part of $\BB$
that is solenoidal in the $\theta$,$\phi$ plane.
Solid lines represent clockwise oriented magnetic field lines and dotted lines
counter-clockwise ones.
{\it Middle panel: } The arrows show $(U_{\theta},U_{\phi})$
and colors show $U_r$ (blue corresponds to downflows).
{\it Right panel:} Color-scale representation of the density $\rho$; dark
blue stands for low and light yellow for high values.
{\it Lower row, left panel:} Color-scale representation of specific
entropy $s$; dark
blue stands for low and light yellow for high values.
{\it Middle panel: } Color-scale representation of the current density
squared $\JJ^2$; dark
blue stands for low and light yellow for high values.
{\it Right panel:} current helicity $\JJ\cdot\BB$, color-scale as in
\Fig{jb}.
The dashed line indicates the equator at $\theta=\pi/2=90^{\circ}$.
}
\label{eje_pt}
\end{figure*}

The ejection causes also a strong variation in the density.
If the time-averaged density profile is subtracted from instantaneous
ones, the density fluctuations are obtained. 
After removing the density stratification one obtains
$\Delta\overline{\rho(t)}=\overline{\rho(t)}-\bra{\overline{\rho}}_t$.
We plot these density fluctuations, $\Delta\overline{\rho(t)}$,
in the right column of \Fig{eje} to visualize the effect of the
ejection on the density.
The density in the ejection is much lower than in the rest of the
coronal part.
However, the density variations are also associated with fluctuations
in the specific entropy ($\Delta s/c_{\rm p}\approx0.01$), which suggests
that thermal buoyancy also plays a role.
One interpretation could be that the strong magnetic field reduces
the density to achieve total pressure equilibrium and the ejection rises
partly because of magnetic buoyancy.
Such an effect is also seen by inspecting other ejections.

To characterize the emergence we plot different properties of the
ejection in the $\theta$,$\phi$ plane; see \Fig{eje_pt}.
The magnetic field shows a strong concentration in its radial and azimuthal
components.
The concentration is associated with a downflow in spite of it being a
low-density region.
It is interesting to note that in this case the gas velocity
does not reflect the actual pattern speed.
From the time evolution of the low-density region shown in Fig.~\ref{eje}, 
we know that this region is moving radially upwards in a way that is consistent
with a motion expected from buoyancy forces.
In particular, the specific entropy has a high value in this region.
In visualizations of the current density,
we see the formation of two current sheets.
This leads to two current helicity regions of opposite sign.

When discussing coronal ejections, one is usually interested in the
plasma $\beta$ parameter to characterize the corona.
In our simplified coronal part, the plasma $\beta$ does not decrease with
radius, but it stays rather high, which is due to the
low magnetic field strength, especially in the coronal part,
even though $\Brms^2/\Beq^2=$0.1--0.4 in the convection zone.
The time-averaged value is always above $5\times 10^4$, and is therefore not
comparable with the values in the solar corona,
where the plasma $\beta$ is very low because of the low density.
There the magnetic field can drag dense plasma from the lower corona
to its upper part.
In our simulations the density stratification of the convection zone
is much lower than in the Sun. 
Therefore, the density in the corona in our model is much higher and
is closer to the density of the photosphere or the chromosphere.
A rising magnetic flux tube has formed a low-density region in its
interior due to a higher magnetic pressure.
As the tube rises further into the coronal part, the density inside
the tube is still lower than that outside because the coronal density is
rather high in our model.

The simplification of a high plasma $\beta$ corona might not be
suitable to describe properly the mass flux of the plasma dragged by
the magnetic field of the CME in the corona.
However, the early work of \cite{MBS88}, 
\cite{OS93}, and \cite{W08} has shown that an
isothermal force-free approach (not to be confused with force-free
magnetic equilibria) can describe the coronal magnetic field and even
plasmoid ejections rather well.
Note that in those papers the pressure gradient term was omitted, just
like in the coronal part of WB.
How important this really is remains unclear, because the pressure
gradient term was not omitted in the work of WBM,
which still showed ejections similar to those of WB.
It would therefore be useful to compare our present model with one where
the pressure gradient term is ignored in the coronal part, just like in WB.

\begin{figure}[t!]
\begin{center}
\includegraphics[width=0.49\columnwidth]{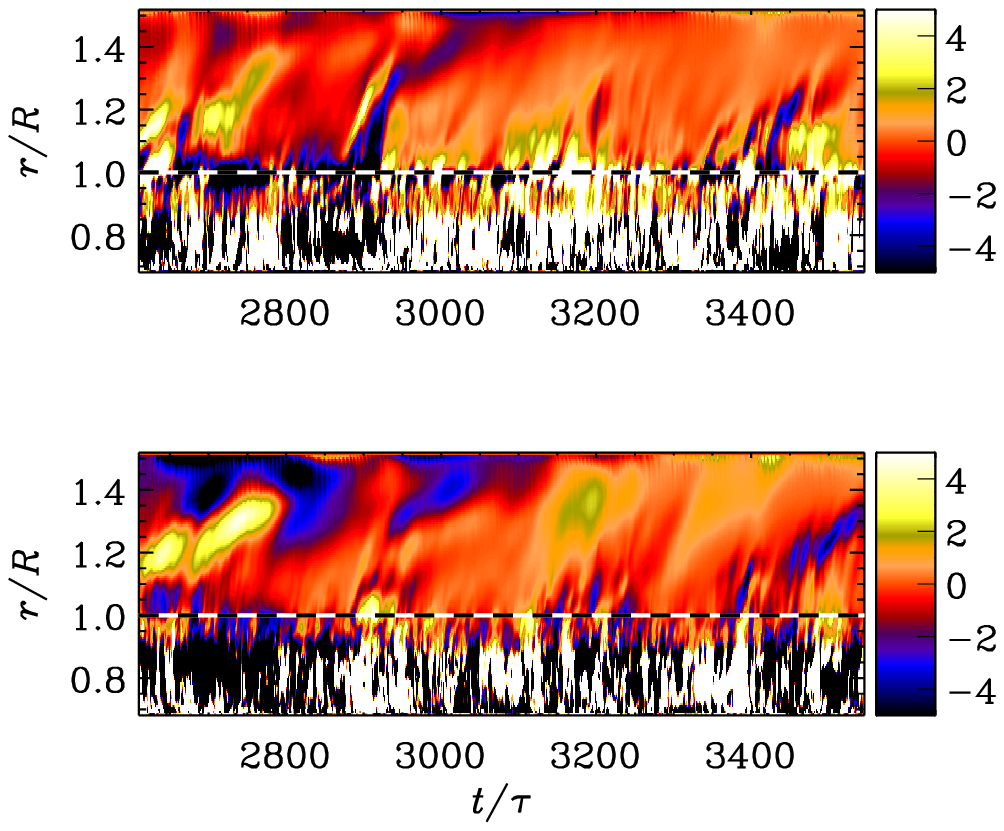}
\includegraphics[width=0.49\columnwidth]{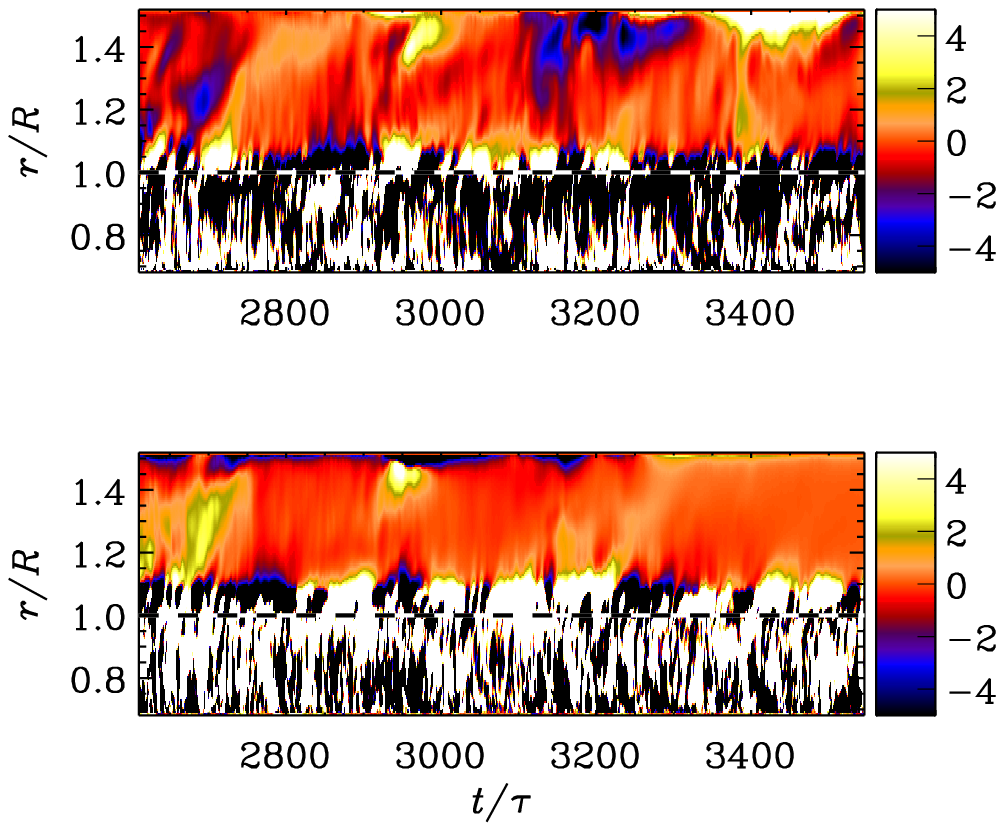}
\end{center}\caption[]{
Recurrence of ejections shown by plotting the dependence of the dimensionless ratio
$\mu_0 R\,\overline{\JJ\cdot\BB} / \bra{\overline{\BB^2}}_t$
on time $t/\tau$ and radius $r$ in terms of the solar radius,
taken from Run~A5.
The top panels show a narrow band in $\theta$ in the northern
hemisphere and the bottom ones in the southern hemisphere.
We have also averaged in latitude from
$4.1^{\circ}$ to $19.5^{\circ}$ (left panel) and $32.5^{\circ}$ to
$45.5^{\circ}$ (right).
Dark blue shades represent negative and light yellow positive values.
The dashed horizontal lines show the location of the surface at $r=\Rsun$.
Recurrence of ejections shown by plotting the dependence
}
\label{pjb}
\end{figure}

The ejection seen in \Figss{jb}{eje_pt} is not a single
event---others follow in a recurrent fashion.
However, the periodicity is not as clear as in previous
work (WB,WBM).
For Run~A5, for example, we observe around five ejections during
a time interval of about $1000$ turnover times.
A clearer indication for the recurrence of the ejections can be seen in
\Fig{pjb}, where the normalized current density is averaged over two
narrow latitude bands in each hemisphere.
The slope of structures in the outer parts in these $rt$ diagrams gives
an indication about the ejection speed $V_{\rm ej}$ which turns out to be
around one solar radius in 200--250 turnover times.
This translates to $V_{\rm ej}/\urms\approx0.1$, which is somewhat
less than the values 0.2--0.5 found for the simulations of WBM.
However, the mechanism which sets the time scale of ejections is
at present still unclear.

Given that gravity decreases with radius, there is in principle the
possibility of a radial wind with a critical point at $r_*=GM/2\cs^2$
\citep{Cho98}, which would be at $r_*=9.3\,\Rsun$, i.e.\ well outside our coronal part.
Because of this and the fact that
we use closed boundary conditions with no mass flux out
of the domain, no such wind can occur in our simulations.
Using a boundary condition that would allow a mass flux in the radial
direction could change the speed and the ejection properties
significantly.
Including a solar-like wind in a model can have two major effects, which
require a much higher amount of computational resources.
The radial variation of gravity applied in these simulations implies
the presence of a critical point rather close
to the surface of the convection zone.
Therefore, if a wind were to develop,
the resulting velocity in the
convection zone would be too high for a dynamo to develop; the magnetic
field would be blown out too quickly.
Using instead a more realistic profile for the solar wind with a
position of the critical point around $r_*=10\Rsun$, the corresponding
density stratification would be too strong to be stably resolved.

\section{Conclusions}
\label{concl}

In the present paper we have presented an extension of
the two-layer approach of WB and WBM by including a
self-consistent rotating convection zone into the model.
We find a large-scale magnetic field generated by the convective
turbulent motion in the convection zone.
At moderate rotation rates, for a Coriolis number larger than 3, we
obtain a differential rotation pattern showing super-rotation,
i.e., an equator rotating faster than the poles.
The dynamo solutions we find are different and some of them have a
periodic oscillatory behavior, where the large-scale magnetic field does
not change sign; only the strength is varying.
At the maxima, the velocity is suppressed due to the backreaction via
the Lorentz force.
Small-scale magnetic structures seem to show an equatorward migration
near the equator and a poleward one near the poles.

Using a convectively driven dynamo complicates the generation of
ejections into a coronal part due to lower relative kinetic helicity.
However, it was possible to produce ejections in two of the runs.
The shape and the bipolar helicity structure are comparable with
those of WBM.
Due to the relatively high plasma $\beta$ in the outer parts of
our model (compared with the solar corona), the ejections produce
local minima of density which are carried along and ejected out to the
top of the domain.
The ejections occur recurrently, but not clearly periodically, which is
similar to the Sun.

Note that our results have to be interpreted cautiously, given the use of a
simplistic solar atmosphere. 
We neglect the effects of high temperature and low plasma $\beta$.
However, we feel that the mechanism of emergence of magnetic
structures driven by dynamo action from self-consistent convection may
not strongly depend on these two conditions.
This suggestion has to be proven in more detail in forthcoming work.

An extension of the present work would require a detailed parameter study of
cause and properties of the ejections.
This also includes an advanced model for the solar corona with a lower
plasma $\beta$ and more efficient convection, which has a
stronger stratification and is cooled by radiation.
Another important aspect would be the generation of a self-consistent
solar wind which supports and interacts with the ejections.

%
 \begin{acks}
The authors thank Hardi Peter for discussion of
the dynamics and rotation behavior of the solar corona.
We also thank the anonymous referee for many useful suggestions.
We acknowledge the allocation of computing resources provided by the
Swedish National Allocations Committee at the Center for
Parallel Computers at the Royal Institute of Technology in
Stockholm, the National Supercomputer Centers in Link\"oping and the
High Performance Computing Center North in Ume\aa.
Part of the computations have been carried out in 
the facilities hosted by the CSC  -- IT Center for Science in Espoo, Finland, 
which are financed by the Finnish ministry of education.
This work was supported in part by
the European Research Council under the AstroDyn Research Project No.\ 227952, 
the Swedish Research Council Grant No.\ 621-2007-4064, and the Academy 
of Finland grants 136189, 140970 (PJK) and 218159, 141017 (MJM) as well
as the HPC-Europa2 project, funded by the European Commission - DG
Research in the Seventh Framework Programme under grant agreement No.\ 228398.

 \end{acks}
%
%
%

\end{article}
\end{document}